\documentclass[a4paper,fleqn,usenatbib,useAMS]{mnras}

\usepackage{graphicx}	
\usepackage{amsmath}	
\usepackage{amssymb}	
\usepackage{multicol}        
\usepackage{bm}		
\usepackage{pdflscape}	

\newcommand{\angstrom}{\text{\normalfont\AA}}
\newcommand{\RN}[1]{%
  \textup{\uppercase\expandafter{\romannumeral#1}}%
}

\usepackage[T1]{fontenc}
\usepackage{ae,aecompl}

\usepackage{newtxtext,newtxmath}

\usepackage{multirow}

\title[Two submm-bright protoclusters]{Two sub-millimetre bright protoclusters bounding the epoch of peak star-formation activity}

\author[K. M. Lacaille et al.]{Kevin M. Lacaille,$^{1,2}$\thanks{Contact e-mail: \href{mailto:kevinlacaille@dal.ca}{kevinlacaille@physics.mcmaster.ca}} Scott C. Chapman,$^{2,3}$ Ian Smail,$^{4}$ C. C. Steidel,$^{5}$ A. W. Blain,$^{6}$ 
\newauthor{ J. Geach,$^{7}$ A. Golob,$^{8}$ M. Gurwell,$^{9}$ R. J. Ivison,$^{10,11}$ N. Reddy,$^{12}$ M. Sawicki$^{8}$}\\
$^{1}$Department of Physics and Astronomy, McMaster University, Hamilton, ON L8S 4M1 Canada\\
$^{2}$Department of Physics and Atmospheric Science, Dalhousie University, Halifax, NS, B3H 4R2, Canada\\
$^3$Institute of Astronomy, Madingley Road, Cambridge, CB3 0HA, UK\\
$^{4}$Centre for Extragalactic Astronomy, Department of Physics, Durham University, South Road, Durham DH1 3LE \\
$^{5}$California Institute of Technology, MS 249-17, Pasadena, CA 91125, USA \\
$^6$Department of Physics \& Astronomy, University of Leicester, Leicester. LE1 7RH, UK\\
$^7$Centre for Astrophysics Research, University of Hertfordshire, Hatfield, Hertfordshire, AL10 9AB, UK\\
$^{8}$Department of Astronomy and Physics, Saint Mary's University, 923 Robie Street, Halifax, NS, B3H 3C3, Canada\\
$^9$Harvard-Smithsonian Center for Astrophysics, 60 Garden Street, Cambridge, MA 02138, USA\\
$^{10}$European Southern Observatory, Karl Schwarzschild Strasse 2, D-85748 Garching, Germany\\
$^{11}$Institute for Astronomy, University of Edinburgh, Royal Observatory, Blackford Hill, Edinburgh, EH9 3HJ, UK\\
$^{12}$Department of Physics \& Astronomy, UC Riverside, 900 University Ave, Riverside, CA 92521, USA\\}

\pubyear{2019}
\begin{document}
\label{firstpage}
\pagerange{\pageref{firstpage}--\pageref{lastpage}}
\maketitle

\begin{abstract}
We present  James Clerk Maxwell Telescope Submillimetre Common-User Bolometer Array 2 (SCUBA-2) 850 \& 450 $\mu$m observations ($\sigma_{850}\sim0.5$ mJy, $\sigma_{450}\sim5$ mJy) 
of the HS1549+19  and HS1700+64 survey fields containing two of the largest known galaxy over-densities at $z=2.85$ and $2.30$, respectively.
We detect 56 sub-millimetre galaxies (SMGs) with SNR $> 4$ over $\sim 50$ arcmin$^2$ at 850 $\mu$m with flux densities of 3 -- 17 mJy. 
The number counts indicate over-densities in the $3$-arcmin diameter core region ($\sim 1.5$ Mpc at $z=2.5$) of $6^{+4}_{-2}\times$ (HS1549) and $4^{+6}_{-2}\times$ (HS1700) compared to blank field surveys.
Within these core regions, we spectroscopically confirm that approximately one third of the SMGs lie at the protocluster redshifts for both HS1549 and HS1700.
We use statistical identifications of other SMGs in the wider fields to constrain an additional four candidate protocluster members in each system. 
We combine multi wavelength estimates of the star-formation rates (SFRs) from Lyman-break dropout- and narrowband-selected galaxies, and the SCUBA-2 SMGs, to estimate total SFRs of 12,$500\pm2800$ M$_\odot$ yr$^{-1}$ ($4900\pm1200$ M$_\odot$ yr$^{-1}$) in HS1549 (HS1700),
and SFR densities (SFRDs) within the central 1.5-Mpc diameter of each protocluster to be $3000\pm900$ M$_\odot$ yr$^{-1}$ Mpc$^{-3}$ ($1300\pm400$ M$_\odot$ yr$^{-1}$ Mpc$^{-3}$) in the HS1549 (HS1700) protocluster, $\sim10^4\times$ larger than the global SFRDs found at their respective epochs, due to the concentration of star-forming galaxies in the small volume of the dense cluster cores.
Our results suggest centrally concentrated starbursts within protoclusters may be a relatively common scenario for the build up of mass in rich clusters assembling at $z\gtrsim2$.
\end{abstract}




\section{Introduction}



In the local Universe, galaxy clusters represent some of the densest 
environments where mostly passiv e galaxies reside, 
with the most massive elliptical galaxies known residing  at their centres 
(e.g., \citealt{Bower1992}; \citealt{Ellis1997}; \citealt{Stanford1997}; \citealt{VanDokkum2001}; \citealt{Holden2005}).
Therefore, massive galaxy clusters must have formed the bulk of their member galaxies' stellar mass at early times (e.g., \citealt{Archibald2001}; \citealt{Hopkins2006}).
At high-redshift, a significant amount of enhancement of star formation is observed in galaxies residing in candidate protoclusters -- the progenitors of massive local galaxy clusters  
(e.g., \citealt{Steidel1999b}; \citealt{Geach2005}; \citealt{Chapman2009}; \citealt{Capak2011}; \citealt{Muzzin2011}; \citealt{Clements2014}; \citealt{Smail2013};
\newpage 
\noindent
\citealt{Webb2013}; \citealt{Casey2015}; \citealt{Ma2015}; \citealt{Daddi2017}; \citealt{Umehata2016}; \citealt{Stach2017,Stach2018a}; \citealt{Strazzullo2018}).
Thus, there is some evidence that at high redshift, galaxy evolution is accelerated at early times in regions of strong over-density. 

Lyman break galaxies (LBGs) have been used to identify 
 some of the densest known protoclusters at $z>2$, both from  blind redshift surveys  (e.g., \citealt{Steidel1998, Steidel1999b, Steidel2005, Steidel2011}) and from pointed followup to other beacons of early structures such as luminous Radio Galaxies (e.g., \citealt{Kurk2000}; \citealt{Pentericci2000}; \citealt{Hayashino2004}).
Sub-millimetre (submm) surveys of high-redshift galaxy protoclusters traced by LBGs or Radio Galaxies have shown evidence of enhanced star-formation rates (SFRs) within protoclusters (e.g., 
\citealt{Stevens2003a}; \citealt{Chapman2009};  \citealt{Dannerbauer2014a}; \citealt{Capak2011}; \citealt{Smail2013}; \citealt{Casey2015}). 
Recent Submillimetre Common-User Bolometer Array 2 (SCUBA-2) observations of protoclusters have shown that sub-millimetre galaxies (SMGs) in these systems tend to be centrally concentrated and reside in over-dense environments, suggesting rapid galaxy evolution in high-density environments in the early Universe (e.g., \citealt{Casey2015}; \citealt{Ma2015}; \citealt{Umehata2015}; \citealt{Alexander2016}; \citealt{Stach2017}).

In this paper, we present deep Sub-Millimetre Common-User Bolometer Array 2 (SCUBA-2, \citealt{Holland2013}) observations of two protocluster fields that were uncovered through the identification of redshift over-densities of LBGs: HS1549+19 (\citealt{Steidel2011}) and HS1700+64 (\citealt{Steidel2005}) (hereafter HS1549 and HS1700). These target fields were chosen from the 15 survey fields of the Keck Baryonic Structure Survey (KBSS -- e.g., \citealt{Rudie2012}), 
a field galaxy survey at $z > 2$ aimed at understanding the relation of galaxies to their circumgalactic medium. The HS1549 and HS1700 fields contain the strongest redshift over-densities found in the survey. In HS1549, the galaxy over-density is even larger than that of HS1700, exhibiting  a factor of 13 times that of the field in a volume of $\sim 5000$ Mpc$^3$, and has almost 10$\times$ the surface density of Ly$\alpha$ emitters (LAEs) compared to the average among 20 fields covered to a similar depth, representing one of the richest fields of Ly$\alpha$-selected objects ever observed, at any redshift (\citealt{Steidel2011} \& in preparation). 

In \S~\ref{sec:obs} we summarize the details of our submm, and optical/infrared (IR) observations for each  field.
In \S~\ref{sec:results} we present our submm detections, quantify the submm number counts, and present candidate IR counterparts. 
In \S~\ref{sec:discuss} we determine the cluster membership for a selection of submm detections using spectroscopic and photometric redshifts, and then estimate the total star-formation rate for each protocluster.
Finally, we present our conclusions in \S~\ref{sec:conclusions}.
Throughout this paper, we use the Vega magnitude system, and assume a flat concordance cosmology with $\rm (\Omega_m,\Omega_\Lambda,H_0) = (0.3,0.7,70$ km\,s$^{-1}$\,Mpc$^{-1}$). 
 In this cosmology, the Universe is 2.9 and 2.3 Gyr old and 1.0$^{\prime\prime}$ corresponds to 8.4 and 8.0 kpc in physical length at $z=2.3$ and 2.85, respectively.


\section{Observations and Data Reduction}
\label{sec:obs}

%
This work is primarily based on submm imaging using SCUBA-2 mounted on the James Clerk Maxwell Telescope (JCMT). 
We  also make use of existing 
multi-wavelength imagery and photometry within our fields from various archives (e.g. {\it Spitzer} Infrared Array Camera (IRAC) and Multiband Imaging Photometer (MIPS), and \emph{Hubble Space Telescope} (\emph{HST})) (see \S~\ref{sec:Ancillary}).

\subsection{SCUBA-2 Observations}

Observations were conducted in Band 1-2 weather conditions ($\tau_{225\text{ GHz}}$ $\sim$ 0.04--0.07) over four nights between 26$^\text{th}$ May 2012 and 20$^\text{th}$ April 2013 totalling 9.8 hours of on-sky integration time (HS1549), and over five nights between 10$^\text{th}$ April and 18$^\text{th}$ September 2013, totalling 19.7 hours of on-sky integration (HS1700). The mapping centre of the SCUBA-2 H1549+19 field was $(\text{RA},\text{Dec}) = (15^{\text{h}}51^{\text{m}}53^{\text{s}}, +19^{\mathrm{o}}11^\prime04^{\prime\prime})$, and the mapping centre for HS1700+64 field was $(\text{RA},\text{Dec}) = (17^{\text{h}}01^{\text{m}}00.6^{\text{s}}, +64^{\mathrm{o}}12^\prime09^{\prime\prime})$, and both fields are surveyed over $\sim 155$ arcmin$^2$. 
The mapping centres for the SCUBA-2 observations correspond to the position of the most luminous QSOs in the corresponding KBSS field.
A standard 3-arcmin diameter \textsc{daisy} mapping pattern was used (e.g., \citealt{Kackley2010a}), which keeps the pointing centre on one of the four SCUBA-2 sub-arrays at all times during exposure.
The coverage of the SCUBA-2 observations in HS1549 is wider than the optical coverage, and thus the SMG distribution is wider than the LBG distribution in the field.

Individual 30 min scans are reduced using the dynamic iterative map-maker of the \textsc{smurf} package (\citealt{Jenness2013}; \citealt{Chapin2013}). Maps from independent scans are co-added in an optimal stack using the variance of the data contributing to each pixel to weight spatially aligned pixels. Finally, since we are interested in (generally faint) extragalactic point sources, we apply a beam matched filter to improve point source detectability, resulting in a map that is convolved with an estimate of the 450 $\mu$m beam. The average exposure time over the nominal 3-arcmin \textsc{daisy} mapping region (in practice there is usable data beyond this) is approximately 9 ksec per $2^{\prime\prime}\times2^{\prime\prime}$ pixel for HS1549 and 18 ksec per $2^{\prime\prime}\times2^{\prime\prime}$ pixel for HS1700.

The sky opacity at JCMT has been obtained by fitting extinction models to hundreds of standard calibrators observed since the commissioning of SCUBA-2 (\citealt{Dempsey2012}). These maps have been converted from pW to Jy using the standard flux conversion factors (FCFs) of FCF$_{450} = 491$ Jy beam$^{-1}$ pW$^{-1}$ and FCF$_{850} = 547$ Jy beam$^{-1}$ pW$^{-1}$, with effective 450 and 850 $\mu$m beam sizes of 10\arcsec and 15\arcsec (\citealt{Dempsey2013}), respectively.

To determine the accuracy of the absolute pointing accuracy of the 850 $\mu$m catalogue, we compare the coordinates of 850 $\mu$m sources (detected at $>4\sigma$, see Tables \ref{table:h1549} \& \ref{table:h1700}) with the closest matched counterparts at 24 $\mu$m (similar as to what is described in \citealt{Ma2015}). Upon removing 9 (7) sources in HS1549 (HS1700) (outliers, non-detections, or foreground galaxy contamination), the mean offset of the remaining 20 (20) HS1549 (HS1700) counterparts at 24 $\mu$m are $\Delta \text{RA} = 0.8^{\prime\prime} \pm 0.6^{\prime\prime}$ ($1.1^{\prime\prime} \pm 0.8^{\prime\prime}$) for HS1549 (HS1700) and $\Delta \text{Dec} = 0.4^{\prime\prime}\pm0.4 ^{\prime\prime}$ ($0.7^{\prime\prime}\pm0.6 ^{\prime\prime}$) for HS1549 (HS1700). Hence, there are no significant offsets between the SCUBA-2 850 $\mu$m and infrared astrometry

The variance map was derived from the pixel integration (e.g., \citealt{Koprowski2015}). The RMS within the central 3-arcmin diameter regions are 0.6 mJy beam$^{-1}$ and 4.6 mJy beam$^{-1}$ (HS1549) and 0.45 mJy beam$^{-1}$ and 4.3 mJy beam$^{-1}$ (HS1700) at 850 $\mu$m and 450 $\mu$m, respectively. Our depths reached at both 850 $\mu$m and 450 $\mu$m and smaller beam sizes allow us to probe sources at these redshifts more effectively than the confusion limited \emph{Herschel} maps (e.g., \citealt{Kato2016})

\subsection{SMA Observations}
\label{sec:SMA}

The SMA was used to resolve the central elongated submm source \emph{1549.1} (Table \ref{table:h1549}). The details of these observations are given in Chapman et al. (in preparation), and are briefly discussed  here. A mosaic of three SMA pointings were obtained on August 15$^\text{th}$ and September 7$^\text{th}$, 2013 in the compact configuration (synthesized beam size $\sim2^{\prime\prime}$ with natural weighting) in good weather ($\tau_{225 \text{ GHz}} \sim 0.08$) with a total on-source integration time of approximately 12 hr through the three tracks, yielding a final RMS$_{870}$ of 0.8 mJy beam$^{-1}$ in the central regions. The astrometric uncertainties are $\Delta\alpha=0.24^{\prime\prime}$ (0.20$^{\prime\prime}$ systematic; 0.13$^{\prime\prime}$ statistical) and $\Delta\delta=0.22^{\prime\prime}$ (0.19$^{\prime\prime}$ systematic; 0.10$^{\prime\prime}$ statistical). The data were calibrated using the {\sc mir} software package (\citealt{Scoville1993}), modified for the SMA. 

\subsection{Gemini spectroscopy}
\label{sec:GNIRS}


Near-infrared spectra of thirteen of our identifications for proto-cluster SMGs were obtained using the cross-dispersed mode of the Gemini Near-Infrared Spectrograph (GNIRS) on the Gemini North 8.1 m telescope (\emph{1700.4}, \emph{1700.5\_2}, \emph{1700.7\_1,\_2,\_3,\_4}, \emph{1700.8\_1}, \emph{1700.9}, \emph{1700.14}, \emph{1700.15\_1,\_2}, and \emph{1700.16}, Table \ref{table:h1700}).  This GNIRS configuration provides a continuous spectral coverage from $\sim0.84$ to 2.48 $\mu$m at a spectral resolution of $\sim1500$ with a spatial scale of 0.15$^{\prime\prime}$ per pixel. The $1.0^{\prime\prime} \times 7^{\prime\prime}$ slit centred on the peak of the 2.2 $\mu$m emission. The seeing during the  observations was $\sim$0.8$^{\prime\prime}$ as measured from the telluric A1V standard HIP 58616, observed right before the science targets at a similar airmass. The strongest emission features we expect to observe are $[\text{O}\RN{2}]$ 3727, $[\text{O}\RN{3}]$ 5007, [\text{N}\RN{2}] 6548, [\text{N}\RN{2}] 6583, and H$\alpha$.

The observations used an object-sky-sky-object pattern, with the sky position 50$^{\prime\prime}$ away from the science target, free of extended emission or background stars. Eight individual on-source integrations of 240 seconds each were carried out.

The spectral reduction, extraction, and wavelength and flux calibration procedures were performed using version 1.9 of the
``\textsc{xdgnirs}'' code detailed in \cite{Mason2015}. Briefly, the processing consists of removing cosmic ray-like features, dividing by a flat field, subtracting sky emission, and rectifying the tilted, curved spectra. Wavelength calibration is achieved using argon arc spectra, and then a spectrum of each order is extracted, divided by a standard star to cancel telluric absorption lines, and roughly flux-calibrated using the telluric standard star spectrum. The pipeline merges the different spectral orders for each extraction window into a single 1D spectrum from 0.84 $\mu$m to 2.48 $\mu$m. In all cases the agreement in flux between the overlapping regions of two consecutive orders was very good, and scaling factors of $< 3\%$ were necessary.
See \S~\ref{sec:NIR_spec} for the spectroscopic redshifts obtained.

\subsection{Ancillary Data}
\label{sec:Ancillary}

We also draw on previous imaging taken in these fields. 
For the HS1700 field we obtained additional data from:
The William Herschel 4.2m Telescope (WHT) on La Palma with the Prime Focus Imager (see \citealt{Steidel2004a}), 
{\it Spitzer}-IRAC images (see \citealt{Shapley2005}),
{\it Spitzer}-MIPS images (see \citealt{Reddy2006}),
{\it HST}/ACS images (see \citealt{Peter2007}),
{\it HST}/WFC3 F160W images (see \citealt{Law2012}),
Palomar/WIRC for near-IR photometry (see \citealt{Erb2006}),
and Ly$\alpha$ and H$\alpha$ narrow-band images from Palomar LFC (see \citealt{Steidel2011} and \citealt{Bogosavljevic2010}).
Additionally, for the HS1549 field we obtained data from:          
\emph{Spitzer}-IRAC 24 $\mu$m images (see \citealt{Reddy2010}),
\emph{Spitzer}-MIPS images  (see \citealt{Reddy2006}),
{\it HST}/F160W images (see \citealt{Law2012}),
Palomar/WIRC for near-IR photometry (see \citealt{Erb2006}),
and Ly$\alpha$ imaging from Keck/LRIS (see \citealt{Steidel2011} and \citealt{Mostardi2013}).
%
Finally, we obtain 250 $\mu$m, 350 $\mu$m, and 500 $\mu$m fluxes from \citealt{Kato2016} archival \emph{Herschel}-SPIRE maps for 21/27 SCUBA-2 sources in the HS1700 field. We estimate the uncertainties for each flux measurement from the quadrature sum of the confusion noise and instrumental noise.

\begin{table*}
\caption{SCUBA-2 850 $\mu$m source catalogue of $>4\sigma$ sources in HS1549. Sources with 450 $\mu$m counterparts have been included for $>3\sigma$ if they reside within the 850 $\mu$m beam and $>2.5\sigma$ if they match their IR-counterpart.
Additionally, we present the IRAC counterparts for all SCUBA-2 identifications. The secure identifications at 4.5 $\mu$m or 24 $\mu$m with $P\leq0.05$ are shown in bold and tentative associations ($0.05<P\leq 0.10$) are presented in italics. SCUBA-2 IDs without any identifiable IRAC counterparts have been left out of the table.} 
\begin{center}
\begin{tabular}{ccccccccccc}
        \hline
        ID$_{850}$ & RA$_{850}$ & Dec$_{850}$ & S$_{850}$(SNR) & S$_{450}$(SNR) & ID$_\text{IRAC}$ & RA$_\text{IRAC}$ & Dec$_\text{IRAC}$ & $\theta$ & $P_\text{IRAC}$ & $z$\\
	~	&	(J2000)	&	(J2000)	&	(mJy)	&	(mJy)	&	~	&	(J2000)	&	(J2000)	&	(arcsec)	&	~&~\\
        \hline
        \hline
1\_1$^{a}$	&	$15^{\mathrm{h}}51^{\mathrm{m}}53.8^{\mathrm{s}}$	&	$+19^{\mathrm{o}}11'09.9''$	&	9.4$^a$(8.9)$^a$	&	15(3.2) & --  & --  & --  & --  & -- & 2.856$^{\dagger\|}$\\
1\_2$^{a}$	&	$15^{\mathrm{h}}51^{\mathrm{m}}53.2^{\mathrm{s}}$	&	$+19^{\mathrm{o}}10'59.1''$	&	5.6$^a$(5.3)$^a$	&	$<13.8$ & --  & --  & --  & --  & -- & 2.851$^{\mathsection\|}$\\
1\_3$^{a}$	&	$15^{\mathrm{h}}51^{\mathrm{m}}52.5^{\mathrm{s}}$	&	$+19^{\mathrm{o}}11'03.9''$	&	8.8$^a$(8.4)$^a$	&	30(6.3) & --  & --  & --  & --  & -- & 2.847$^{\dagger\|}$\\
\hline
2	&	$15^{\mathrm{h}}52^{\mathrm{m}}03.6^{\mathrm{s}}$	&	$+19^{\mathrm{o}}12'52.3''$	&	7.0(8.2)	&	19(3.0) & \textbf{2}	&	$\boldsymbol{+15^{\mathrm{h}}52^{\mathrm{m}}3.3^{\mathrm{s}}}$	&	$\boldsymbol{+19^{\mathrm{o}}12'51.5''}$	&	\textbf{3.31}	&			\textbf{0.042} & 2.85$^\ddagger$\\
\hline
\multirow{2}{*}{3}	&	\multirow{2}{*}{$15^{\mathrm{h}}51^{\mathrm{m}}49.6^{\mathrm{s}}$}	&	\multirow{2}{*}{$+19^{\mathrm{o}}10'45.2''$}	&	\multirow{2}{*}{4.9(7.7)}	&	\multirow{2}{*}{$<13.8$} & \emph{3\_1}	&	\emph{+15}$^h$\emph{51}$^m$\emph{49.4}$^s$	&	\emph{+19}$^o$\emph{10'40.8''}	&	\emph{5.04}	&					\emph{0.095} & \multirow{2}{*}{2.918$^\|$}\\
~&~&~&~&~&	3\_2	&	+$15^{\mathrm{h}}51^{\mathrm{m}}49.4^{\mathrm{s}}$	&	+$19^{\mathrm{o}}10'50.3''$	&	6.20	& 0.140 & ~\\
\hline
4	&	$15^{\mathrm{h}}52^{\mathrm{m}}03.5^{\mathrm{s}}$	&	$+19^{\mathrm{o}}10'02.8''$	&	7.2(7.7)	&	18(2.5) & \textbf{4}	&	$\boldsymbol{+15^{\mathrm{h}}52^{\mathrm{m}}03.4^{\mathrm{s}}}$	&	$\boldsymbol{+19^{\mathrm{o}}10'01.5''}$	&	\textbf{1.64}	&			\textbf{0.010} & 2.388$^\ddagger$\\
\hline
5	&	$15^{\mathrm{h}}52^{\mathrm{m}}15.0^{\mathrm{s}}$	&	$+19^{\mathrm{o}}10'17.7''$	&	8.3(7.5)	&	20(2.7)& --  & --  & --  & --  & -- &--\\
\hline
6	&	$15^{\mathrm{h}}52^{\mathrm{m}}12.9^{\mathrm{s}}$	&	$+19^{\mathrm{o}}11'18.9''$	&	7.5(7.3)	&	19(2.7)& --  & --  & --  & --  & --  &--\\
\hline
7	&	$15^{\mathrm{h}}51^{\mathrm{m}}30.7^{\mathrm{s}}$	&	$+19^{\mathrm{o}}10'56.7''$	&	7.4(7.0)	&	38(4.9)& --  & --  & --  & --  & --  &--\\
\hline
8	&	$15^{\mathrm{h}}51^{\mathrm{m}}37.3^{\mathrm{s}}$	&	$+19^{\mathrm{o}}09'14.0''$	&	5.9(6.6)	&	32(4.5) & \textbf{8}	&	$\boldsymbol{+15^{\mathrm{h}}51^{\mathrm{m}}37.1^{\mathrm{s}}}$	&	$\boldsymbol{+19^{\mathrm{o}}09'13.3''}$	&	\textbf{2.36}	&						\textbf{0.022} &--\\
\hline
9	&	$15^{\mathrm{h}}51^{\mathrm{m}}48.2^{\mathrm{s}}$	&	$+19^{\mathrm{o}}11'36.4''$	&	4.2(6.2)	&	$<13.8$ & \emph{9}	&	\emph{+15}$^h$\emph{51}$^m$\emph{48.0}$^s$	&	\emph{+19}$^o$\emph{11'39.3''}	&	\emph{4.13}	&						\emph{0.065} &--\\
\hline
10	&	$15^{\mathrm{h}}51^{\mathrm{m}}47.8^{\mathrm{s}}$	&	$+19^{\mathrm{o}}08'33.1''$	&	5.2(6.1)	&	$<13.8$ & \emph{10}	&	\emph{+15}$^h$\emph{51}$^m$\emph{47.5}$^s$	&	\emph{+19}$^o$\emph{08'33.2''}	&	\emph{4.27}	&						\emph{0.069} & $\neq2.85^\|$\\
\hline
11	&	$15^{\mathrm{h}}51^{\mathrm{m}}50.0^{\mathrm{s}}$	&	$+19^{\mathrm{o}}11'41.1''$	&	3.9(6.0)	&	$<13.8$ & \textbf{11}	&	$\boldsymbol{+15^{\mathrm{h}}51^{\mathrm{m}}49.9^{\mathrm{s}}}$	&	$\boldsymbol{+19^{\mathrm{o}}11'40.7''}$	&	\textbf{1.54}	&						\textbf{0.009} &--\\
\hline
\multirow{4}{*}{12}	&	\multirow{4}{*}{$15^{\mathrm{h}}51^{\mathrm{m}}52.0^{\mathrm{s}}$}	&	\multirow{4}{*}{$+19^{\mathrm{o}}13'48.9''$}	&	\multirow{4}{*}{5.4(5.9)}	&	\multirow{4}{*}{$<13.8$} & \emph{12\_1}	&	\emph{+15}$^h$\emph{51}$^m$\emph{52.0}$^s$	&	\emph{+19}$^o$\emph{13'53.4''}	&	\emph{4.56}	&					\emph{0.078} &--\\
~&~&~&~&~&	\textbf{12\_2}	&	$\boldsymbol{+15^{\mathrm{h}}51^{\mathrm{m}}52.0^{\mathrm{s}}}$	&	$\boldsymbol{+19^{\mathrm{o}}13'45.8''}$	&	\textbf{3.12}	&					\textbf{0.038} &--\\
~&~&~&~&~&	12\_3	&	+$15^{\mathrm{h}}51^{\mathrm{m}}51.8^{\mathrm{s}}$	&	+$19^{\mathrm{o}}13'43.8''$	&	5.59	&						0.115 &--\\
~&~&~&~&~&	12\_4	&	+$15^{\mathrm{h}}51^{\mathrm{m}}52.2^{\mathrm{s}}$	&	+$19^{\mathrm{o}}13'41.7''$	&	8.15	&						0.230 &--\\
\hline
\multirow{2}{*}{13}	&	\multirow{2}{*}{$15^{\mathrm{h}}51^{\mathrm{m}}45.7^{\mathrm{s}}$}	&	\multirow{2}{*}{$+19^{\mathrm{o}}11'15.7''$}	&	\multirow{2}{*}{4.2(5.7)}	&	\multirow{2}{*}{$<13.8$} & \textbf{13\_1}	&	$\boldsymbol{+15^{\mathrm{h}}51^{\mathrm{m}}45.7^{\mathrm{s}}}$	&	$\boldsymbol{+19^{\mathrm{o}}11'16.1''}$	&	\textbf{0.87}	&					\textbf{0.003} &--\\
~&~&~&~&~& 13\_2	&	+$15^{\mathrm{h}}51^{\mathrm{m}}45.2^{\mathrm{s}}$	&	+$19^{\mathrm{o}}11'16.2''$	&	6.96	&						0.173 &--\\
\hline
14	&	$15^{\mathrm{h}}51^{\mathrm{m}}33.2^{\mathrm{s}}$	&	$+19^{\mathrm{o}}08'03.9''$	&	7.0(5.3)	&	$<13.8$ & --  & --  & --  & --  & --  &--\\
\hline
\multirow{3}{*}{15}	&	\multirow{3}{*}{$15^{\mathrm{h}}51^{\mathrm{m}}57.0^{\mathrm{s}}$}	&	\multirow{3}{*}{$+19^{\mathrm{o}}11'34.3''$}	&	\multirow{3}{*}{3.3(4.9)}	&	\multirow{3}{*}{16(3.1)} & \textbf{15\_1}	&	$\boldsymbol{+15^{\mathrm{h}}51^{\mathrm{m}}56.9^{\mathrm{s}}}$	&	$\boldsymbol{+19^{\mathrm{o}}11'32.9''}$	&	\textbf{1.46}	&					\textbf{0.008} &--\\
~&~&~&~&~&	15\_2	&	+$15^{\mathrm{h}}51^{\mathrm{m}}56.8^{\mathrm{s}}$	&	+$19^{\mathrm{o}}11'39.7''$	&	5.90	&						0.128 &--\\
~&~&~&~&~&	15\_3	&	+$15^{\mathrm{h}}51^{\mathrm{m}}56.5^{\mathrm{s}}$	&	+$19^{\mathrm{o}}11'33.1''$	&	7.42	&						0.194 &--\\
\hline
16	&	$15^{\mathrm{h}}52^{\mathrm{m}}08.6^{\mathrm{s}}$	&	$+19^{\mathrm{o}}13'27.3''$	&	4.9(4.8)	&	$<13.8$& --  & --  & --  & --  & --  &--\\
\hline
\multirow{2}{*}{17}	&	\multirow{2}{*}{$15^{\mathrm{h}}51^{\mathrm{m}}45.5^{\mathrm{s}}$}	&	\multirow{2}{*}{$+19^{\mathrm{o}}07'07.6''$}	&	\multirow{2}{*}{5.8(4.7)}	&	\multirow{2}{*}{$<13.8$} & 17\_1	&	+$15^{\mathrm{h}}51^{\mathrm{m}}45.2^{\mathrm{s}}$	&	+$19^{\mathrm{o}}07'09.1''$	&	5.46	&						0.110 &--\\
~&~&~&~&~&	17\_2	&	+$15^{\mathrm{h}}51^{\mathrm{m}}46.0^{\mathrm{s}}$	&	+$19^{\mathrm{o}}07'05.9''$	&	7.12	&						0.180 &--\\
\hline
18	&	$15^{\mathrm{h}}52^{\mathrm{m}}13.1^{\mathrm{s}}$	&	$+19^{\mathrm{o}}14'09.2''$	&	7.8(4.6)	&	$<13.8$& --  & --  & --  & --  & --  &--\\
\hline
19	&	$15^{\mathrm{h}}51^{\mathrm{m}}32.8^{\mathrm{s}}$	&	$+19^{\mathrm{o}}10'27.6''$	&	4.7(4.6)	&	$<13.8$& --  & --  & --  & --  & --  &--\\
\hline
20	&	$15^{\mathrm{h}}51^{\mathrm{m}}59.7^{\mathrm{s}}$	&	$+19^{\mathrm{o}}16'24.3''$	&	6.1(4.5)	&	$<13.8$& --  & --  & --  & --  & --  &--\\
\hline
21	&	$15^{\mathrm{h}}51^{\mathrm{m}}53.6^{\mathrm{s}}$	&	$+19^{\mathrm{o}}12'29.6''$	&	3.1(4.5)	&	$<13.8$ & \textbf{21}	&	$\boldsymbol{+15^{\mathrm{h}}51^{\mathrm{m}}53.4^{\mathrm{s}}}$	&	$\boldsymbol{+19^{\mathrm{o}}12'28.1''}$	&	\textbf{3.13}	&						\textbf{0.038} &--\\
\hline
\multirow{2}{*}{22}	&	\multirow{2}{*}{$15^{\mathrm{h}}52^{\mathrm{m}}04.1^{\mathrm{s}}$}	&	\multirow{2}{*}{$+19^{\mathrm{o}}08'25.6''$}	&	\multirow{2}{*}{3.9(4.4)}	&	\multirow{2}{*}{$<13.8$} & \textbf{22\_1}	&	$\boldsymbol{+15^{\mathrm{h}}52^{\mathrm{m}}04.1^{\mathrm{s}}}$	&	$\boldsymbol{+19^{\mathrm{o}}08'25.5''}$	&	\textbf{1.04}	&					\textbf{0.004} &--\\
~&~&~&~&~&	\textbf{22\_2}	&	$\boldsymbol{+15^{\mathrm{h}}52^{\mathrm{m}}04.1^{\mathrm{s}}}$	&	$\boldsymbol{+19^{\mathrm{o}}08'28.9''}$	&	\textbf{3.28}	&					\textbf{0.041} &--\\
\hline
23	&	$15^{\mathrm{h}}51^{\mathrm{m}}35.0^{\mathrm{s}}$	&	$+19^{\mathrm{o}}12'10.9''$	&	4.3(4.3)	&	$<13.8$ & \emph{23}	&	\emph{+15}$^h$\emph{51}$^m$\emph{35.3}$^s$	&	\emph{+19}$^o$\emph{12'09.2''}	&	\emph{4.22}	&						\emph{0.068} &--\\
\hline
\multicolumn{10}{l}{$^a$ SMA detection, therefore IRAC ID not needed.}\\
\multicolumn{10}{l}{$^\dagger$ Optical spectroscopic redshift. }\\
\multicolumn{10}{l}{$^\|$ IRAM-NOEMA CO(3-2). }\\
\multicolumn{10}{l}{$^\mathsection$ Hamburg Quasar Survey. }\\
\multicolumn{10}{l}{$^\ddagger$ NB-imaging associated redshift}\\
\end{tabular}
\end{center}
\label{table:h1549}
\end{table*}

\begin{table*}
\contcaption{.}
\begin{center}
\begin{tabular}{ccccccccccc}
        \hline
        ID$_{850}$ & RA$_{850}$ & Dec$_{850}$ & S$_{850}$(SNR) & S$_{450}$(SNR) & ID$_\text{IRAC}$ & RA$_\text{IRAC}$ & Dec$_\text{IRAC}$ & $\theta$ & $P_\text{IRAC}$ & $z$\\
	~	&	(J2000)	&	(J2000)	&	(mJy)	&	(mJy)	&	~	&	(J2000)	&	(J2000)	&	(arcsec)	&	~&~\\
        \hline
        \hline
\multirow{4}{*}{24}	&	\multirow{4}{*}{$15^{\mathrm{h}}51^{\mathrm{m}}54.0^{\mathrm{s}}$}	&	\multirow{4}{*}{$+19^{\mathrm{o}}10'15.0''$}	&	\multirow{4}{*}{2.7(4.2)}	&	\multirow{4}{*}{$<13.8$} & 24\_1	&	+$15^{\mathrm{h}}51^{\mathrm{m}}53.5^{\mathrm{s}}$	&	+$19^{\mathrm{o}}10'17.3''$	&	7.07	&						0.178 &--\\
~&~&~&~&~&	\textbf{24\_2}	&	$\boldsymbol{+15^{\mathrm{h}}51^{\mathrm{m}}53.8^{\mathrm{s}}}$	&	$\boldsymbol{+19^{\mathrm{o}}10'17.3''}$	&	\textbf{3.38}	&					\textbf{0.044} &--\\
~&~&~&~&~&	\textbf{24\_3}	&	$\boldsymbol{+15^{\mathrm{h}}51^{\mathrm{m}}54.2^{\mathrm{s}}}$	&	$\boldsymbol{+19^{\mathrm{o}}10'15.8''}$	&	\textbf{2.42}	&					\textbf{0.023} &--\\
~&~&~&~&~&	24\_4	&	+$15^{\mathrm{h}}51^{\mathrm{m}}54.5^{\mathrm{s}}$	&	+$19^{\mathrm{o}}10'18.3''$	&	7.63	&						0.204 &--\\
\hline
\multirow{2}{*}{25}	&	\multirow{2}{*}{$15^{\mathrm{h}}52^{\mathrm{m}}03.3^{\mathrm{s}}$}	&	\multirow{2}{*}{$+19^{\mathrm{o}}14'13.0''$}	&	\multirow{2}{*}{4.1(4.2)}	&	\multirow{2}{*}{$<13.8$} & \textbf{25\_1}	&	$\boldsymbol{+15^{\mathrm{h}}52^{\mathrm{m}}03.1^{\mathrm{s}}}$	&	$\boldsymbol{+19^{\mathrm{o}}14'14.3''}$	&	\textbf{2.75}	&					\textbf{0.029} &--\\
~&~&~&~&~&	25\_2	&	+$15^{\mathrm{h}}52^{\mathrm{m}}03.4^{\mathrm{s}}$	&	+$19^{\mathrm{o}}14'08.2''$	&	5.28	&						0.104 &--\\
\hline
26	&	$15^{\mathrm{h}}51^{\mathrm{m}}58.4^{\mathrm{s}}$	&	$+19^{\mathrm{o}}09'31.6''$	&	3.3(4.2)	&	$<13.8$ & \textbf{26\_1}	&	$\boldsymbol{+15^{\mathrm{h}}51^{\mathrm{m}}58.4^{\mathrm{s}}}$	&	$\boldsymbol{+19^{\mathrm{o}}09'29.6''}$	&	\textbf{2.07}	&					\textbf{0.017} &--\\
\hline
\multirow{2}{*}{27}	&	\multirow{2}{*}{$15^{\mathrm{h}}51^{\mathrm{m}}46.7^{\mathrm{s}}$}	&	\multirow{2}{*}{$+19^{\mathrm{o}}13'13.0''$}	&	\multirow{2}{*}{3.3(4.1)}	&	\multirow{2}{*}{$<13.8$} & \textbf{27\_1}	&	$\boldsymbol{+15^{\mathrm{h}}51^{\mathrm{m}}46.6^{\mathrm{s}}}$	&	$\boldsymbol{+19^{\mathrm{o}}13'13.3''}$	&	\textbf{1.74}	&					\textbf{0.012} &--\\
~&~&~&~&~&	\emph{27\_2}	&	+$15^{\mathrm{h}}51^{\mathrm{m}}46.6^{\mathrm{s}}$	&	+$19^{\mathrm{o}}13'08.5''$	&	4.76	&					\emph{0.085} &--\\
\hline
\multirow{2}{*}{28}	&	\multirow{2}{*}{$15^{\mathrm{h}}51^{\mathrm{m}}50.3^{\mathrm{s}}$}	&	\multirow{2}{*}{$+19^{\mathrm{o}}15'15.0''$}	&	\multirow{2}{*}{4.8(4.1)}	&	\multirow{2}{*}{$<13.8$} & \textbf{28\_1}	&	$\boldsymbol{+15^{\mathrm{h}}51^{\mathrm{m}}50.2^{\mathrm{s}}}$	&	$\boldsymbol{+19^{\mathrm{o}}15'12.9''}$	&	\textbf{2.29}	&					\textbf{0.020} &--\\
~&~&~&~&~&	28\_2	&	+$15^{\mathrm{h}}51^{\mathrm{m}}50.3^{\mathrm{s}}$	&	+$19^{\mathrm{o}}15'09.6''$	&	5.35	&						0.106 &--\\
\hline
29	&	$15^{\mathrm{h}}51^{\mathrm{m}}48.2^{\mathrm{s}}$	&	$+19^{\mathrm{o}}15'37.6''$	&	4.8(4.1)	&	$<13.8$& --  & --  & --  & --  & --  &--\\
\hline
\end{tabular}
\end{center}
\end{table*}



\begin{table*}
\caption{The same as Table \ref{table:h1549}, but for HS1700.}
\begin{center}
\begin{tabular}{ccccccccccc}
        \hline
        ID$_{850}$ & RA$_{850}$ & Dec$_{850}$ & S$_{850}$(SNR) & S$_{450}$(SNR) & ID$_\text{IRAC}$ & RA$_\text{IRAC}$ & Dec$_\text{IRAC}$ & $\theta$ & $P_\text{IRAC}$ & $z$\\
	~	&	(J2000)	&	(J2000)	&	(mJy)	&	(mJy)	&	~	&	(J2000)	&	(J2000)	&	(arcsec)	&	~&~\\
        \hline
        \hline
1	&	$17^{\mathrm{h}}01^{\mathrm{m}}17.8^{\mathrm{s}}$	&	$+64^{\mathrm{o}}14'37.3''$	&	16.9(24.7)	&	30(4.2) & \textbf{1}	&	$\boldsymbol{+17^{\mathrm{h}}01^{\mathrm{m}}17.6^{\mathrm{s}}}$	&	$\boldsymbol{+64^{\mathrm{o}}14'37.850''}$	&	\textbf{1.11}	&						\textbf{0.005} & 2.816$^\|$\\
\hline
\multirow{2}{*}{2}	&	\multirow{2}{*}{$17^{\mathrm{h}}01^{\mathrm{m}}13.3^{\mathrm{s}}$}	&	\multirow{2}{*}{$+64^{\mathrm{o}}12'02.8''$}	&	\multirow{2}{*}{6.8(13.6)}	&	\multirow{2}{*}{18(3.7)} & \textbf{2\_1}	&	$\boldsymbol{+17^{\mathrm{h}}01^{\mathrm{m}}13.1^{\mathrm{s}}}$	&	$\boldsymbol{+64^{\mathrm{o}}12'01.980''}$	&	\textbf{1.82}	&						\textbf{0.013} &--\\
~&~&~&~&~&	\emph{2\_2}	&	\emph{+17}$^h$\emph{01}$^m$\emph{12.8}$^s$	&	\emph{+64}$^o$\emph{12'05.420''}	&	\emph{4.33}	&								\emph{0.071} &--\\
\hline
\multirow{3}{*}{3}	&	\multirow{3}{*}{$17^{\mathrm{h}}01^{\mathrm{m}}05.6^{\mathrm{s}}$}	&	\multirow{3}{*}{$+64^{\mathrm{o}}11'42.1''$}	&	\multirow{3}{*}{6.2(12.9)}	&	\multirow{3}{*}{14(2.9)} & \textbf{3}\textbf{\_1}	&	$\boldsymbol{+17^{\mathrm{h}}01^{\mathrm{m}}5.7^{\mathrm{s}}}$	&	$\boldsymbol{+64^{\mathrm{o}}11'43.690''}$	&	\textbf{1.64}	&	\textbf{0.011} & --\\
~&~&~&~&~&	\emph{3\_2}	&	\emph{+17}$^h$\emph{01}$^m$\emph{6.2}$^s$	&	\emph{+64}$^o$\emph{11'39.730''}	&	\emph{4.31}	&							\emph{0.070} &--\\
~&~&~&~&~&	\textbf{3\_3}	&	$\boldsymbol{+17^{\mathrm{h}}01^{\mathrm{m}}5.1^{\mathrm{s}}}$	&	$\boldsymbol{+64^{\mathrm{o}}11'42.100''}$	&	\textbf{3.47}	&							\textbf{0.046} &--\\
\hline
4	&	$17^{\mathrm{h}}01^{\mathrm{m}}10.7^{\mathrm{s}}$	&	$+64^{\mathrm{o}}07'20.7''$	&	10.9(12.9)	&	30(3.9) & \textbf{4}	&	$\boldsymbol{+17^{\mathrm{h}}01^{\mathrm{m}}10.8^{\mathrm{s}}}$	&	$\boldsymbol{+64^{\mathrm{o}}07'20.750''}$	&	\textbf{0.49}	&						\textbf{0.001} &$2.318^\mathparagraph$\\
\hline
\multirow{3}{*}{5}	&	\multirow{3}{*}{$17^{\mathrm{h}}00^{\mathrm{m}}58.3^{\mathrm{s}}$}	&	\multirow{3}{*}{$+64^{\mathrm{o}}13'08.3''$}	&	\multirow{3}{*}{6.3(12.8)}	&	\multirow{3}{*}{21(4.5)} & \textbf{5\_1}	&	$\boldsymbol{+17^{\mathrm{h}}00^{\mathrm{m}}57.9^{\mathrm{s}}}$	&	$\boldsymbol{+64^{\mathrm{o}}13'10.310''}$	&	\textbf{3.30}	&					\textbf{0.042} & 2.3$^\times$\\
~&~&~&~&~&	\textbf{5\_2}	&	$\boldsymbol{+17^{\mathrm{h}}00^{\mathrm{m}}58.5^{\mathrm{s}}}$	&	$\boldsymbol{+64^{\mathrm{o}}13'06.160''}$	&	\textbf{2.28}	&	\textbf{0.020} & 2.303$^\mathparagraph$\\
~&~&~&~&~&	\emph{5\_3}	&	\emph{+17}$^h$\emph{00}$^m$\emph{58.2}$^s$	&	\emph{+64}$^o$\emph{13'03.680''}	&	\emph{4.62}	&				\emph{0.080} & 2.3$^\times$\\
\hline
\multirow{2}{*}{6}	&	\multirow{2}{*}{$17^{\mathrm{h}}01^{\mathrm{m}}07.6^{\mathrm{s}}$}	&	\multirow{2}{*}{$+64^{\mathrm{o}}12'45.3''$}	&	\multirow{2}{*}{5.1(10.9)}	&	\multirow{2}{*}{$<12.9$} & \textbf{6\_1}	&	$\boldsymbol{+17^{\mathrm{h}}01^{\mathrm{m}}7.2^{\mathrm{s}}}$	&	$\boldsymbol{+64^{\mathrm{o}}12'44.240''}$	&	\textbf{3.28}	&						\textbf{0.041} &--\\
~&~&~&~&~&	\emph{6\_2}	&	\emph{+17}$^h$\emph{01}$^m$\emph{8.3}$^s$	&	\emph{+64}$^o$\emph{12'47.360''}	&	\emph{4.76}	&						\emph{0.085} &--\\
\hline
\multirow{4}{*}{7}	&	\multirow{4}{*}{$17^{\mathrm{h}}00^{\mathrm{m}}38.7^{\mathrm{s}}$}	&	\multirow{4}{*}{$+64^{\mathrm{o}}14'58.2''$}	&	\multirow{4}{*}{6.6(9.7)}	&	\multirow{4}{*}{23(3.2)} & \textbf{7\_1}	&	$\boldsymbol{+17^{\mathrm{h}}00^{\mathrm{m}}39.0^{\mathrm{s}}}$	&	$\boldsymbol{+64^{\mathrm{o}}14'58.300''}$	&	\textbf{1.91}	&				\textbf{0.014} &2.313$^\mathparagraph$\\
~&~&~&~&~&	\textbf{7\_2}	&	$\boldsymbol{+17^{\mathrm{h}}00^{\mathrm{m}}38.8^{\mathrm{s}}}$	&	$\boldsymbol{+64^{\mathrm{o}}15'01.270''}$	&	\textbf{3.09}	&						\textbf{0.037} &$\neq2.3^\mathparagraph$\\
~&~&~&~&~&	7\_3	&	+$17^{\mathrm{h}}00^{\mathrm{m}}37.7^{\mathrm{s}}$	&	+$64^{\mathrm{o}}14'55.430''$	&	6.78	&						0.165 &$\neq2.3^\mathparagraph$\\
~&~&~&~&~&	7\_4	&	+$17^{\mathrm{h}}00^{\mathrm{m}}37.1^{\mathrm{s}}$	&	+$64^{\mathrm{o}}14'53.240''$	&	11.26	&					0.392 &$\neq2.3^\mathparagraph$\\
\hline
\multirow{2}{*}{8}	&	\multirow{2}{*}{$17^{\mathrm{h}}00^{\mathrm{m}}56.1^{\mathrm{s}}$}	&	\multirow{2}{*}{$+64^{\mathrm{o}}12'02.3''$}	&	\multirow{2}{*}{4.6(9.6)}	&	\multirow{2}{*}{15(3.2)} & \emph{8\_1}	&	\emph{+17}$^h$\emph{00}$^m$\emph{56.8}$^s$	&	\emph{+64}$^o$\emph{12'03.430''}	&	\emph{4.94}	&		\emph{0.091} & $\neq2.3^\mathparagraph$\\
~&~&~&~&~&	8\_2	&	+$17^{\mathrm{h}}00^{\mathrm{m}}55.4^{\mathrm{s}}$	&	+$64^{\mathrm{o}}12'08.800''$	&	7.67	&						0.206 &--\\
\hline
9	&	$17^{\mathrm{h}}01^{\mathrm{m}}48.6^{\mathrm{s}}$	&	$+64^{\mathrm{o}}12'57.0''$	&	7.3(8.5)	&	35(3.6) & \textbf{9}	&	$\boldsymbol{+17^{\mathrm{h}}01^{\mathrm{m}}48.7^{\mathrm{s}}}$	&	$\boldsymbol{+64^{\mathrm{o}}12'58.700''}$	&	\textbf{1.73}	&					\textbf{0.012} &$\neq2.3^\mathparagraph$\\
\hline
\multirow{2}{*}{10}	&	\multirow{2}{*}{$17^{\mathrm{h}}00^{\mathrm{m}}56.7^{\mathrm{s}}$}	&	\multirow{2}{*}{$+64^{\mathrm{o}}16'30.5''$}	&	\multirow{2}{*}{5.0(7.3)}	&	\multirow{2}{*}{21(3.0)} & \textbf{10}\textbf{\_1}	&	$\boldsymbol{+17^{\mathrm{h}}00^{\mathrm{m}}56.6^{\mathrm{s}}}$	&	$\boldsymbol{+64^{\mathrm{o}}16'31.310''}$	&	\textbf{0.99}	&			\textbf{0.004} &--\\
~&~&~&~&~&	\textbf{10}\_\textbf{2}	&	$\boldsymbol{+17^{\mathrm{h}}00^{\mathrm{m}}57.1^{\mathrm{s}}}$	&	$\boldsymbol{+64^{\mathrm{o}}16'29.810''}$	&	\textbf{3.30}	&					\textbf{0.042} &--\\
\hline
11	&	$17^{\mathrm{h}}00^{\mathrm{m}}38.6^{\mathrm{s}}$	&	$+64^{\mathrm{o}}13'42.6''$	&	5.0(7.3)	&	30(4.5) & --  & --  & --  & --  & -- & --\\
\hline
12	&	$17^{\mathrm{h}}01^{\mathrm{m}}00.8^{\mathrm{s}}$	&	$+64^{\mathrm{o}}12'06.4''$	&	2.6(5.6)	&	15(3.3) & \textbf{12}	&	$\boldsymbol{+17^{\mathrm{h}}01^{\mathrm{m}}0.5^{\mathrm{s}}}$	&	$\boldsymbol{+64^{\mathrm{o}}12'09.090''}$	&	\textbf{3.46}	&			\textbf{0.046} & 2.72$^\mathsection$\\
\hline
13	&	$17^{\mathrm{h}}00^{\mathrm{m}}59.3^{\mathrm{s}}$	&	$+64^{\mathrm{o}}14'57.4''$	&	3.3(5.3)	&	$<12.9$ & \textbf{13}	&	$\boldsymbol{+17^{\mathrm{h}}00^{\mathrm{m}}59.2^{\mathrm{s}}}$	&	$\boldsymbol{+64^{\mathrm{o}}14'58.180''}$	&	\textbf{1.24}	&					\textbf{0.006} &--\\
\hline
14	&	$17^{\mathrm{h}}00^{\mathrm{m}}48.2^{\mathrm{s}}$	&	$+64^{\mathrm{o}}13'26.2''$	&	3.1(5.3)	&	$<12.9$ & \textbf{14}	&	$\boldsymbol{+17^{\mathrm{h}}00^{\mathrm{m}}48.3^{\mathrm{s}}}$	&	$\boldsymbol{+64^{\mathrm{o}}13'26.040''}$	&	\textbf{0.34}	&	\textbf{0.001} & $\neq2.3^\mathparagraph$\\
\hline
\multirow{2}{*}{15}	&	\multirow{2}{*}{$17^{\mathrm{h}}00^{\mathrm{m}}14.5^{\mathrm{s}}$}	&	\multirow{2}{*}{$+64^{\mathrm{o}}14'50.5''$}	&	\multirow{2}{*}{4.7(5.1)}	&	\multirow{2}{*}{$<12.9$} & \textbf{15}\textbf{\_1}	&	$\boldsymbol{+17^{\mathrm{h}}00^{\mathrm{m}}14.6^{\mathrm{s}}}$	&	$\boldsymbol{+64^{\mathrm{o}}14'51.340''}$	&	\textbf{1.46}	&			\textbf{0.008} & $\neq2.3^\mathparagraph$\\
~&~&~&~&~&	\emph{15\_2}	&	\emph{+17}$^h$\emph{00}$^m$\emph{13.7}$^s$	&	\emph{+64}$^o$\emph{14'51.030''}	&	\emph{5.06}	&					\emph{0.096} &$\neq2.3^\mathparagraph$\\
\hline
16	&	$17^{\mathrm{h}}01^{\mathrm{m}}34.8^{\mathrm{s}}$	&	$+64^{\mathrm{o}}14'54.4''$	&	4.1(5.1)	&	30(3.3) & \textbf{16}	&	$\boldsymbol{+17^{\mathrm{h}}01^{\mathrm{m}}34.9^{\mathrm{s}}}$	&	$\boldsymbol{+64^{\mathrm{o}}14'52.810''}$	&	\textbf{1.82}	&	\textbf{0.013} & $1.575^\mathparagraph$\\
\hline
17	&	$17^{\mathrm{h}}01^{\mathrm{m}}29.0^{\mathrm{s}}$	&	$+64^{\mathrm{o}}09'10.5''$	&	3.7(4.9)	&	$<12.9$ & \textbf{17}	&	$\boldsymbol{+17^{\mathrm{h}}01^{\mathrm{m}}29.1^{\mathrm{s}}}$	&	$\boldsymbol{+64^{\mathrm{o}}09'07.470''}$	&	\textbf{3.15}	&						\textbf{0.038} &2.306$^\mathparagraph$\\
\hline
18	&	$17^{\mathrm{h}}01^{\mathrm{m}}44.0^{\mathrm{s}}$	&	$+64^{\mathrm{o}}08'36.0''$	&	4.6(4.8)	&	$<12.9$ & \textbf{18}	&	$\boldsymbol{+17^{\mathrm{h}}01^{\mathrm{m}}44.2^{\mathrm{s}}}$	&	$\boldsymbol{+64^{\mathrm{o}}08'36.190''}$	&	\textbf{1.47}	&					\textbf{0.008} &--\\
\hline
19	&	$17^{\mathrm{h}}00^{\mathrm{m}}54.5^{\mathrm{s}}$	&	$+64^{\mathrm{o}}17'47.5''$	&	4.4(4.6)	&	$<12.9$ & \textbf{19}	&	$\boldsymbol{+17^{\mathrm{h}}00^{\mathrm{m}}54.3^{\mathrm{s}}}$	&	$\boldsymbol{+64^{\mathrm{o}}17'45.830''}$	&	\textbf{2.00}	&					\textbf{0.016} &--\\
\hline
20	&	$17^{\mathrm{h}}01^{\mathrm{m}}19.0^{\mathrm{s}}$	&	$+64^{\mathrm{o}}13'43.3''$	&	2.7(4.6)	&	$<12.9$ & \textbf{20}	&	$\boldsymbol{+17^{\mathrm{h}}01^{\mathrm{m}}19.1^{\mathrm{s}}}$	&	$\boldsymbol{+64^{\mathrm{o}}13'45.270''}$	&	\textbf{1.98}	&					\textbf{0.015} &--\\
\hline
21	&	$17^{\mathrm{h}}00^{\mathrm{m}}46.3^{\mathrm{s}}$	&	$+64^{\mathrm{o}}14'38.8''$	&	3.0(4.5)	&	$<12.9$  & --  & --  & --  & --  & --  &--\\
\hline
22	&	$17^{\mathrm{h}}01^{\mathrm{m}}48.3^{\mathrm{s}}$	&	$+64^{\mathrm{o}}10'12.9''$	&	3.9(4.5)	&	$<12.9$ & \textbf{22}	&	$\boldsymbol{+17^{\mathrm{h}}01^{\mathrm{m}}48.5^{\mathrm{s}}}$	&	$\boldsymbol{+64^{\mathrm{o}}10'11.110''}$	&	\textbf{2.10}	&					\textbf{0.017} &--\\
\hline
23	&	$17^{\mathrm{h}}01^{\mathrm{m}}06.4^{\mathrm{s}}$	&	$+64^{\mathrm{o}}08'23.3''$	&	3.2(4.5)	&	$<12.9$ & \emph{23}	&	\emph{+17}$^h$\emph{01}$^m$\emph{6.0}$^s$	&	\emph{+64}$^o$\emph{08'20.090''}	&	\emph{4.13}	&						\emph{0.065} &--\\
\hline
24	&	$17^{\mathrm{h}}00^{\mathrm{m}}40.6^{\mathrm{s}}$	&	$+64^{\mathrm{o}}16'53.3''$	&	3.5(4.5)	&	$<12.9$ & \textbf{24}	&	$\boldsymbol{+17^{\mathrm{h}}00^{\mathrm{m}}40.6^{\mathrm{s}}}$	&	$\boldsymbol{+64^{\mathrm{o}}16'53.560''}$	&	\textbf{0.64}	&					\textbf{0.002} &--\\
\hline
25	&	$17^{\mathrm{h}}00^{\mathrm{m}}16.2^{\mathrm{s}}$	&	$+64^{\mathrm{o}}12'23.0''$	&	3.7(4.4)	&	$<12.9$ & \textbf{25}	&	$\boldsymbol{+17^{\mathrm{h}}00^{\mathrm{m}}16.1^{\mathrm{s}}}$	&	$\boldsymbol{+64^{\mathrm{o}}12'20.710''}$	&	\textbf{2.40}	&					\textbf{0.022} &--\\
\hline
\multirow{2}{*}{26}	&	\multirow{2}{*}{$17^{\mathrm{h}}01^{\mathrm{m}}00.3^{\mathrm{s}}$}	&	\multirow{2}{*}{$+64^{\mathrm{o}}07'01.4''$}	&	\multirow{2}{*}{3.8(4.4)}	&	\multirow{2}{*}{$<12.9$} & \textbf{26}\_1	&	$\boldsymbol{+17^{\mathrm{h}}01^{\mathrm{m}}0.3^{\mathrm{s}}}$	&	$\boldsymbol{+64^{\mathrm{o}}06'58.290''}$	&	\textbf{3.11}	&					\textbf{0.037} &--\\
~&~&~&~&~&	\textbf{26}\textbf{\_2}	&	$\boldsymbol{+17^{\mathrm{h}}01^{\mathrm{m}}0.3^{\mathrm{s}}}$	&	$\boldsymbol{+64^{\mathrm{o}}07'00.760''}$	&	\textbf{0.67}	&			\textbf{0.002} &--\\
\hline
\multirow{2}{*}{27}	&	\multirow{2}{*}{$17^{\mathrm{h}}00^{\mathrm{m}}32.1^{\mathrm{s}}$}	&	\multirow{2}{*}{$+64^{\mathrm{o}}10'25.2''$}	&	\multirow{2}{*}{3.0(4.2)}	&	\multirow{2}{*}{$<12.9$} & \textbf{27}\textbf{\_1}	&	$\boldsymbol{+17^{\mathrm{h}}00^{\mathrm{m}}32.1^{\mathrm{s}}}$	&	$\boldsymbol{+64^{\mathrm{o}}10'25.140''}$	&	\textbf{0.14}	&			\textbf{0.001} &--\\
~&~&~&~&~&	\emph{27\_2}	&	\emph{+17}$^h$\emph{00}$^m$\emph{31.5}$^s$	&	\emph{+64}$^o$\emph{10'24.760''}	&	\emph{4.05}	&					\emph{0.062} &--\\
	\hline
\multicolumn{10}{l}{$^\|$ IRAM-NOEMA CO(3-2)}\\
\multicolumn{10}{l}{$^\mathparagraph$ Gemini GNIRS. }\\
\multicolumn{10}{l}{$^\times$ Photometric redshift.  }\\
\multicolumn{10}{l}{$^\mathsection$ Hamburg Quasar Survey.}\\
\end{tabular}
\end{center}
\label{table:h1700}
\end{table*}

\begin{figure*}
	\centering
	\includegraphics[width=\textwidth]{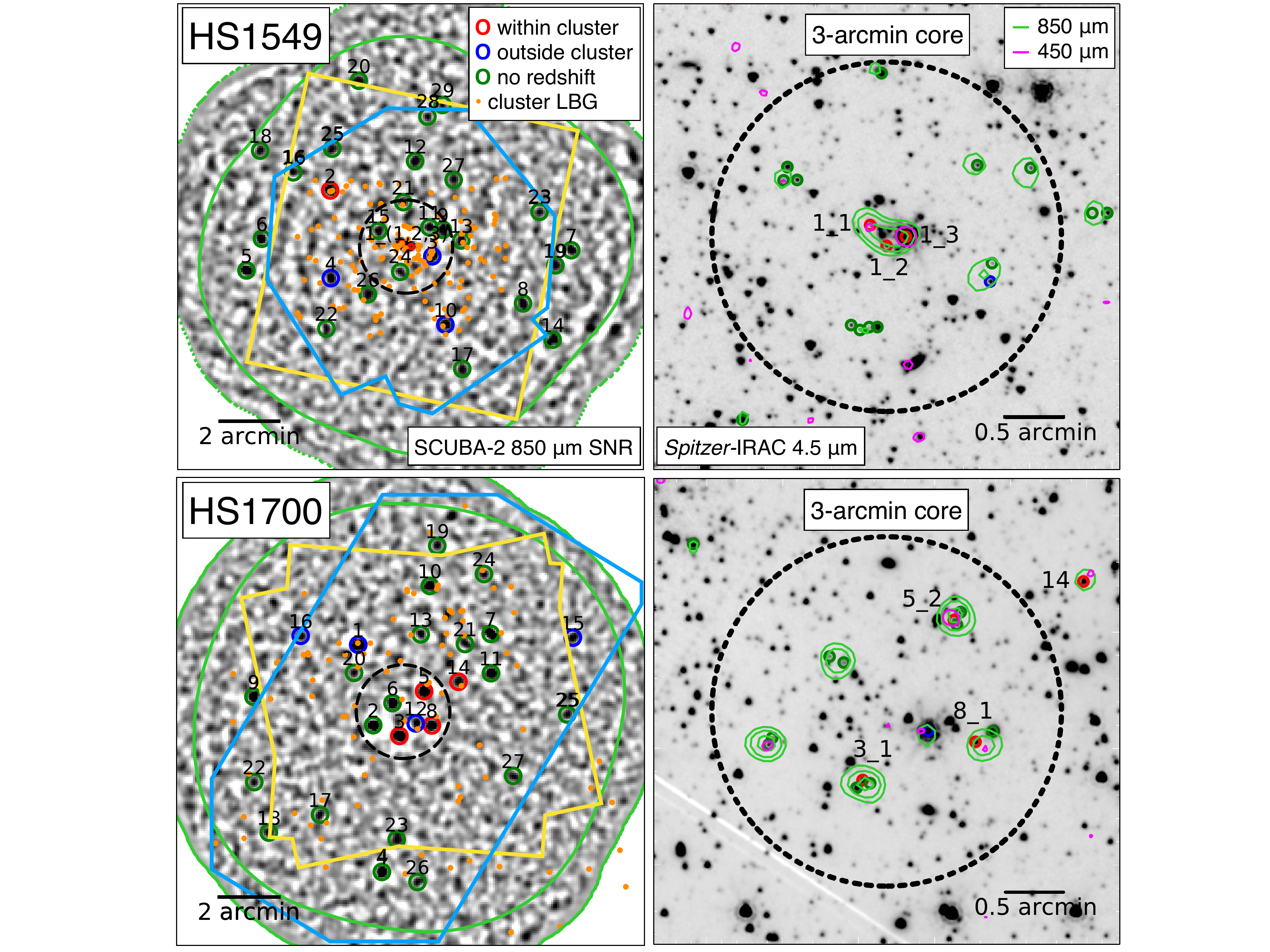}
	\caption{Images of the HS1549 field and the HS1700 field. \textbf{Left:} SCUBA-2 850 $\mu$m SNR images. 
	The coloured open circles represent $>4\sigma$ SCUBA-2 850 $\mu$m identified sources. The red circles are submm sources within the protoclusters, blue circles are submm sources residing outside of the protoclusters, and green circles are submm sources that have no spectroscopically or photometrically confirmed redshifts. The orange filled circles represent all LBGs that reside within the protoclusters ($z_\text{protocluster} \pm 0.05$). 
	The light green outer contour is the boundary in which we detected all of our SCUBA-2 850 $\mu$m IDs. The yellow and light blue contours show the \emph{Spitzer}-MIPS 24 $\mu$m and the \emph{Spitzer}-IRAC 4.5 $\mu$m coverage, respectively.
	\textbf{Right:} \emph{Spitzer}-IRAC 4.5 $\mu$m images of the 3-arcmin core region of each protocluster field. The open circles represent IRAC IDs associated with the SCUBA-2 detections with the same colour coding as the panels on the left. The light green contours of SNR$_{850}=(4,7,11)$ and magenta contours of SNR$_{450} = 3$.}
	\label{fig:fields}
\end{figure*}


\section{Analysis \& Results}
\label{sec:results}

\subsection{Submm Source Detection}
\label{sec:submm_det}
We have generated submm source catalogues from our 850 $\mu$m SCUBA-2 maps, 
using the data and variance maps produced from \textsc{oracdr}'s pipeline process via the \textsc{Starlink} package. We removed 
the high noise regions near the  edge of the map, where the on-sky exposure time was  less than 20\% that of the centre of the map. The regions we selected sources from  had ${\rm RMS}_{850} \leq 1.5$ mJy beam$^{-1}$ (HS1549) and ${\rm RMS}_{850} \leq 0.9$ mJy beam$^{-1}$ (HS1700).

We used a peak-finding algorithm to choose sources whose flux per beam  exceeded $4\times$ the local RMS noise in the variance map. We then found the position of these sources by fitting centroids to their peak position. Within our 850 $\mu$m SCUBA-2 maps we have detected 29 $>$4$\sigma$ sources (HS1549) and 27 $>$4$\sigma$ sources (HS1700), as shown in Tables \ref{table:h1549} and \ref{table:h1700}, listed in decreasing SNR. 
Figure \ref{fig:fields} shows the 850 $\mu$m SCUBA-2 images for each field and  \emph{Spitzer}-IRAC 4.5 $\mu$m image of the 3-arcmin diameter core regions.
The core regions are defined as the central 3-arcmin regions of our SCUBA-2 maps, the deepest regions of our maps where there is uniform noise.
%
The source at the centre of the HS1549 field has been resolved by the SMA at 870$\mu$m into three components with fluxes of (9, 9, and 6 mJy). These are sufficiently separated to allow for the smaller 7$^{\prime\prime}$ beam of SCUBA-2 at 450 $\mu$m to provide useful photometry at each SMA position (see \S~\ref{sec:counterparts}).

\subsection{Number Counts}

\begin{figure*}
	\centering
	\includegraphics[width=\textwidth]{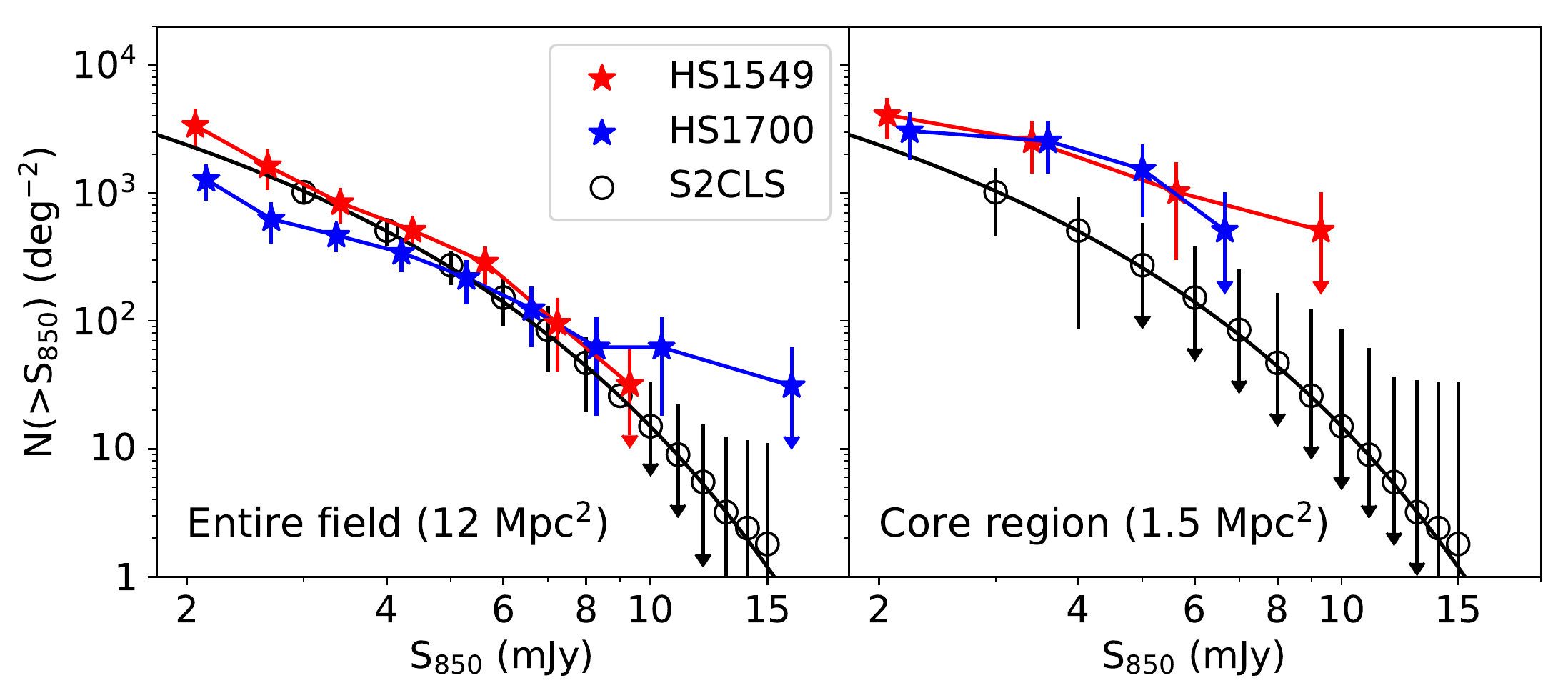}
	\caption[]{Cumulative number counts of $>4\sigma$ 850 $\mu$m sohurces for each protocluster's field (\textbf{top}) and 3-arcmin core (\textbf{bottom}). The solid line shows a Schechter function fit to the cumulative counts of the SCUBA-2 Cosmology Legacy Survey (S2CLS, \citealt{Geach2017}). The error bars for each sample represent the uncertainties derived using Monte Carlo error analysis. The uncertainties shown for the field sample are derived from the entire S2CLS field ($\simeq 4$ deg$^2$). The uncertainties shown for the core sample are derived from S2CLS-UDS field ($\simeq 0.9$ deg$^2$).
We find that neither protocluster field is over dense over their entire SCUBA-2 fields, however, the HS1700 field suggests a possible overdensity between 5 -- 100$\times$ for sources brighter than 8 mJy. In addition, we find that both field's $\sim 1.5$ Mpc core regions show overdensities of $6^{+4}_{-2}\times$ (HS1549) and $4^{+6}_{-2}\times$ (HS1700) compared to blank field surveys.	
	}
	\label{fig:number_counts}
\end{figure*}

When estimating the cumulative number counts we only consider flux measurements that have been corrected for statistically boosting effects, such as a submm source being boosted by positive noise and faint line of sight and background galaxies.
We correct for such selection biases by deboosting measured flux densities according to an empirical measure of flux boosting from the SCUBA-2 Cosmology Legacy Survey (S2CLS) (\citealt{Geach2017}).
We show the cumulative number counts for all $>4\sigma$ 850 $\mu$m sources within each protocluster field in Figure \ref{fig:number_counts}. We compute the counts 
within each deboosted flux bin and divide by the area over which we find these sources. 
If the flux bin is less than the average RMS of the entire field then the area is adjusted to the part of the map which is sensitive enough to detect these sources, and only sources that reside within these adjusted areas are counted.

We compare our counts to that of the S2CLS, the largest 850 $\mu$m survey to date with $\sim3000$ detected submm sources over $\sim5$ deg$^2$ with an average $1\sigma_{850}$ sensitivity of 1.2 mJy beam$^{-1}$. 
We estimate sampling uncertainties on the S2CLS counts (which are larger than published Poisson uncertainties) by randomly sampling ($N=10^5$) the deboosted source catalogs for sources which lie within the inner 50-arcmin diameter of S2CLS-UDS field with apertures the size of the field and core, 12 and 3-arcmin diameter, respectively, and measuring the  1$\sigma$ deviation from the mean. The S2CLS-UDS field was chosen for measuring uncertainties, as it was generated from PONG observations, where the RMS remains relatively constant throughout the map (\citealt{Dempsey2013}).
Over the entire HS1549 SCUBA-2 field we do not find any over-density for sources brighter than 4 mJy (see Figure \ref{fig:number_counts}), compared to S2CLS sampled on the same scale of 12-arcmin diameter ($\sim6$ Mpc at $z\sim2.5$).
Sampled on the same scale, we discover a possible overdensity between 5 -- 100$\times$ for sources brighter than 8 mJy in the HS1700 SCUBA-2 field, compared to the S2CLS field. However, these measures are highly uncertain, as there are few sources per bin.

We then quantify the number counts only in a {\it core} region of each protocluster field to assess whether the protoclusters show a more centrally concentrated over density in SMGs. We select a 3-arcmin diameter region where the deepest sensitivity limit is reached in the DAISY mode of SCUBA-2, and where the noise is relatively constant over the full area.
%
%
The core regions were selected based on the mean optical positions of the LBGs within each field.
The number counts over the 3-arcmin diameter core regions shown in Figure \ref{fig:number_counts} indicate that both fields are significantly over-dense (at $>3\sigma$ considering the full flux range of the counts) compared to the average counts of the S2CLS, even considering our Monte Carlo derived uncertainties in the 3-arcmin aperture, which are significantly larger than the Poisson errors on the S2CLS counts themselves. Our protocluster regions show overdensities within their $\sim 1.5$ Mpc core regions of $6^{+4}_{-2}\times$ (HS1549) and $4^{+6}_{-2}\times$ (HS1700).

%

This analysis reveals that above survey depth the large 12-arcmin diameter SCUBA-2 fields are likely diluted by line of sight SMGs and are not statistically over-dense, similar to many known protocluster fields (e.g., SSA22 \citealt{Steidel1998,Steidel2010a}, HDF1.99 \citealt{Chapman2009}, COSMOS \citealt{Casey2013}). However, there appears to be a clear active (SMG luminous) core region in both protoclusters.

Since SCUBA-2 has a relatively large beam compared to the size of a high-redshift galaxy, there is the possibility that the submm emission from multiple galaxies is being blended and identified as a single submm source (e.g., \citealt{Younger2009}; \citealt{Hodge2013}; \citealt{Karim2013}; \citealt{Simpson2015}; \citealt{Hill2018}; \citealt{Stach2017,Stach2018a}). Therefore, the number counts presented in Figure \ref{fig:number_counts} are upper limits at bright 850 $\mu$m fluxes and lower limits at fainter  fluxes. 
This could lead to an even stronger centrally dominated over-density of fainter sources once the sources are resolved by interferometry.

The SMG over-density in HS1700 has a distribution that peaks near the mean optical galaxy cluster centre (\citealt{Steidel2005}), but is offset $\sim40^{\prime\prime}$ (320 kpc)  to the north compared to the optical galaxies. 
\cite{Kato2016} have suggested a central protocluster position  $\sim2.1^\prime$ north using seven \emph{Herschel}-SPIRE sources having colours consistent with $z\sim2.3$, 
although only one of their sources confirmed to lie within the protocluster, \emph{1700.5}, which has a spectroscopic line detections and photometric redshifts locating the source at $z=2.3$ (see \S \ref{sec:NIR_spec} and \S \ref{sec:photoz}).
When applying our SCUBA-2 over-density analysis at the \cite{Kato2016} offset position, we find no such evidence for excess. 
The optical centroid of the protocluster is likely the most reliable tracer of the mass distribution, as the optical sources have longer  duty cycles than the more active SMGs, even though the latter are likely to be more massive on average (e.g., \citealt{Chapman2009}). However, the submm observations can be useful tracers of peak activity periods within the protocluster, and can thus reveal the formation modes.

\subsection{Counterpart Identification}
\label{sec:counterparts}
To make progress understanding the protocluster membership of the SMGs in each field, we need to first attempt to identify the SMGs at other wavelengths, and then assess their redshifts.
As we lack interferometric followup to the majority of our SMGs, we attempt to determine the IR counterparts to our 850 $\mu$m-identified sources  using \emph{Spitzer}-IRAC 4.5 $\mu$m and 8.0 $\mu$m. Multi-wavelength cutouts for each SCUBA-2 source are found in the Appendix.
The IRAC counterpart identification of SMGs can be performed using a statistical approach based on the relative source densities at the identifying wavelength. We determine the chance of an IRAC source being associated with the submm source using Poisson statistics to calculate the probability of finding a source at random within some area given by:
\begin{equation}
P_\text{IRAC} = 1-\exp\left(-\pi\eta_{\left(>S\right)}\theta^2\right)
\end{equation}
%
where $\eta_{\left(>S\right)}$ is the surface density of IRAC sources above a flux density level $S$ per unit solid angle, and $\theta$ is the angular offset between IRAC source and the SCUBA-2 ID. 
The probability of a random association includes a correction factor which takes into consideration the depth of the IRAC map as described in \cite{Downes1986} and \cite{Dunlop1989}.
We present all IRAC counterparts found within 15-arcseconds (the size of the SCUBA-2 beam at 850 $\mu$m) in Tables \ref{table:h1549} and \ref{table:h1700}.
A counterpart is considered a reliable match if $P\leq0.05$, while we consider a counterpart to be a tentative match if $0.05<P\leq0.1$ (consistent with e.g., \citealt{Ivison2002,Pope2006,Chapin2009,Wardlow2010,Yun2011,Hodge2013,An2018}).
Using this procedure we have found likely ($P\leq0.05$) IRAC counterparts for 19 of 37 (HS1549) and 28 of 39 (HS1700) SMGs, and an additional 6 (HS1549) and 8 (HS1700) tentative counterparts ($0.05<P\leq0.1$). Any remaining IRAC identifications listed are not statistically reliable ($P>0.1$). 
We note that the search for IRAC counterparts has limitations, and we expect as many as 50\% of the IDs may not be correct (e.g., \citealt{Hodge2013}). We also complement the IRAC analysis with an assessment of the SPIRE colours of the SCUBA-2 sources in Appendix A. We reiterate that interferometry is required for a complete study of the cluster membership.

There are ten (HS1549) and two (HS1700) SCUBA-2 sources lacking any viable IRAC IDs due to either the lack of coverage of the IRAC data, obscuration by a foreground star or galaxy, or there being no IRAC-detected galaxies residing within the SCUBA-2 error circle.
%
These sources have been excluded from Tables \ref{table:h1549} and \ref{table:h1700}.
%
There are four (one) SCUBA-2 850 $\mu$m sources in HS1549 (HS1700) which appear significantly elongated, even with the large 15$^{\prime\prime}$ beam (\emph{1549.1, 1549.3, 1549.7, 1549.25, 1700.15}), suggesting they may have multiple submm counterparts.  In the first case,  \emph{1549.1}, the SMA observations (described in \S~2) directly resolve three  components at $>5\sigma$, 
consistent with the elongated SCUBA-2 morphology (fluxes and positions in Table \ref{table:h1549}).

We also make use of MIPS 24 $\mu$m data which provides an alternative route to identify star-forming galaxies, and can be a useful aid to confirming the identification to the SMGs.
However, there are four cases in HS1549 (\emph{1549.12}, \emph{1549.22}, \emph{1549.24}, \emph{1549.27}) and five cases in HS1700 (\emph{1700.2}, \emph{1700.3}, \emph{1700.7} \emph{1700.10}, \emph{1700.27}) where IRAC identifications cannot be reliably measured with MIPS 24 $\mu$m because the large point spread function of a nearby bright 24 $\mu$m source overlaps the IRAC position. 
In HS1700 we do not have MIPS 24 $\mu$m
data for 7 of 27 SCUBA-2 sources, making ID's more uncertain. Furthermore, in HS1549 there is no coverage with MIPS or IRAC  for 8 of 29 sources, making IR identification impossible for these sources. We exclude these sources from our protocluster membership analysis.

Finally, we consider whether any of the SCUBA-2 sources are gravitationally lensed by searching for counterpart IDs that appear to be very low redshift, massive foreground galaxies. While there are two foreground galaxy clusters (of modest mass) in the HS1700 field, we find no evidence for strong galaxy-galaxy gravitational lensing for our SCUBA-2 sources or IRAC IDs (consistent with the expected rate of lensing of typical SMGs of $\simeq$ few percent, \citealt{Chapman2002}).
The counterpart to \emph{1549.16} shows a bright IRAC and optical ID, offset 2$^{\prime\prime}$ from the 850/450 $\mu$m centroids, which may suggest weak gravitational lensing of our SCUBA-2 ID.

\subsection{Cluster Membership}
\label{sec:clustermembership}

To assess the redshifts of the SCUBA-2 source IDs, we used several spectroscopic and photometric data sets. Using one or more of $^{12}$CO(3-2) detections, near-IR and optical spectroscopy, we confirm three (four) SMG-candidates residing in the protocluster of HS1549 (HS1700), one (two) SMG(s) likely being members (the primary ID is not confirmed at the protocluster redshift, but nearby sources in the SCUBA-2 beam are). 
Further, we confirm three (four) SMGs definitely residing outside of the protoclusters. 
We note that although CO is an unambiguous tracer for the the SCUBA-2 IDs, the optical/NIR spectroscopic sources could be uncorrelated with the S2 sources, as they might be other foreground/background sources along the same line of sight, rather than the actual SMG.
Since there are many galaxies at the protoclusters' redshifts in these protocluster fields, finding that an IRAC ID with a statistically reliable P-value at the protocluster's redshift, doesn't guarantee necessarily it is the SMG ID.


\subsubsection{IRAM-NOEMA CO(3-2)}
The most secure way of determining the redshift of an SMG is by detecting the associated molecular reservoir directly (e.g.,  as described in \citealt{Bothwell2012}).
In HS1549, the core region was observed with IRAM-NOEMA detecting CO(3-2) lines at $z\sim2.85$ centred on each of the  three SMA identified sources (Chapman et al. in preparation).
At the redshift 2.85, the 4.5$\sigma$ CO line detection threshold if $L_\text{lim} = 7\times10^9$ K km s$^{-1}$ pc$^2$, comparable with the faintest CO detections reported so far from 3mm observations at these redshifts (\citealt{Bothwell2012}; \citealt{Tacconi2013}).

The positions of the 870 $\mu$m sources are consistent within 0.2$^{\prime\prime}$ with the centroid of the line measurements from NOEMA for all three counterparts. All three sources are clearly in the protocluster:
\emph{1549.1\_1}  ($z=2.856$), \emph{1549.1\_2}  ($z=2.851$), and \emph{1549.1\_3}  ($z=2.847$). 
The same IRAM-NOEMA programme also showed that \emph{1549.3} resides at redshift $z=2.918$, which is close to the protocluster redshift ($\Delta \text{v}_\text{rest} = 20,400$ km s$^{-1}$), but is too far from the peak redshift distribution to be in the same collapsing structure. 
\emph{1549.3} does not reside at a statistically robust redshift peak, but may represent a structure which will eventually merge with the primary $z=2.85$ cluster.
Additionally,  \emph{1549.10} was found to not be detected in CO(3-2), suggesting this source lies at a redshift other than the protocluster's redshift or is gas poor.

In the HS1700, no SMGs have been proven to be physically related to the HS1700 protocluster based on the NOEMA CO(3-2) observations.
The CO(3-2) followups have been undertaken by \cite{Tacconi2013} and \cite{Chapman2015}. In the latter study, the brightest SMG in the HS1700 field, \emph{1700.1}, was identified at $z=2.816$, residing behind the protocluster. The same PdBI programme also showed  \emph{1700.15}  was not detected in CO(3-2), again suggesting this source lies at a different redshift.
The \cite{Tacconi2013} observations include  \emph{1700.11} within the primary beam, failing to detect any CO line, but the narrow bandwidth  of these older observations does not preclude  \emph{1700.11}  still lying within the protocluster.

\subsubsection{Optical Spectroscopy}
\label{sec:opticalspec}

In HS1549,  two of the three central sources  detected in CO(3-2),  \emph{1549.1\_1} and  \emph{1549.1\_3}, had previously known redshifts from the optical. \emph{1549.1\_1}  is associated to an AGN, MD17 in Steidel et al.\ (in preparation) ($z=2.856$), while \emph{1549.1\_3} is the central hyper-luminous QSO itself ($z=2.847$), both detecting the standard type-1 AGN high ionization emission lines. 
The source \emph{1700.12} is associated to the central QSO in HS1700. 
Redshifts for these QSOs were identified at $z_{\rm HS1700}=2.72$ (behind the protocluster) and $z_{\rm HS1549} = 2.85$ (within the protocluster) in the Hamburg Quasar Survey (\citealt{Reimers1989}). 

\subsubsection{Near-IR Spectroscopy}
\label{sec:NIR_spec}

%

We have followed up thirteen submm sources within HS1700 using Gemini GNIRS and the details of these observations are found in \S~\ref{sec:GNIRS}.   
With 5 detections and 8 non-detections, we have confirmed 4 of 13 sources as protocluster members. The remainder do not detect emission lines, but this is not conclusive of the source not being at the protocluster redshift as their counterparts are sufficiently faint in the near-IR continuum that the non-detection of nebular emission lines is still consistent with expected equivalent widths in star forming galaxies. 
We found \emph{1700.4} ($z=2.318$), \emph{1700.5\_2} ($z=2.303$), \emph{1700.7\_1} ($z=2.313$), and \emph{1700.17} ($z=2.306$) show prominent H$\alpha$, [NII] detections and lie in the protocluster, while \emph{1700.16} was found to reside out of the protocluster at $z=1.575$. In all cases, the redshift is for the most probable IRAC identification to the SMG, and is either a likely identification ($P \leq 0.05$) or a tentative association ($0.05 < P \leq 0.10$). 
We note that the GNIRS spectra for \emph{1700.4}, \emph{1700.5\_2}, and \emph{1700.17} have strong sky line residuals surrounding the main spectral features, and as a result they may have systematic offsets of a couple hundred km s$^{-1}$, resulting in somewhat uncertain IDs.
The spectra for the five positive detections are found in Appendix B.

\subsubsection{LBGs and NB-imaging}

Where possible, we determine cluster membership from IRAC IDs discussed in \S~\ref{sec:counterparts}, which we cross-correlate against existing catalogs of spectroscopically identified LBGs, and narrowband sources targeting the protocluster redshift. 

For the SCUBA-2 ID \emph{1549.2} we found three existing spectroscopic detections within the SCUBA-2 $15^{\prime\prime}$ beam which all reside within the protocluster at $z=2.837$, $z=2.835$, and $z=2.842$. Although none of these LBGs are associated to the IRAC ID, the LBGs may still be  associated to the SCUBA-2 ID. 

Using the same technique, we found the SCUBA-2 ID \emph{1549.4} is possibly associated with two different galaxies separated by 3$^{\prime\prime}$, a BX-selected galaxy residing at $z = 2.388$, outside of the protocluster, and an IRAC ID.
Although both galaxies are at similar angular offsets to the SCUBA-2 ID, the density of IRAC galaxies is significantly higher than 
that of BX-selected galaxies for the limiting magnitude of the map. Therefore, the BX-selected galaxy is  more likely associated to the SCUBA-2 ID, with $P = 0.004$, suggesting \emph{1549.4} likely resides outside of the protocluster.

\subsubsection{Photometric Redshifts}
\label{sec:photoz}

For the most part, our IRAC IDs do not have photometric redshifts in the catalogs, due to very faint fluxes in the optical bands. Deeper optical imaging would allow for additional constraints from photometric redshift analysis. However, 
we have identified two colour-selected LBGs which have red infrared colours,  $J-K > 2.3$, that align with IRAC IDs in HS1700 with no recorded redshift (\emph{1700.5\_1}: \emph{DRG46} (also an H$\alpha$-NB detection), \emph{1700.5\_3}: \emph{DRG44}).
As shown in large IR surveys, these photometric detections and colour-selected criteria suggest possible identification with the protocluster redshift (e.g., \citealt{Simpson2014}).
%
While robust photometric redshifts are not feasible at this point, as an additional assessment of the redshifts of our IRAC identifications, we compare the IRAC colours for our IDs with those of SMGs from \cite{Simpson2014} for the ALESS survey. This analysis is detailed in Appendix A, and reveals a reasonable consistency of many of our IDs with the redshifts of the protoclusters.

In addition, we determine photometric redshifts for SCUBA-2 IDs by measuring the SPIRE flux ratio, as shown in Appendix A, Figure \ref{fig:SPIRE-CC}. 
In general, the photometric redshift determined by the SPIRE colours are in agreement with the SCUBA-2 IDs known redshifts, with the exception of \emph{1700.12}. However, it can be said that since \emph{1700.12} is a known QSO at $z=2.84$, therefore raising its dust temperature, thereby shifting its SED peak such that $S_{500}>S_{350}>S_{250}$.
Using the SPIRE colours, we uncover a eight new candidate protocluster members:
\emph{1700.2} ($z = 2.8\pm 0.6$), 
\emph{1700.3} ($z = 2.1\pm 0.5$), 
\emph{1700.9} ($z = 2.2\pm 0.3$),
\emph{1700.10} ($z = 2.0\pm 0.3$),
\emph{1700.13} ($z = 1.3\pm 1.1$),
\emph{1700.15} ($z = 2.3\pm 0.7$),
\emph{1700.23} ($z = 2.4\pm 0.6$),
\emph{1700.26} ($z = 2.3\pm 0.4$).

\subsection{Stacking protocluster galaxies at 850 $\mu$m}

To provide measures of the overall dust obscured star-formation activity in each protocluster, we measure the submm flux at the position of each known cluster member (LBGs and narrowband line emitters) in the 850 $\mu$m beam-convolved map, discarding any  positions too close to individually detected SMGs where there is a peak with $S_{850}$ detected at $\geq +4\sigma$. 
We choose an exclusion radius of twice the SCUBA-2 beam (30$^{\prime\prime}$) 
as we found that the negative flux of the {\it bowling} regions (artifact from the matched filtering) around each SMG was negatively biasing the stacked 850 $\mu$m measurements.
Each 850 $\mu$m flux is weighted by the inverse of the variance, as measured in the noise map. We then stack the weighted fluxes and compute the stack as $\sum_i (S_i/\sigma_i) / \sum_i(1/\sigma_i)$. 

We find the 850 $\mu$m stacks to be $\left< S_{850} \right> = 0.25 \pm 0.05$ mJy (HS1549: 114 protocluster galaxies) and $\left< S_{850} \right> = 0.21 \pm 0.05$ mJy (HS1700: 83 protocluster galaxies); the typical protocluster member galaxy has a star-formation rate of $\sim20$ M$_\odot$ yr$^{-1}$, as derived directly from the rest frame 
$\sim230$ $\mu$m. Multiplied by the number of stacked protocluster members,  this results in integrated 850 $\mu$m derived SFRs in each protocluster of $2100\pm500$ M$_\odot$ yr$^{-1}$ (HS1549) and $1300\pm300$ M$_\odot$ yr$^{-1}$ (HS1700), by using Equation \ref{eq:SMG_SFR}. These results show that there is a statistically significant detection of  submm activity associated with star formation among protocluster galaxies.


\section{Discussion}
\label{sec:discuss}

From our analysis of the 850$\mu$m properties of each protocluster in \S~3, we have ascertained that the over densities defined from LBG redshift spikes in HS1549 and HS1700, are mirrored in the dusty SMG population as well.  We have spectroscopically confirmed that in each protocluster, at least four of these SMGs are at the protocluster redshift, and statistically from the counts (or less robustly with photometric redshift identifications) many other SMGs must lie at the protocluster redshifts. In both protoclusters, the statistically significant SMG over-density is confined to the core ($\sim 1.5$ Mpc diameter) region, consistent with other submm studies of protoclusters that find an overabundance of dusty, luminous galaxies in the cluster core (e.g., \citealt{Stach2017}). 

The HS1549 field also shows a higher count over the entire SCUBA-2 field than the HS1700 field, and higher than the published S2-CLS count in \citealt{Geach2017}. However, compared to our Monte Carlo sampling of the S2CLS-UDS map, the HS1549 field is not significantly over dense. 
Nonetheless, the elevated count of HS1549 over HS1700 is consistent with the HS1549 field showing a higher density contrast than HS1700 in optically selected LBGs and NB-selected line emitting galaxies (Steidel et al.\ in preparation).


Having established that there are over densities of SMGs in both protocluster fields on $\sim3$ arcmin scales, it is then of interest to determine how much the SMGs contribute to each protocluster's total star-formation rate, and thus how active these cores are, relative to other known protoclusters
We calculate the star-formation rate of SCUBA-2 850 $\mu$m detected SMGs with a conversion given by \cite{DaCunha2015}:
\begin{equation}
\text{SFR}_\textsc{MAGPHYS} (M_\odot \text{yr}^{-1}) = (75\pm17) \times S_{850 ~\mu\text{m}} (\text{mJy}),
\label{eq:SMG_SFR}
\end{equation}
which is derived from a large population of ALESS SMGs of $z=2-3$ using the \textsc{MAGPHYS} package (\citealt{DaCunha2008}), including the scatter in dust temperature in SMG populations. This relation is in agreement with the \cite{KennicuttJr.1998} star-formation law.
As mentioned in \S~\ref{sec:clustermembership}, we found four robust SCUBA-2 detected SMGs within each protocluster. Using Equation \ref{eq:SMG_SFR} we find that these protocluster SMGs contribute $2300 \pm500$ M$_\odot$ yr$^{-1}$ (HS1549) and $2100 \pm500$ M$_\odot$ yr$^{-1}$ (HS1700), to each protocluster.

To compute SFRs for all known protocluster galaxies undetected at 850 $\mu$m (e.g. LBGs) we assume the conversion given by \cite{Rieke2008}:
\begin{equation}
\text{SFR}(\text{M}_\odot \text{yr}^{-1}) = 1.18 \times 10^{-9} ~\text{L}(24\mu\text{m}, \text{L}_\odot)
\label{eq:LBG_SFR}
\end{equation}
which assumes the use of the \cite{KennicuttJr.1998} star-formation conversion between $\text{L}_{\text{FIR}(8-1000 ~\mu\text{m})}$ and SFR for sources at $z>1.5$, and where $L(24\mu\text{m},L_\odot)$ is the rest frame \emph{Spitzer}-MIPS 24 $\mu$m luminosity with no bandpass corrections. 
This conversion does not account for AGN heating of dust in this wavelength regime, even though AGN fractional contribution is known to be non-negligible (10\%) (e.g., X-ray spectral survey of SCUBA galaxies, \citealt{Alexander2005}; near and mid-IR spectral studies of SMGs, \citealt{Menendez-Delmestre2009,Menendez-Delmestre2011}; far-IR photometry of SMGs, \citealt{Pope2010}).

By applying Equation \ref{eq:LBG_SFR} to all known protocluster galaxies undetected at 850 $\mu$m we find the total SFR from protocluster members not individually detected at 850 $\mu$m,  to be $10,200 \pm 500$ M$_\odot$ yr$^{-1}$ (HS1549) and $2800 \pm 200$ M$_\odot$ yr$^{-1}$ (HS1700). 
These SFR values are consistent with the analysis previously conducted by \cite{Reddy2006} for both protocluster fields.
The addition of 24 $\mu$m undetected LBGs only add $\sim510\pm70$ M$_\odot$ yr$^{-1}$ (HS1549) and $\sim360\pm60$ M$_\odot$ yr$^{-1}$ (HS1700) to the total SFR, assuming these all have SFR $\sim 10$ M$\odot$ yr$^{-1}$ (consistent with being undetected at 24 $\mu$m).

The uncertainty of 850 $\mu$m and  24 $\mu$m derived SFRs are dominated by the uncertainties in their measured fluxes. However a further systematic uncertainty exists for each wavelength. At 850 $\mu$m, the dust temperature is the dominant uncertainly, yielding a 20\% systematic error on the SFR based on the well-constrained sample of SMGs at similar redshifts in \cite{Swinbank2014}, which is captured by the error in Equation \ref{eq:SMG_SFR}.
At 24 $\mu$m, the location of the polycyclic aromatic hydrocarbon (PAH) emission in the \emph{Spitzer}-MIPS band requires a template fit adjustment. To adjust for the position of the PAH line at $z=2.85$ the 24 $\mu$m derived SFR for HS1549's LBGs has been boosted by a factor of two (\citealt{Rieke2008}).
An additional uncertainly for the brightest 24 $\mu$m sources ($S_{24~\mu\text{m}}>0.2$ mJy) is AGN heating.
However weak or non-detection at 850 $\mu$m for these 24 $\mu$m-bright sources is a good way to ascertain whether they are powered by AGN, rather than driven by star formation, as $S_{24~\mu\text{m}}>0.2$ mJy should be well detected $>3\sigma$ at 850 $\mu$m.
Additionally, faint AGN wouldn't be detected at 850 $\mu$m, unless they had a large starburst accompanying the AGN, as the 850 $\mu$m is too cold on the spectral energy distribution (SED).

We also compare the 24 $\mu$m total SFR calculations with our stacked 850 $\mu$m derived SFRs for protocluster galaxies not individually detected at 850 $\mu$m using Equation \ref{eq:SMG_SFR}. 
For these fainter optical protocluster members, stacked 850 $\mu$m SFRs will be a less precise measure, since the 24 $\mu$m SFRs are based on detecting many more of the sources individually.
We find that for our LBGs, the stacked 850 $\mu$m derived SFRs are systematically lower than our 24 $\mu$m derived SFRs ($\sim5\times$ for HS1549 and $\sim2\times$ for HS1700). While the difference in HS1700 is within the errors of the stacking method, the large difference in the HS1549 may be due to a few known factors specific to this field and redshift. There are more 24$\mu$m bright galaxies (with S$_{24}>0.1$ mJy) in HS1549, which may indicate a higher AGN presence. In addition, the 24$\mu$m band becomes less precise a measure of SFR at the higher redshift of HS1549 as the strong PAH feature is no longer present in the band, and the calibration of the SFR$_{24\mu m}$ is more uncertain and variable with environment and galaxy type. We may therefore have overestimated the total SFR using the 24$\mu$m measurements in HS1549.
Assuming 50\% of the 24 $\mu$m flux is contaminated by AGN heating and other systematics associated with the 24 $\mu$m to SFR conversion, without the PAH lines in the band (\citealt{Rieke2008}), a more conservative estimate of the 24 $\mu$m derived SFR would be $5100\pm300$ M$_\odot$ yr$^{-1}$.

%

\begin{table}
\caption{Integrated star-formation rates for different contributing sources.}
\begin{center}
\begin{tabular}{cccc}
        \hline
        Protocluster	&	LBGs \& NB-emitters	&	SMGs	&	All (24 $\mu$m + 850 $\mu$m)\\
        ~	&	$\left(\text{M}_\odot \text{yr}^{-1}\right)$	&	$\left(\text{M}_\odot \text{yr}^{-1}\right)$	&	$\left(\text{M}_\odot \text{yr}^{-1}\right)$\\
        \hline
        \hline
	HS1549	&	10,$200\pm500$	&	$2300\pm500$	&	12,$500\pm2800$\\
	HS1700	&	$2800\pm200$		&	$2100\pm500$	&	$4900\pm1200$\\
	\hline
\end{tabular}
\end{center}
\label{table:SFRs}
\end{table}

\begin{figure}
	\centering
	\includegraphics[width=.49\textwidth]{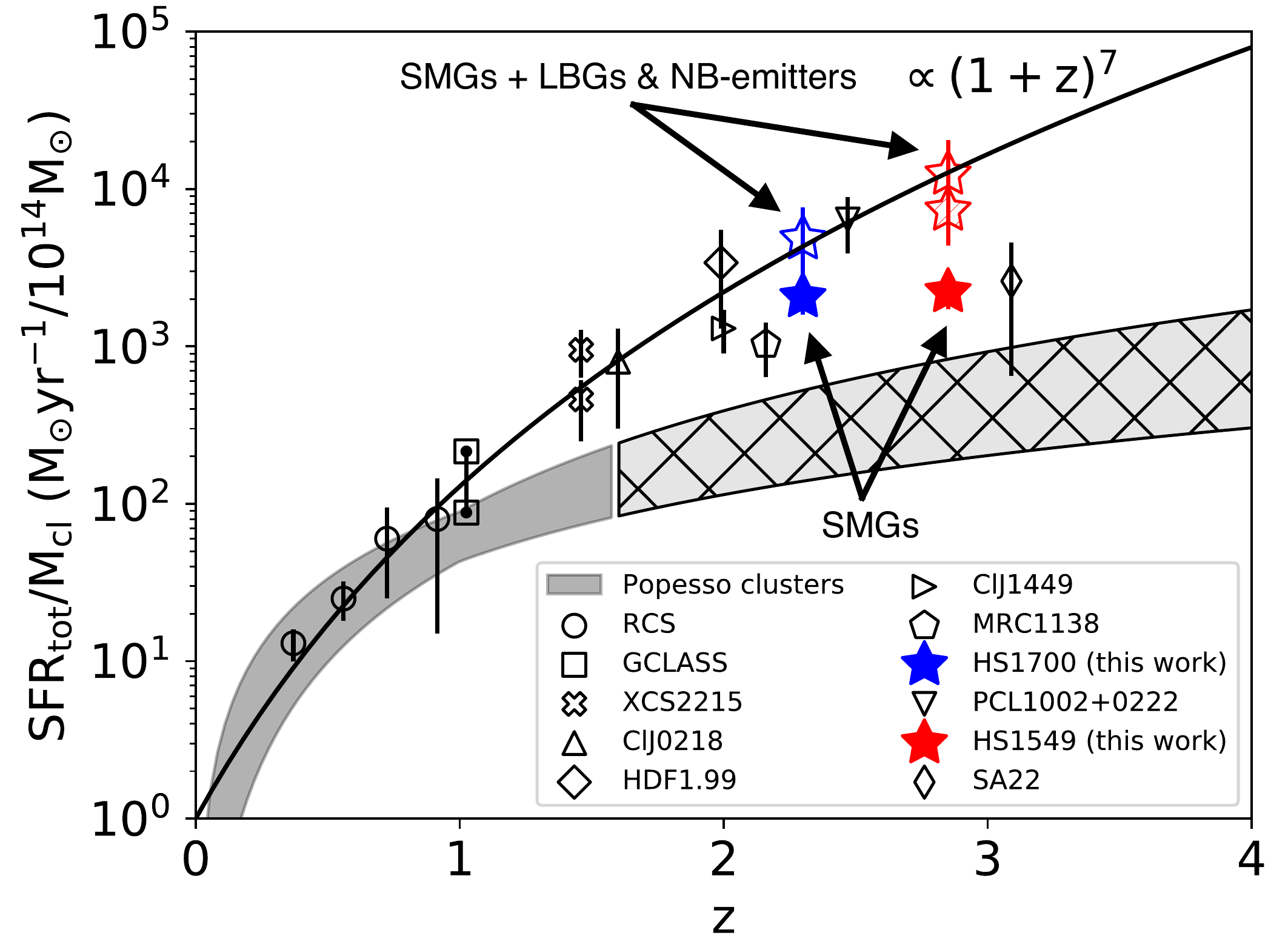}
	\caption[SFR-z figure]
	{Integrated star-formation rate of galaxy clusters and protoclusters normalized by their total mass as a function of redshift.
	We compare field galaxy evolution to protoclusters at $z>1$ and find protoclusters follow a $(1+z)^7$ relation, suggesting much more rapid evolution compared to an extrapolation of the local field galaxy trend.
	The protoclusters considered in this work are represented as star symbols, where the solid stars represent the contribution of only the SMGs, the open stars represent the contribution from the SMGs and the LBGs \& NB-emitters, and the hatched open star for HS1549 represents the contribution from the SMGs and a more conservative estimate for the LBGs \& NB-emitters, where we assume 50\% of the 24 $\mu$m flux is contaminated by AGN heating and other systematic uncertainties associated with the 24 $\mu$m derived SFR.
	All of the other open symbols are clusters and protoclusters from $z \sim 0.4 - 3.1$ and their references may be found in the text. 
	The dark grey filled area is where the \cite{Popesso2011} clusters reach ($z\sim1.6$) and the light grey area is an extrapolation on their trend.}
	\label{fig:SFR}
\end{figure}

In Table \ref{table:SFRs}, we present calculations of the total SFRs of each protocluster. In Figure \ref{fig:SFR} we show the mass normalized integrated SFR as a function of redshift for HS1549 and HS1700, compared to other galaxy clusters from the literature (SA22 \citealt{Steidel1998, Chapman2005, Geach2005}; HDF1.99 \citealt{Chapman2009}; AzTEC-3 \citealt{Capak2011}; GCLASS \citealt{Muzzin2011}; Cl0218.3-0510 \citealt{Smail2013}; RCS \citealt{Webb2013}; MRC1138 \citealt{Dannerbauer2014a}; PCL1002+0222 \citealt{Casey2015}; XCSJ2215 \citealt{Ma2015};  ClJ1449+0856 \citealt{Strazzullo2018}). 
In addition, \cite{Clements2014} find similar trends with their overdensities of dusty, star-forming galaxies uncovered by \emph{Planck} and \emph{Herschel}.

Both protoclusters are estimated to have a total cluster mass of $\sim 10^{14}$ M$_\odot$ \citep{Steidel2005} and Steidel et al.\ in prep. We adopt a 50\% systematic uncertainty in the cluster halo mass in our error bars, which is likely a significantly larger source of uncertainty than our SFR estimates.
Combining SMGs and less luminous protocluster members 
we measure central star-formation rate densities (SFRDs) of $3000\pm900$ M$_\odot \text{yr}^{-1} \text{Mpc}^{-3}$ (HS1549) and $1300\pm400$ M$_\odot \text{yr}^{-1} \text{Mpc}^{-3}$ (HS1700). These SFRDs are $\sim10^4\times$ larger than the global SFRDs found at their respective epochs (\citealt{Madau2014}). The SFRD found for HS1700 is comparable to the SFRD computed by \cite{Kato2016} using their \emph{Herschel}-SPIRE sources.

There must have been large amounts of star formation in the early Universe to have formed the massive galaxies we see today at $z=0$. Star formation activity peaked around $z\sim2$ (e.g. \citealt{Hopkins2006}), thus we'd expect the $\text{SFR}_\text{tot}/$cluster mass $-$ redshift relation to be steeper for higher $z$ and begin to flatten out at $z\sim2$. Our findings show that the HS1549 and HS1700 protoclusters generally seem to follow a trend extending from $z<1$ rich clusters (\citealt{Popesso2011}), with a $(1+z)^7$ relation (e.g., \citealt{Geach2006}; \citealt{Casey2015}; \citealt{Kato2016}).

Finally, 
to  better put our results in context, we consider how these KBSS protoclusters were chosen.
The selection function of these protoclusters is 
not free from bias. While they were uncovered in a uniform and well defined spectroscopic survey, the fields were chosen to be centred on some of the most luminous QSOs in the Universe ($L_{\rm UV}>10^{14}$ L$_\odot$), and as such might serve as beacons to large scale over densities in the galaxy distribution. 
Indeed the HS1549 LBG over-density is at the redshift of the {\it hyper-luminous} QSO (HLQSO), and is the only one of 15 HLQSO fields to show a protocluster at the QSO redshift.
By contrast, the HS1700 HLQSO is not at the protocluster redshift, and is not biasing the selection of this protocluster.
 \cite{Trainor2012} have demonstrated that in general, HLQSOs  do not inhabit especially massive halos, and instead are rare events in comparable mass halos to less luminous QSOs.

Other protoclusters we compare to were found from High redshift Radio Galaxy (HzRG) beacons of over densities (e.g., \citealt{Stevens2003a}; \citealt{Dannerbauer2014a}) or from SMG over densities themselves (e.g., \citealt{Chapman2009}; \citealt{Casey2015}), which have their own associated biases (e.g., \citealt{DeBreuck2004}; \citealt{Ivison2013}).

An interesting aspect of our findings is that our two massive protoclusters (HS1549 and HS1700) reveal mass weighted total SFRs comparable to two systems which show much less significant over densities in LBGs (HDF1.99 \citealt{Chapman2009} and PCL1002 \citealt{Casey2015}). Our protoclusters  are plausibly hosted by much more massive dark matter halos than these latter {\it SMG-dominated} systems, which may highlight more active periods in less massive clusters. 
Two other protoclusters showing comparable LBG over densities to HS1549 and HS1700 ate the SSA22 and MRC1138 protoclusters. Both are apparently less active in ongoing star formation, suggesting that a range of activities and assembly histories are possible for massive protoclusters at $z\sim2-3$.


\section{Conclusions}
\label{sec:conclusions}

We have presented an analysis of a SCUBA-2 sub-millimetre follow up survey for the HS1549 and HS1700 protocluster fields, containing two of the largest know galaxy over-densities at $z>2$. We conclude:
\begin{itemize}
\item We detect 56 SMGs at 850 $\mu$m in the deep SCUBA-2 maps over $\sim 50$ arcmin$^2$ covering both HS1549 and HS1700 survey fields, containing two of the largest galaxy over-densities at $z>2$. The number counts indicate significant over-densities in the $\sim 1.5$ Mpc core region for each protocluster field, $6^{+4}_{-2}\times$ (HS1549) and $4^{+6}_{-2}\times$ (HS1700).
\item IR counterparts are identified for the SMGs via multi-wavelength identification using near- and mid-IR archival data. We employ P-values to each IR counterpart and identify one possible member within each protocluster. Using CO detections, near-IR and optical spectroscopy we determine 3 (4) SMGs solidly identified in the HS1549 (HS1700) protocluster. With the addition \emph{Herschel}-SPIRE colours, we uncover eight additional candidate protocluster members to the HS1700 protocluster.
\item Stacking all known protocluster galaxies and field galaxies we have found a statistically significant increase in submm activity of protocluster galaxies. 
\item Combining SCUBA-2 detected SMGs within the protoclusters and less luminous members we find both protocluster have large integrated mass-normalized star-formation rates that are consistent with a $\propto (1+z)^7$ trend. Both protoclusters have star-formation rate densities $\sim10^4\times$ larger than the global star-formation rate densities found at their respective epochs.
\end{itemize}

\section*{Acknowledgements}
\addcontentsline{toc}{section}{Acknowledgements}
%

KL acknowledges NSGS and OGS for support. 
SCC acknowledges NSERC and CFI for support. 
IRS acknowleges support from the ERC advanced Grant DUSTYGAL (321334), STFC (ST/P000541/1), and a Royal Society/Wolfson Merit Award.
This work is based on observations carried out with the the James Clerk Maxwell Telescope. The James Clerk Maxwell Telescope is operated by the East Asian Observatory on behalf of The National Astronomical Observatory of Japan, Academia Sinica Institute of Astronomy and Astrophysics, the Korea Astronomy and Space Science Institute, the National Astronomical Observatories of China and the Chinese Academy of Sciences (Grant No. XDB09000000), with additional funding support from the Science and Technology Facilities Council of the United Kingdom and participating universities in the United Kingdom and Canada. 
The James Clerk Maxwell Telescope has historically been operated by the Joint Astronomy Centre on behalf of the Science and Technology Facilities Council of the United Kingdom, the National Research Council of Canada and the Netherlands Organization for Scientific Research. 
Additional funds for the construction of SCUBA-2 were provided by the Canada Foundation for Innovation.
IRAM is supported by INSU/CNRS (France), MPG (Germany) and IGN (Spain). 
The Submillimeter Array is a joint project between the Smithsonian Astrophysical Observatory and the Academia Sinica Institute of Astronomy and Astrophysics and is funded by the Smithsonian Institution and the Academia Sinica. 
Based on observations obtained at the Gemini Observatory, which is operated by the Association of Universities for Research in Astronomy, Inc., under a cooperative agreement with the NSF on behalf of the Gemini partnership: the National Science Foundation (United States), the National Research Council (Canada), CONICYT (Chile), Ministerio de Ciencia, Tecnolog\'{i}a e Innovaci\'{o}n Productiva (Argentina), and Minist\'{e}rio da Ci\^{e}ncia, Tecnologia e Inova\c{c}\~{a}o (Brazil).

\emph{Facilities}: James Clerk Maxwell Telescope (JCMT) (\citealt{Holland2013}), Submillimeter Array (SMA), and Gemini.

\emph{Software}: \textsc{Starlink} (\citealt{currie2014}), \textsc{smurf} (\citealt{Chapin2013}; \citealt{Jenness2013}), Python version 2.7, Matplotlib (\citealt{matplotlib}), Astropy (\citealt{astropy}), and APLpy (\citealt{aplpy}).




\bibliographystyle{mnras}
\bibliography{bibliography} 

\section*{Appendix A: Comparison to other SMG samples}
\label{sec:compareSMGs}

\begin{figure*}
	\centering
	\includegraphics[width=\textwidth]{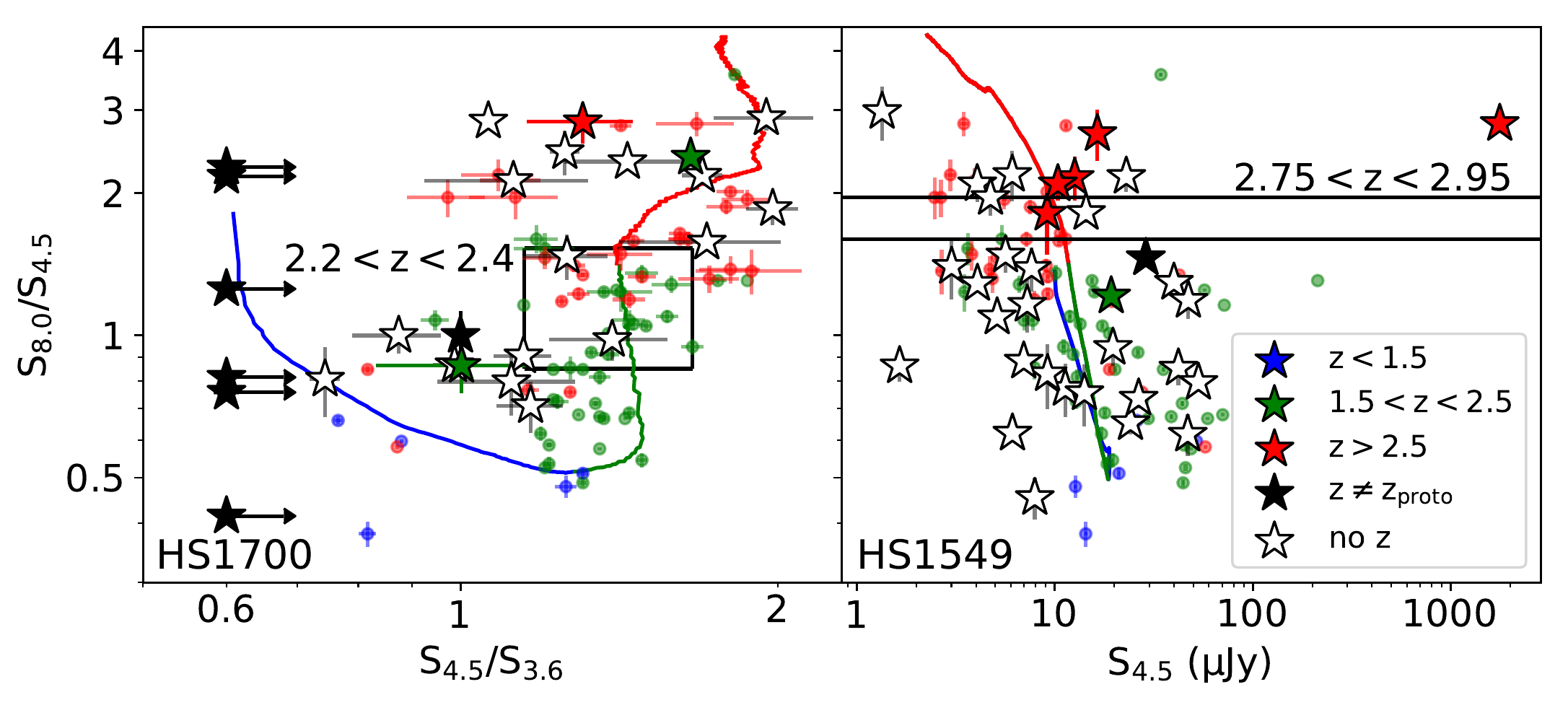}
	\caption[CC-CM figure]
	{The ratio of IRAC fluxes for IRAC IDs for both HS1700 (\emph{left}) and HS1549 (\emph{right}), represented as star symbols. We compare our IDs to ALMA SMGs from \cite{Simpson2014}, represented as filled circles, and an SED track from SMM J2135-0102, a $z = 2.3259$ SMG (\citealt{Swinbank2010}). The data are colour-coded by redshift bins, the open stars have no redshift information, the black filled in stars represent galaxies that explicitly do not reside within the associated protocluster, and the black lines represent regions defined by ALESS SMGs within $z_\text{proto}\pm0.1$.}
	\label{fig:CC-CM}
\end{figure*}

As an additional assessment of the redshifts of our IRAC identifications, we compare the ratio of IRAC fluxes for the IRAC IDs in both of our fields, HS1700 and HS1549, to 77 ALESS SMGs from \cite{Simpson2014} and an SED track from SMM J2135-0102, a $z = 2.3259$ SMG (\citealt{Swinbank2010}). 
In Figure \ref{fig:CC-CM} we show a colour-colour plot for HS1700 and a colour-magnitude diagram for HS1549, since the IRAC 3.6 $\mu$m / 5.6 $\mu$m bands do not cover much of the HS1549 field.
All six IRAC IDs in HS1549 with redshifts have both IRAC fluxes. However, in HS1700 only three of seven submm sources (one of four cluster sources) with spectroscopic redshifts have all measurable fluxes in all three IRAC wavebands (\emph{1700.3\_1} ($z=2.318$), \emph{1700.12} ($z=2.72$), \emph{1700.16} ($z=1.575$)).
We define regions where ALESS SMGs have redshifts of $z_\text{proto}\pm0.1$. 
We measure the contamination rate of ALESS SMGs residing within the region defined by $z_\text{proto}\pm0.1$ and we find $\sim1/3$ of ALESS SMGs are within these regions, while $\sim2/3$ of the ALESS SMGs are either foreground or background sources, for both protocluster redshift bins.
Though we lack redshift data for many of the IRAC IDs in HS1549 and HS1700, we conclude that many of our SMG IDs have colours consistent with the relevant redshift range of ALESS SMGs and the SMM J2135-0102 SED track.

\begin{figure*}
	\centering
	\includegraphics[width=0.6\textwidth]{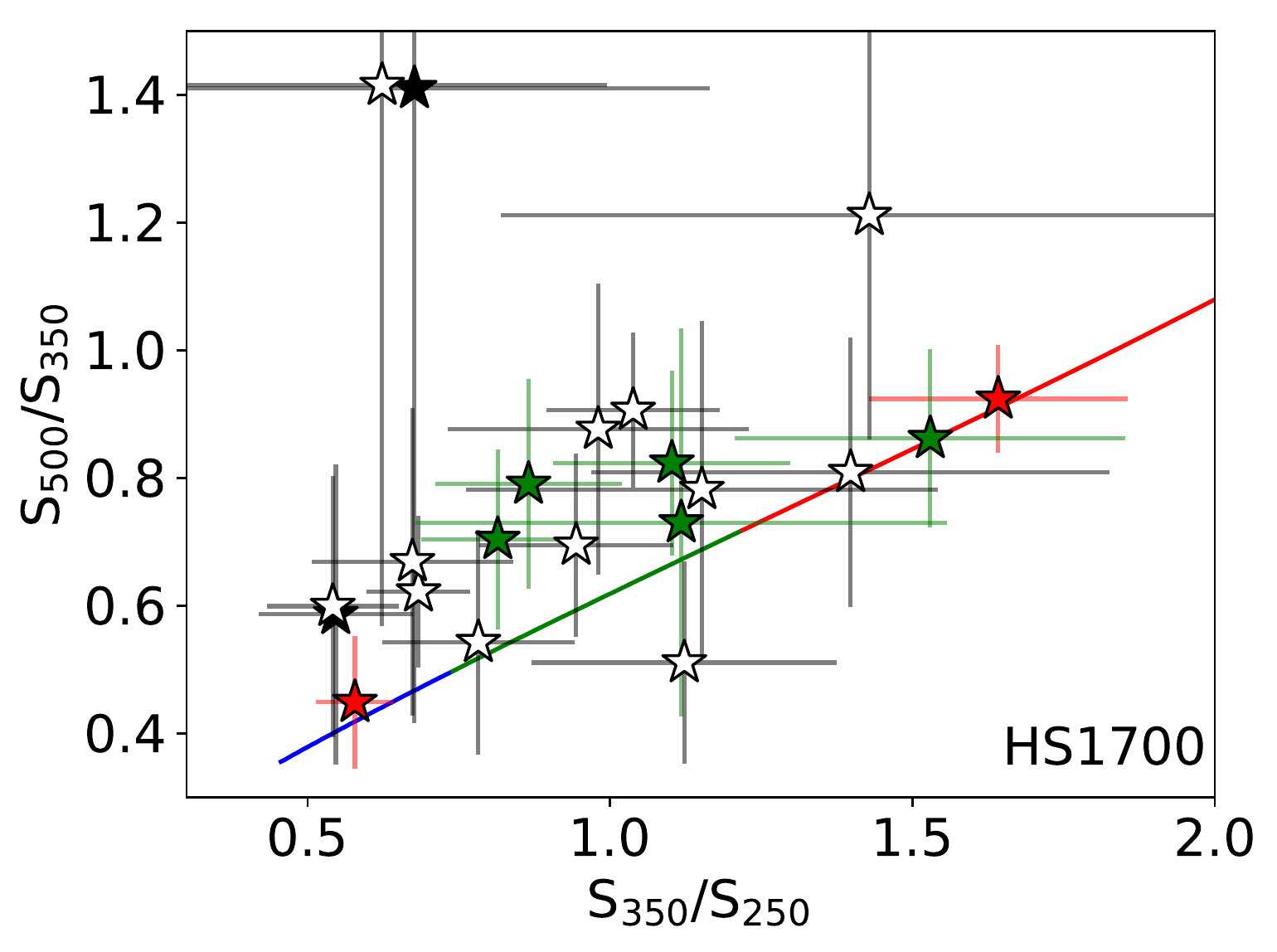}
	\caption[SPIRE-CC figure]
	{
	The ratio of SPIRE fluxes for SCUBA-2 IDs for HS1700. We compare our IDs to an SED track from SMM J2135-0102, a $z = 2.3259$ SMG (\citealt{Swinbank2010}). The data are colour-coded by the same redshift bins as Figure \ref{fig:CC-CM}, the open stars have no redshift information, the black filled in stars represent galaxies that explicitly do not reside within the associated protocluster.
	}
	\label{fig:SPIRE-CC}
\end{figure*}

Using archival \emph{Herschel}-SPIRE data from \cite{Kato2016} we measure SPIRE colours $S_{500}/S_{350}$ and $S_{350}/S_{250}$ for 21/27 SCUBA-2 IDs in the HS1700 field (see Figure \ref{fig:SPIRE-CC}).
The photometric redshifts determined by the SPIRE colours correlate to their known redshifts, with the exception of  are in agreement with the known redshifts, for the exception of the QSO \emph{1700.12}.
Using the SPIRE data, we uncover a 8 new candidate protocluster members.

\section*{Appendix B: Gemini GNIRS Spectra}
\label{sec:Gemini_spectra}

In this section we present Gemini GNIRS spectra for \emph{1700.4}, \emph{1700.5\_2}, \emph{1700.7\_1}, \emph{1700.17}, and \emph{1700.16}.
We find that \emph{1700.7\_1} and \emph{1700.16} have very strong 1D spectra. Although \emph{1700.4}, \emph{1700.5\_2}, and \emph{1700.17} have $\sim$3--4 sigma line detections at the expected $\lambda_\text{obs} \sim$21660 $\angstrom$, the strong sky line residuals in the surrounding regions mean that our IDs are still somewhat uncertain, and as a result the three IDs may have systematic offsets of a couple hundred km s$^{-1}$ due to sky line residuals. 

\begin{figure*}
\includegraphics[width=\textwidth]{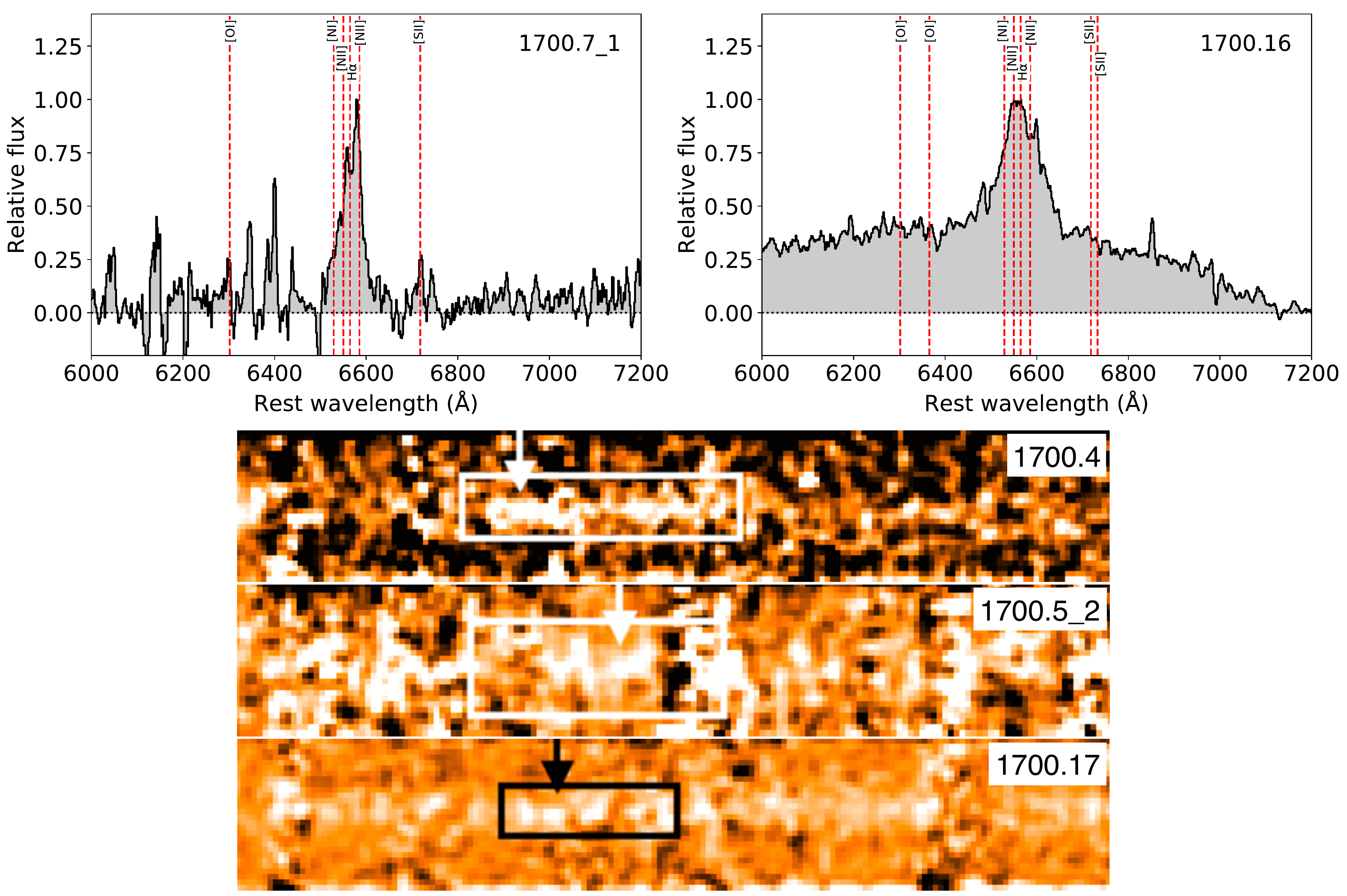}
\caption[Gemini-GNIRS spectra]
{
Gemini GNIRS spectra of the five IRAC IDs with positive detections. 
\textbf{Top:} 1D spectra for \emph{1700.7\_1} and \emph{1700.16}. The vertical axis represents the flux relative to the maximum flux, and the horizontal axis is the observed wavelength normalized by a factor of $(1+z)$. Each spectra is continuum subtracted and the vertical dashed lines represent prominent spectral lines fit to each of the spectra.
\textbf{Bottom:} 2D spectra for \emph{1700.4}, \emph{1700.5\_2}, and \emph{1700.17}. The boxes represent the same range in wavelengths as plotted in the 1D spectra. There are strong sky line residuals in the regions surrounding the spectral lines.}
\end{figure*}

\section*{Appendix C: Multi-wavelength cutouts}

$30\times30$-arcsec$^2$ multi-wavelength cutouts of each $850\mu$m SCUBA-2 source. From left to right: SCUBA-2 850 $\mu$m, SCUBA-2 450 $\mu$m, \emph{Spitzer}-MIPS 24 $\mu$m, \emph{Spitzer}-IRAC 8.0 $\mu$m, \emph{Spitzer}-IRAC 4.5 $\mu$m, 1.2 $\mu$m $J$-band.
We use \emph{Spitzer}-IRAC 5.6 $\mu$m and \emph{Spitzer}-IRAC 3.6 $\mu$m when 8.0 $\mu$m and 4.5 $\mu$m are unavailable.
LBGs are shown as red circles if they reside within the protocluster, otherwise they are shown as cyan coloured circles.
The IRAC ID galaxies are represented as green circles.
The white contours represent SNR$_{850} = (4,5,6,8,10,12)$.
The large black dashed circle represents SCUBA-2's 15$^{\prime\prime}$ beam at 850 $\mu$m.
The yellow contours shown in \emph{1549\_1} represent SNR$_\text{SMA} = (3.5,4.5,5.5)$.

\noindent
\begin{figure*}
\emph{1549.1} \hspace{3in} \emph{1549.2}\\
\includegraphics[width=.49\textwidth]{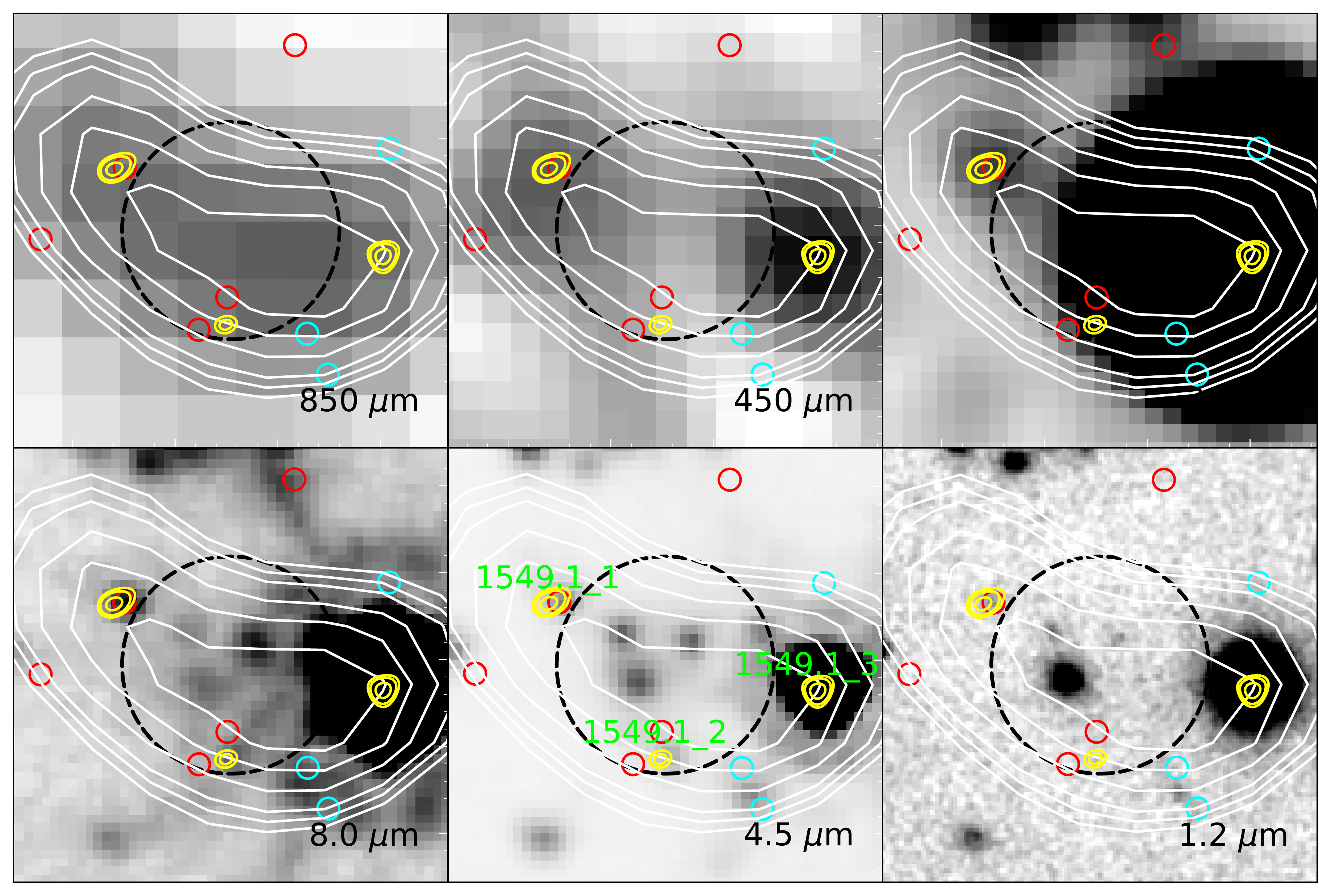}
\includegraphics[width=.49\textwidth]{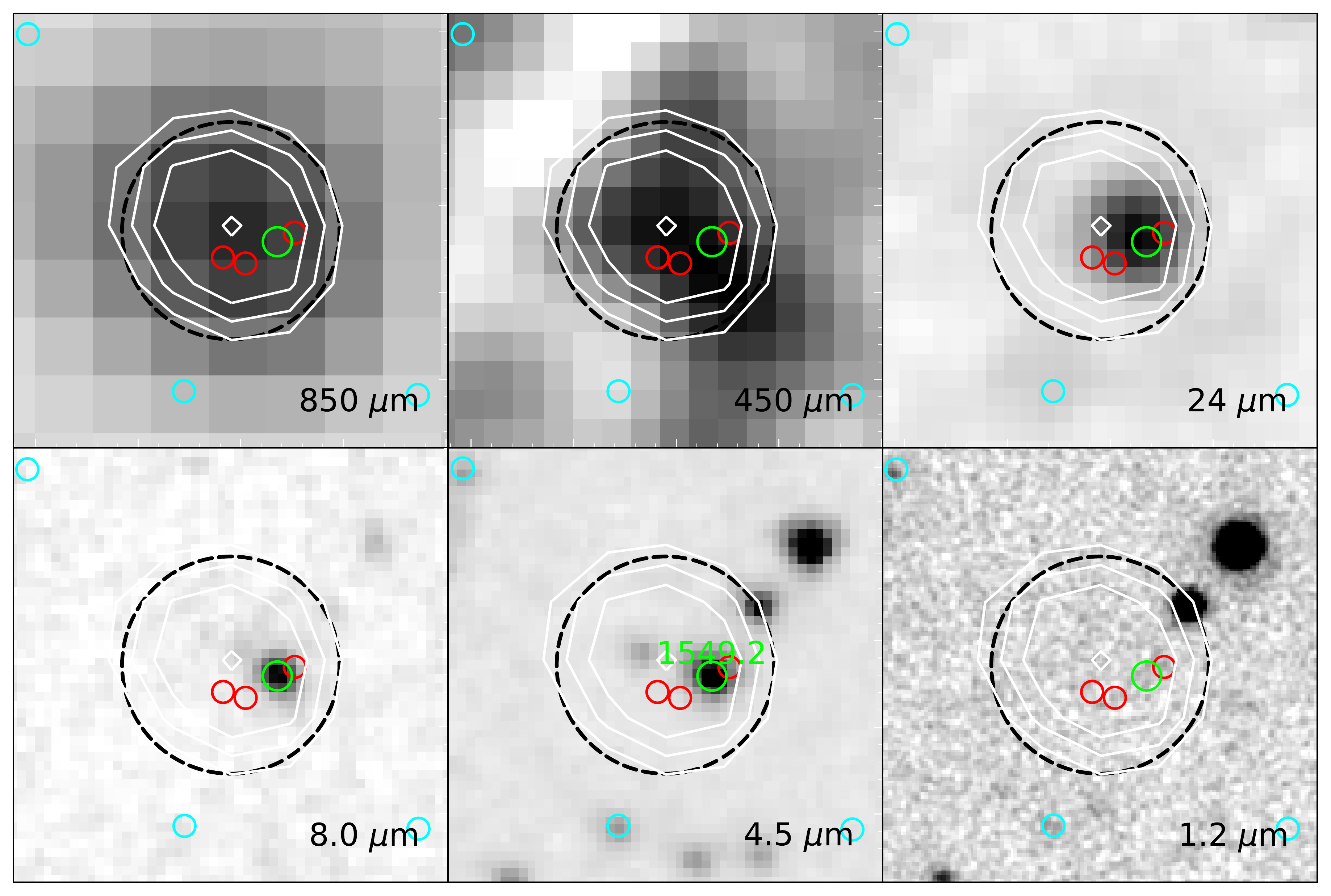}
\emph{1549.3} \hspace{3in} \emph{1549.4}\\
\includegraphics[width=.49\textwidth]{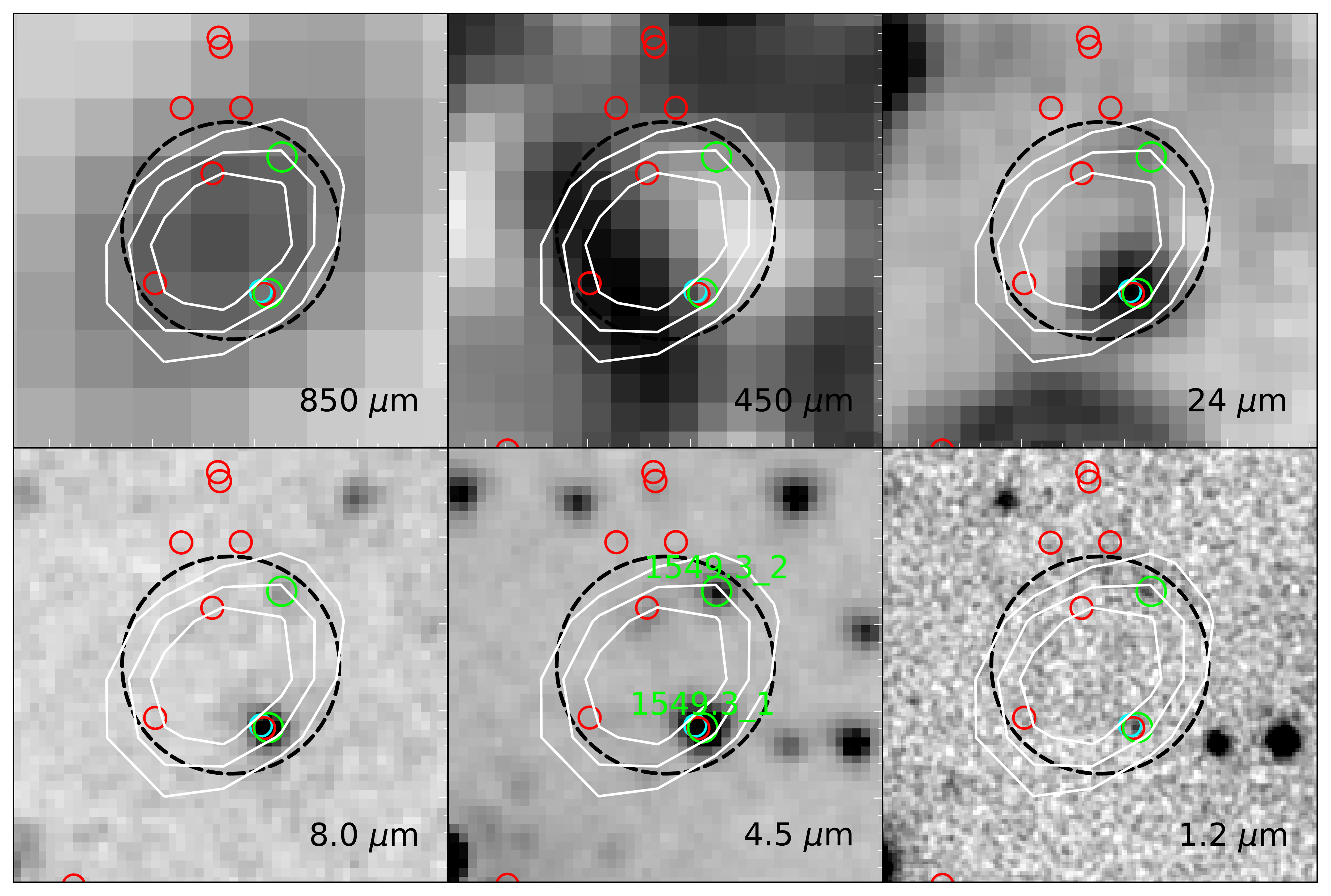}
\includegraphics[width=.49\textwidth]{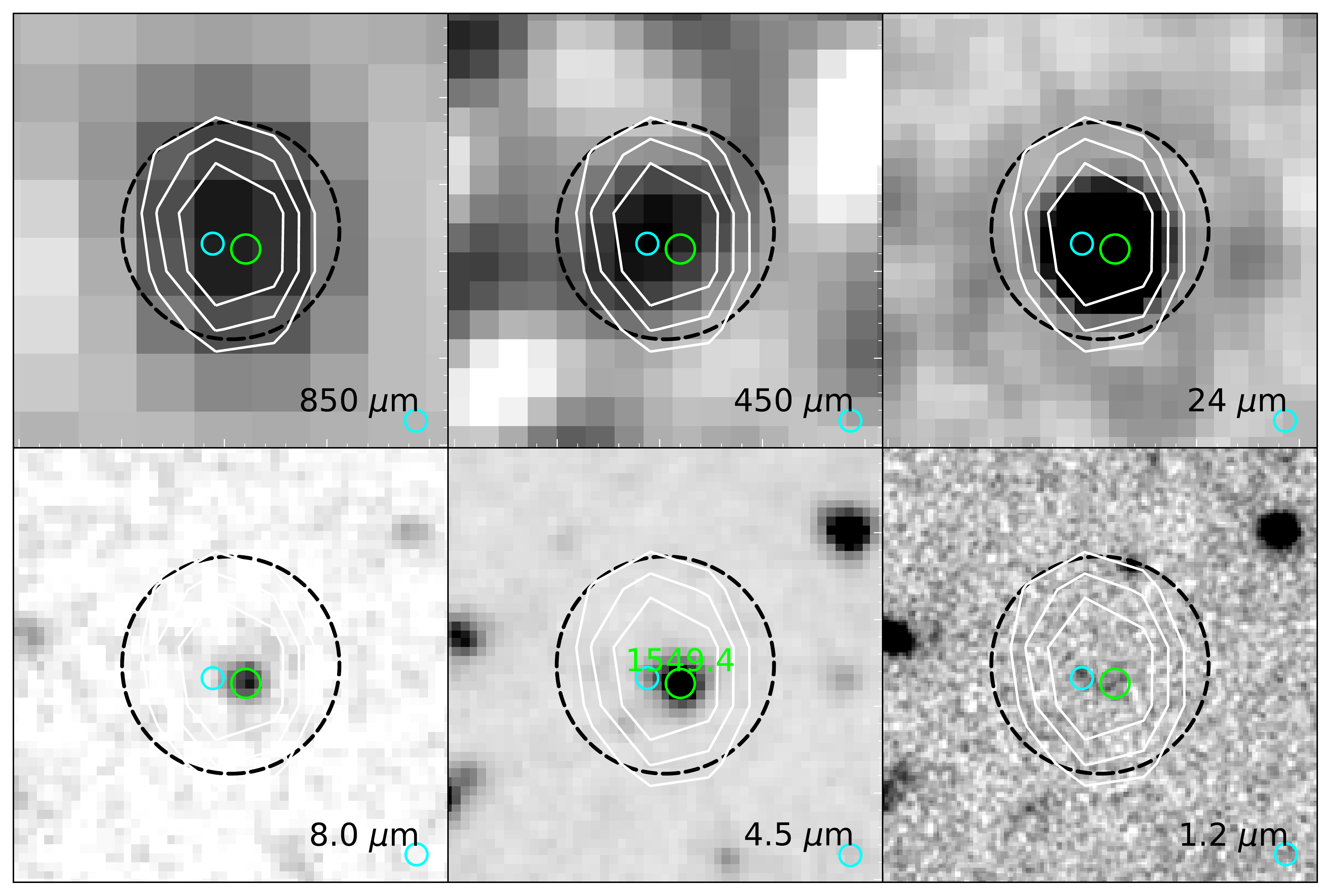}
\emph{1549.5} \hspace{3in} \emph{1549.6}\\
\includegraphics[width=.49\textwidth]{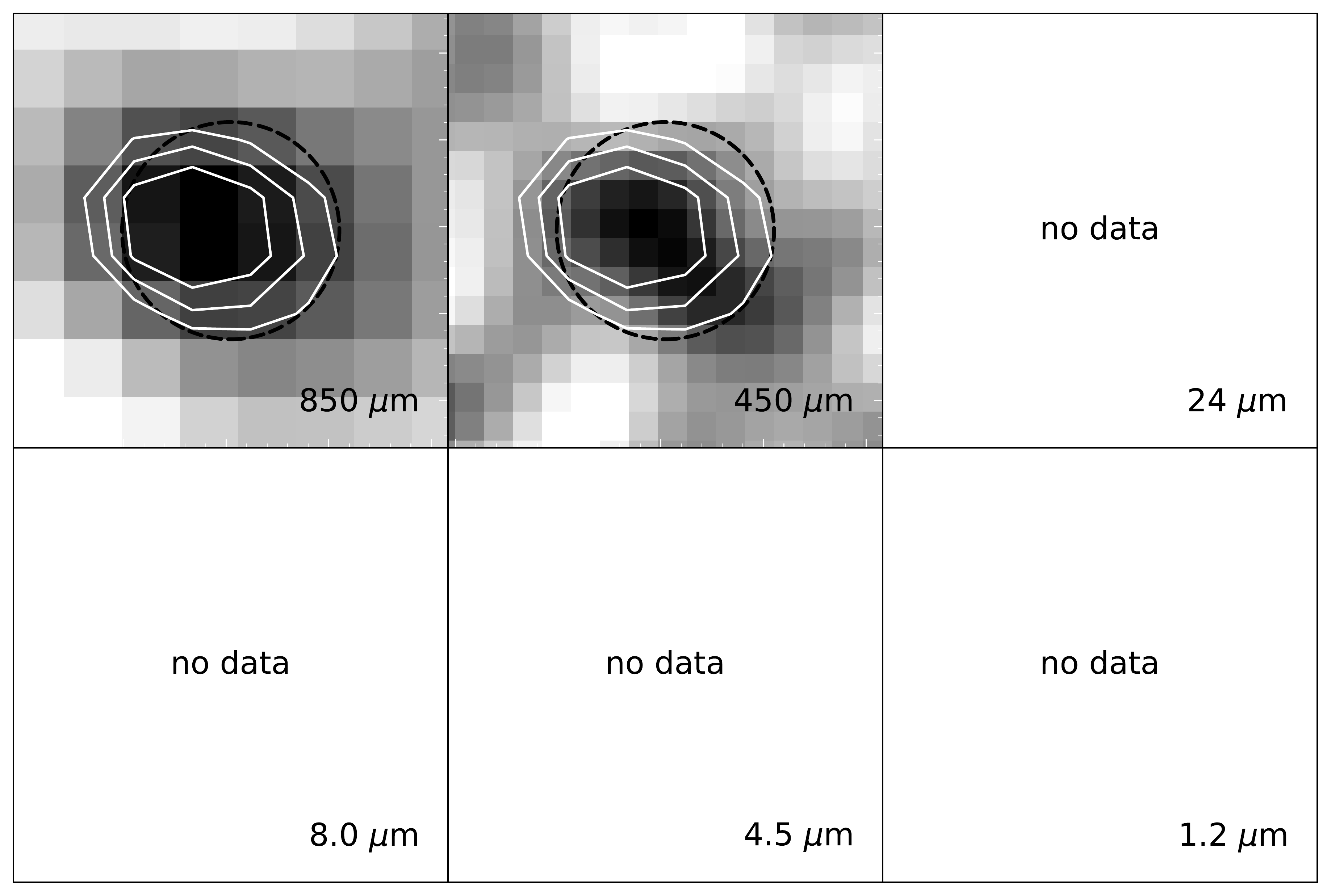}
\includegraphics[width=.49\textwidth]{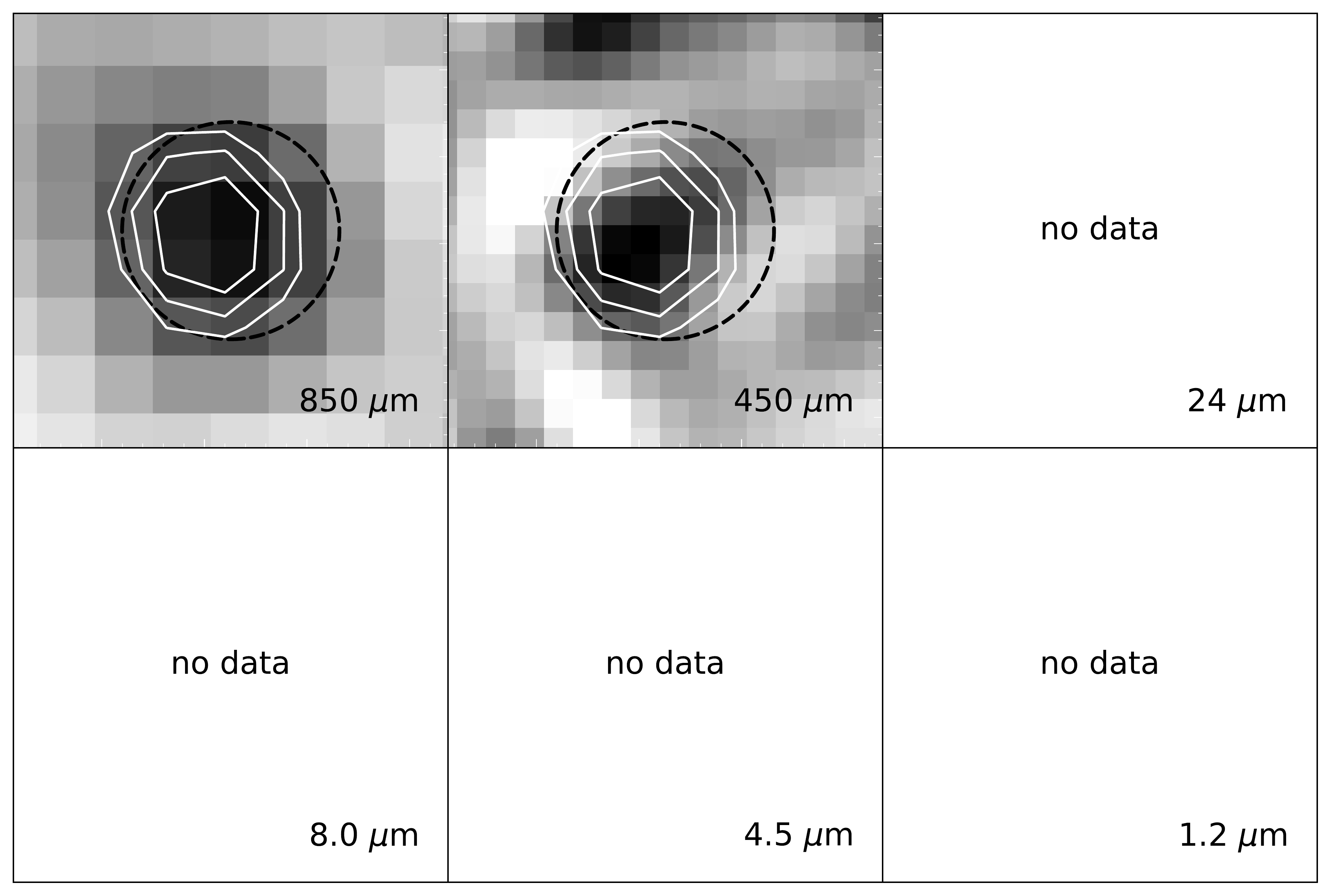}
\emph{1549.7} \hspace{3in} \emph{1549.8}\\
\includegraphics[width=.49\textwidth]{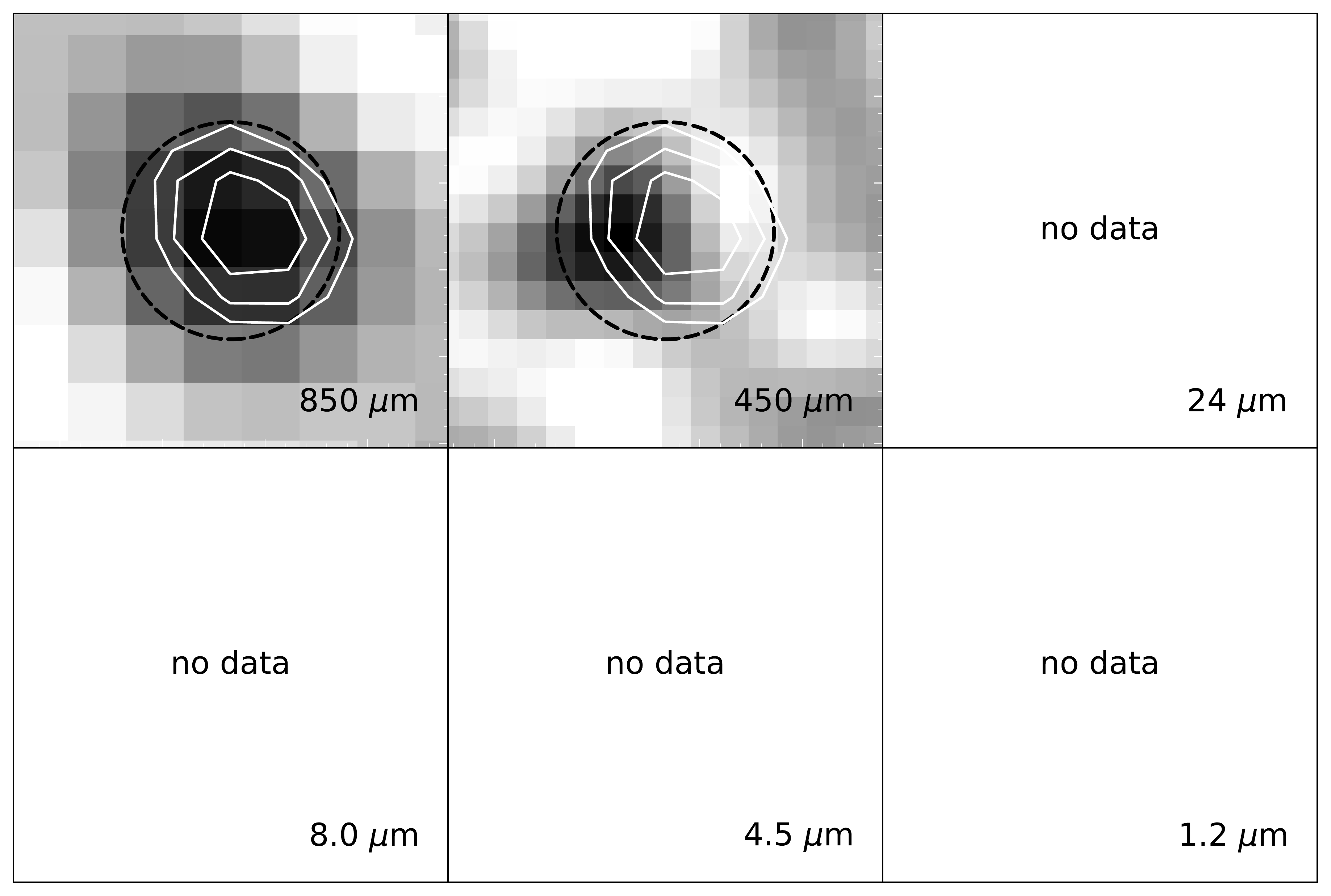}
\includegraphics[width=.49\textwidth]{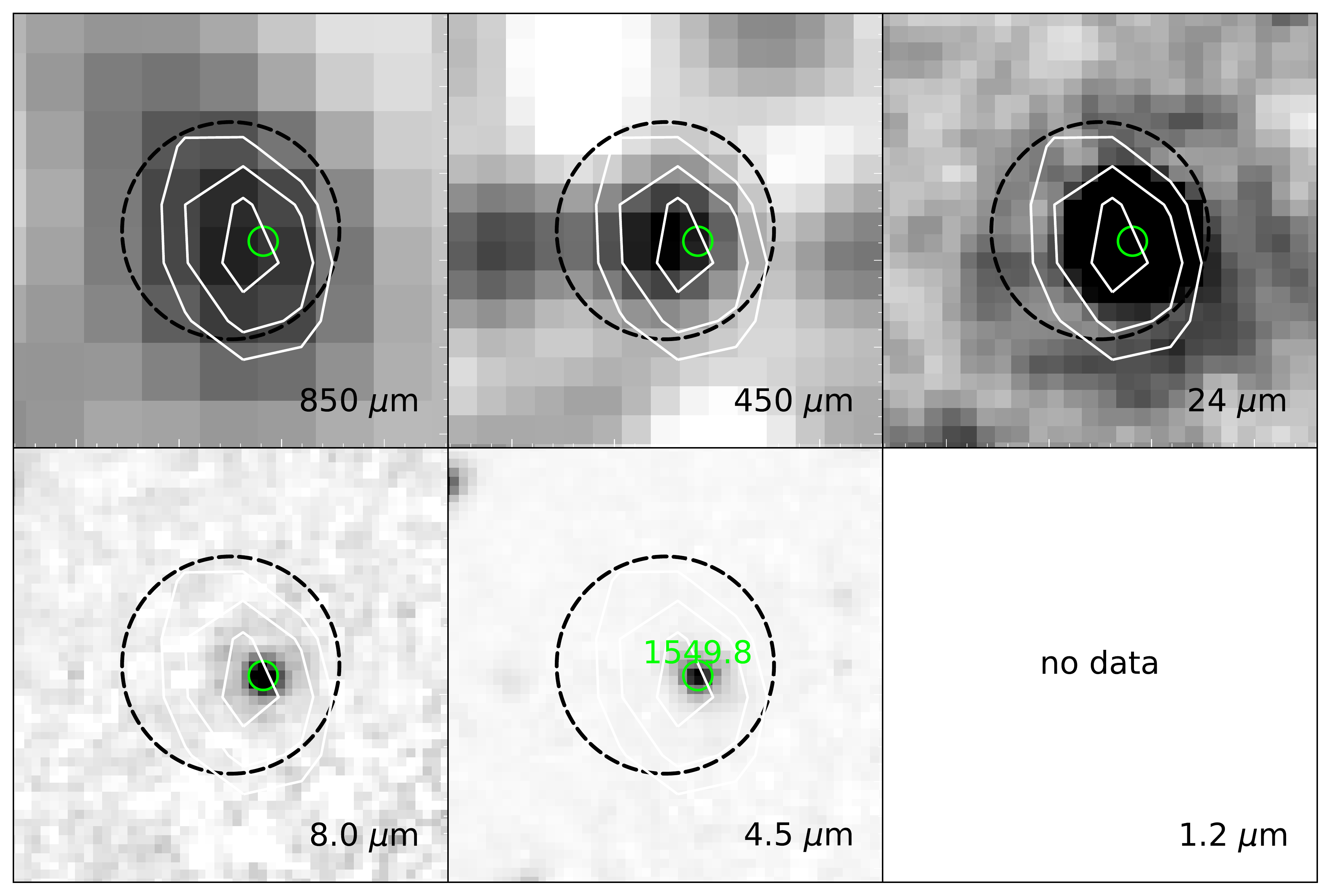}

\end{figure*}
\noindent
\begin{figure*}
\emph{1549.9} \hspace{3in} \emph{1549.10}\\
\includegraphics[width=.49\textwidth]{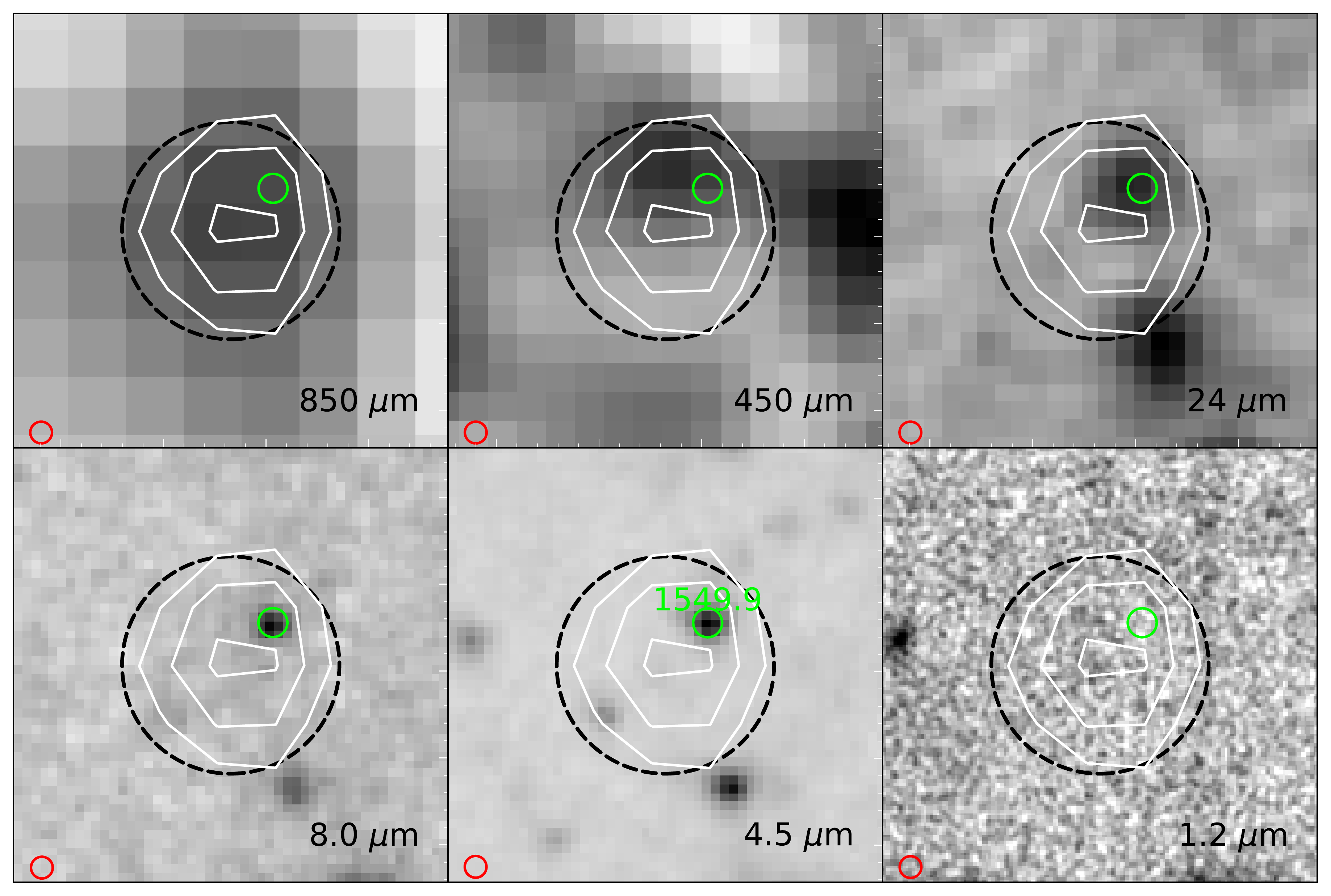}
\includegraphics[width=.49\textwidth]{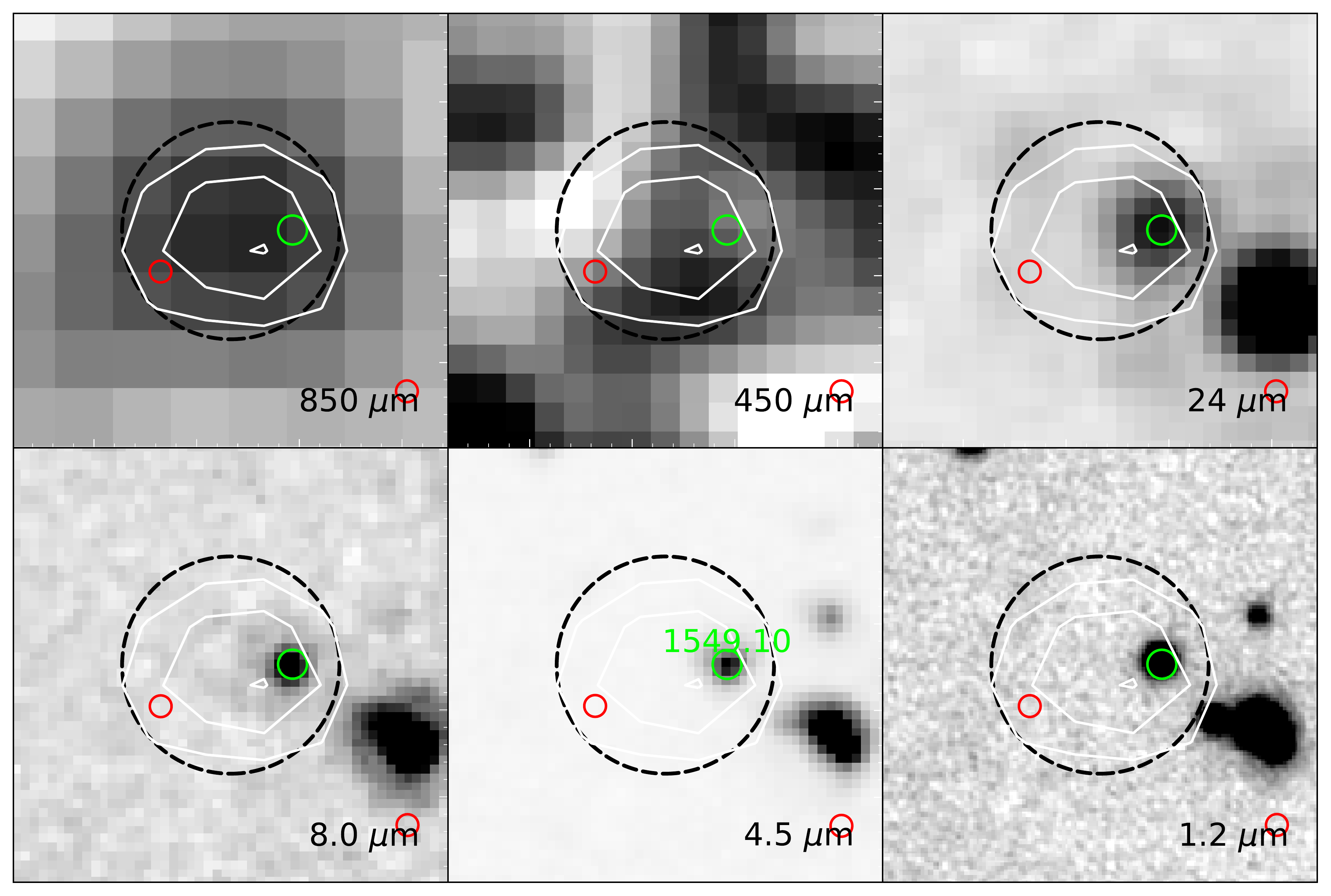}
\emph{1549.11} \hspace{3in} \emph{1549.12}\\
\includegraphics[width=.49\textwidth]{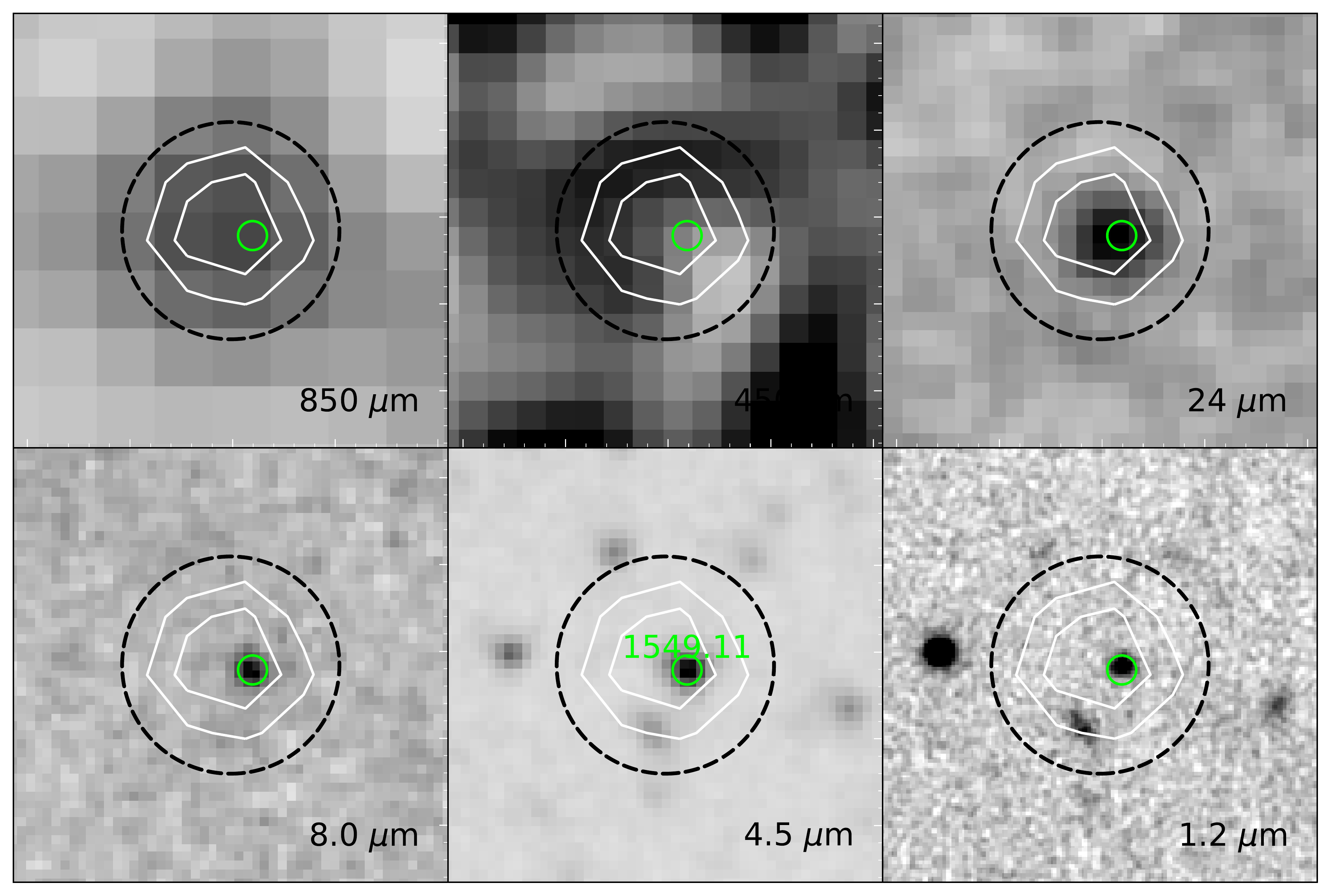}
\includegraphics[width=.49\textwidth]{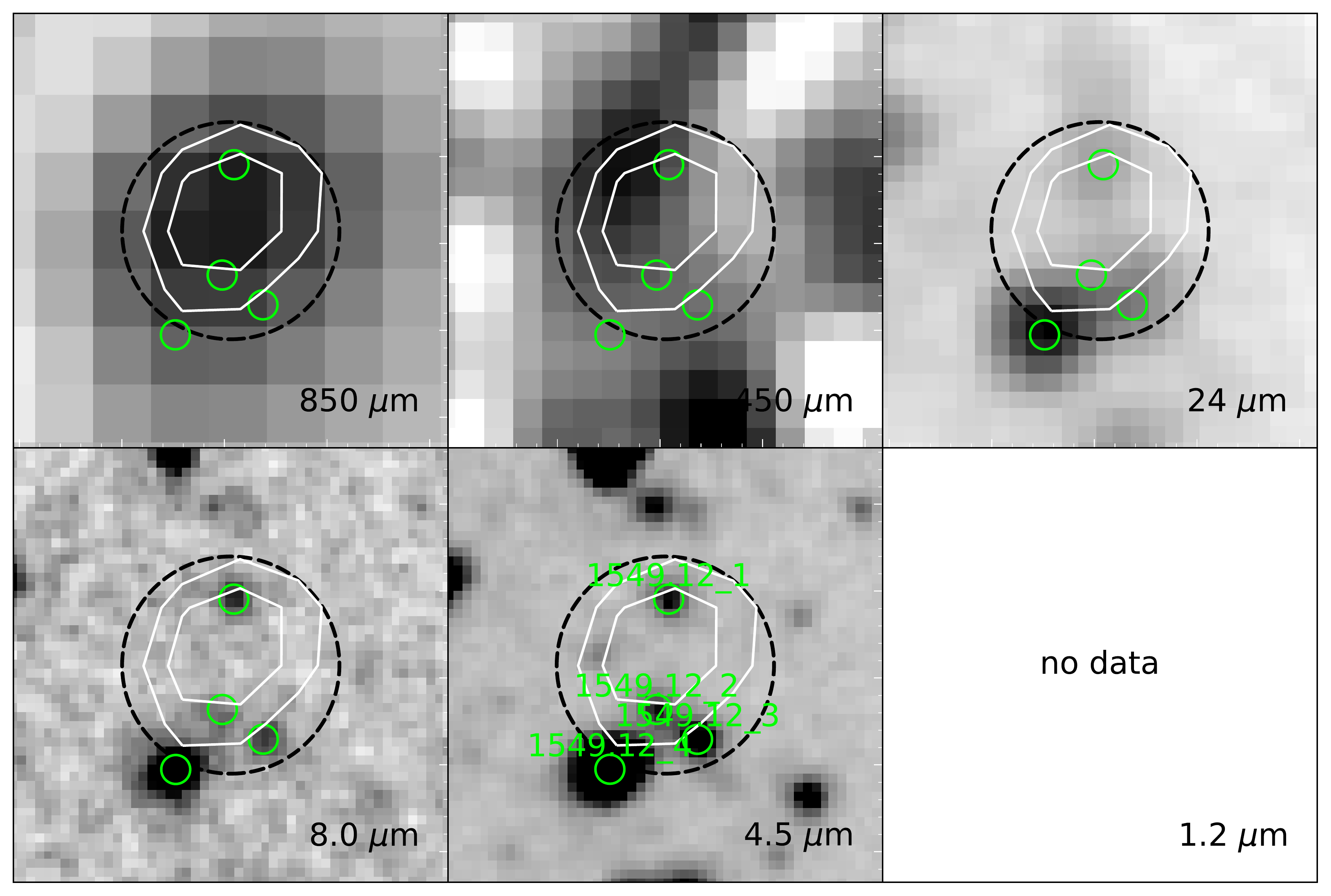}
\emph{1549.13} \hspace{3in} \emph{1549.14}\\
\includegraphics[width=.49\textwidth]{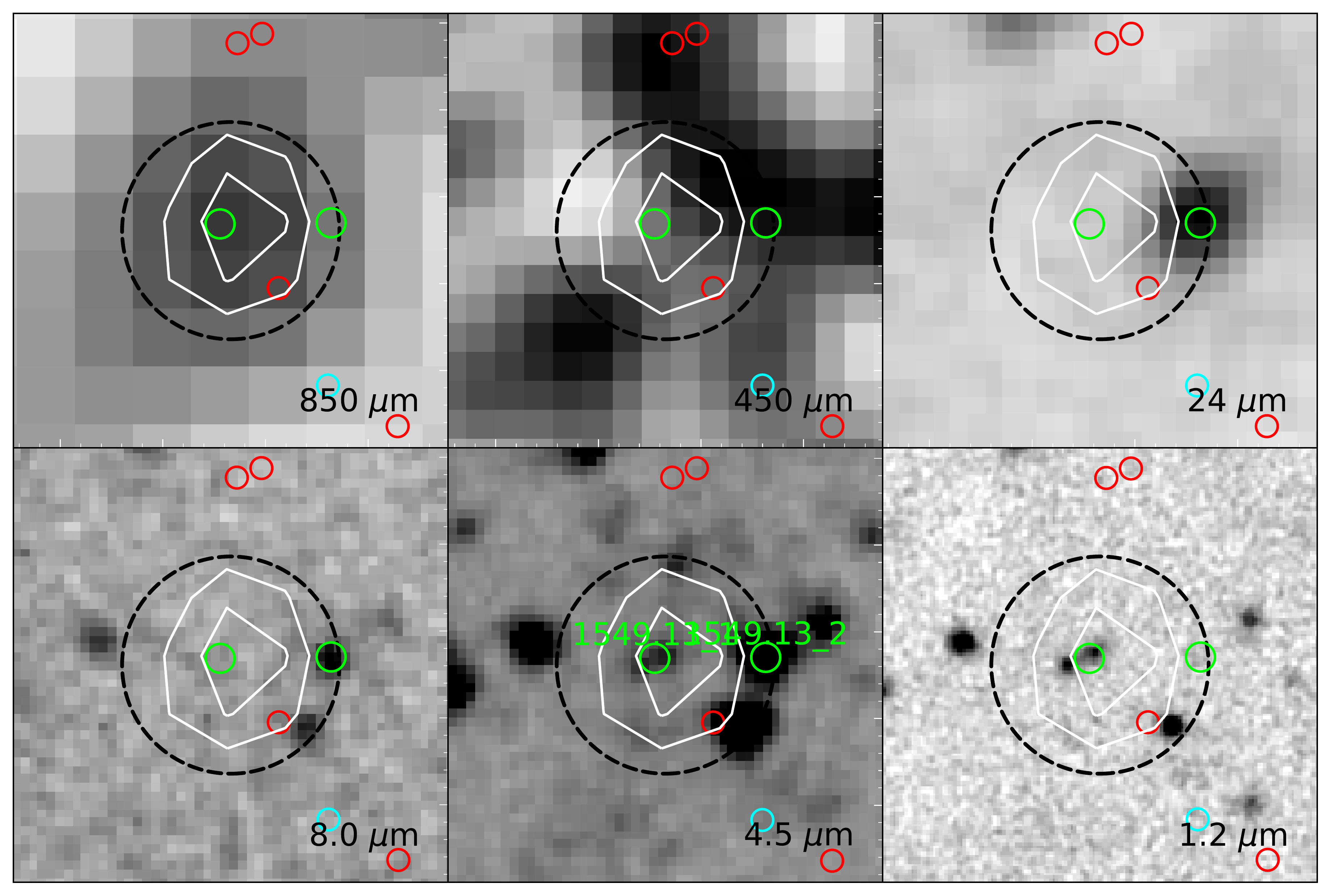}
\includegraphics[width=.49\textwidth]{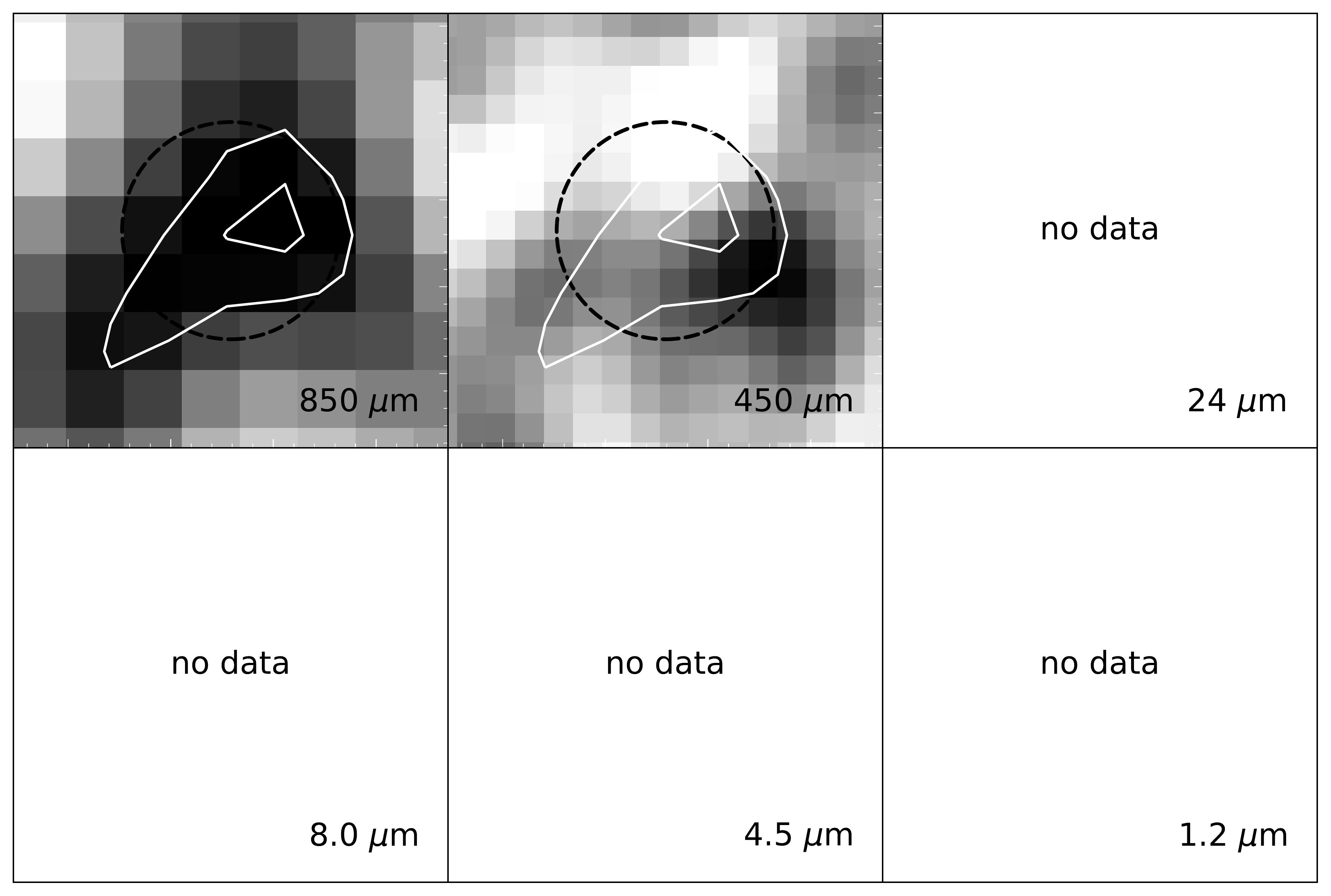}
\emph{1549.15} \hspace{3in} \emph{1549.16}\\
\includegraphics[width=.49\textwidth]{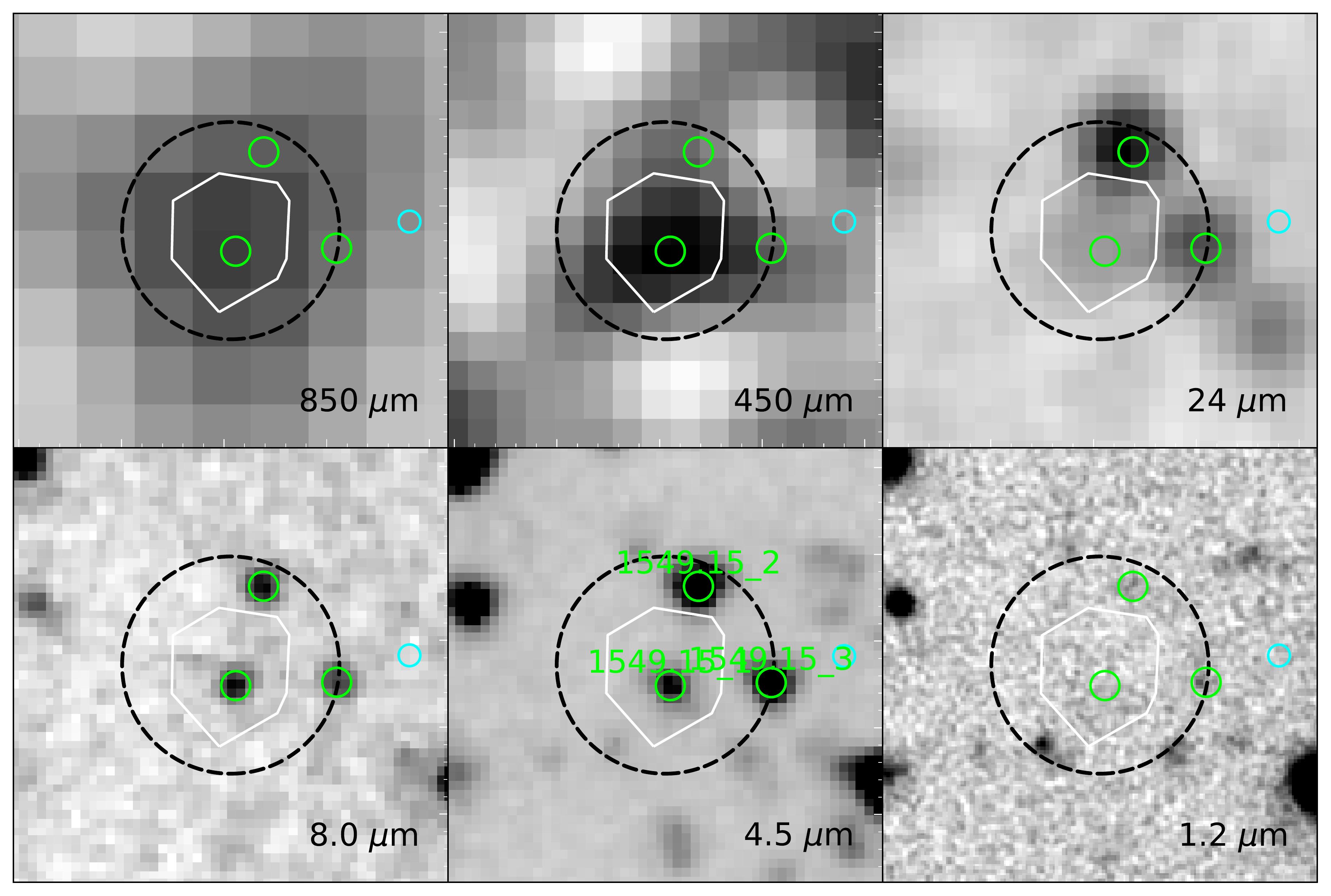}
\includegraphics[width=.49\textwidth]{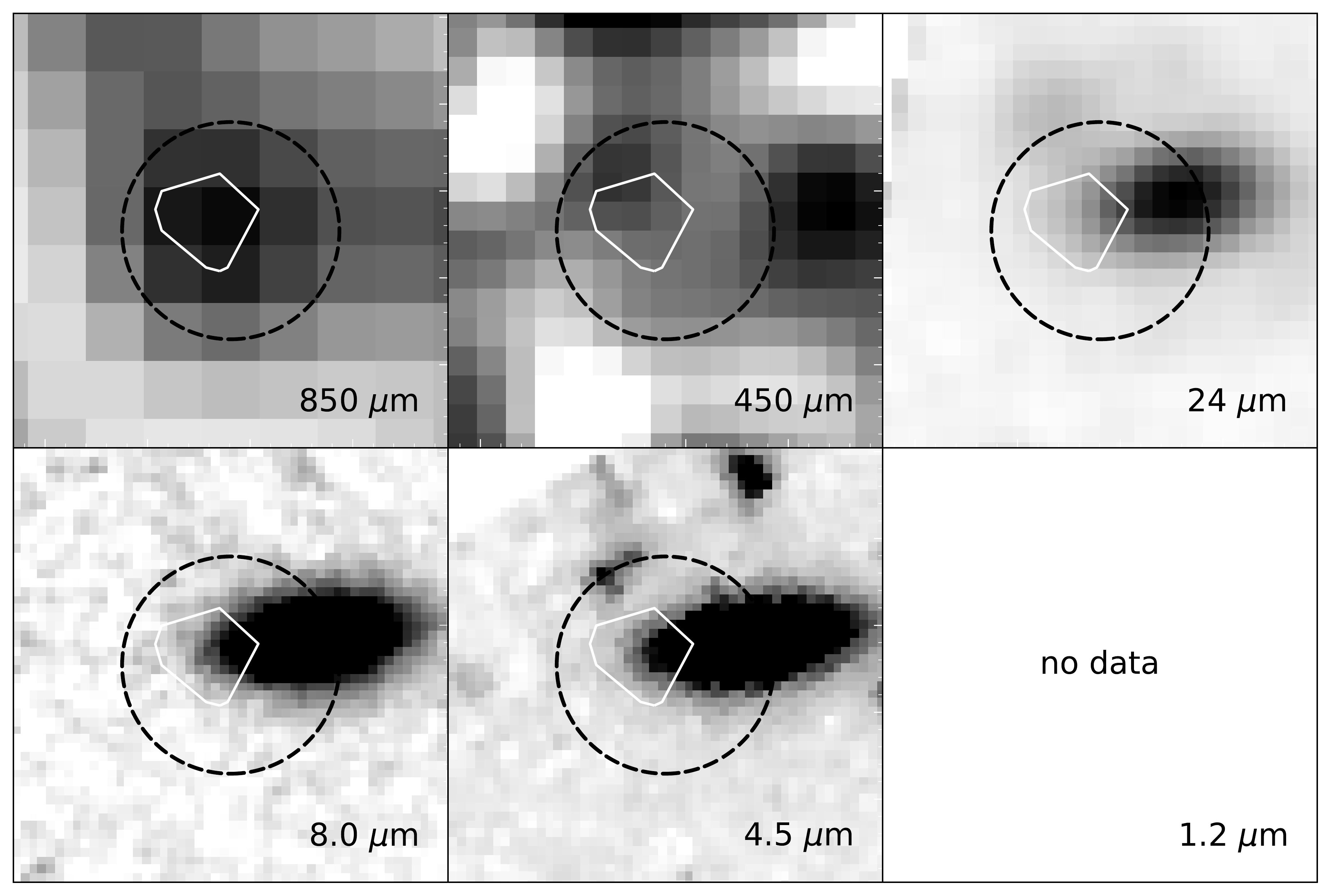}

\end{figure*}
\noindent
\begin{figure*}
\emph{1549.17} \hspace{3in} \emph{1549.18}\\
\includegraphics[width=.49\textwidth]{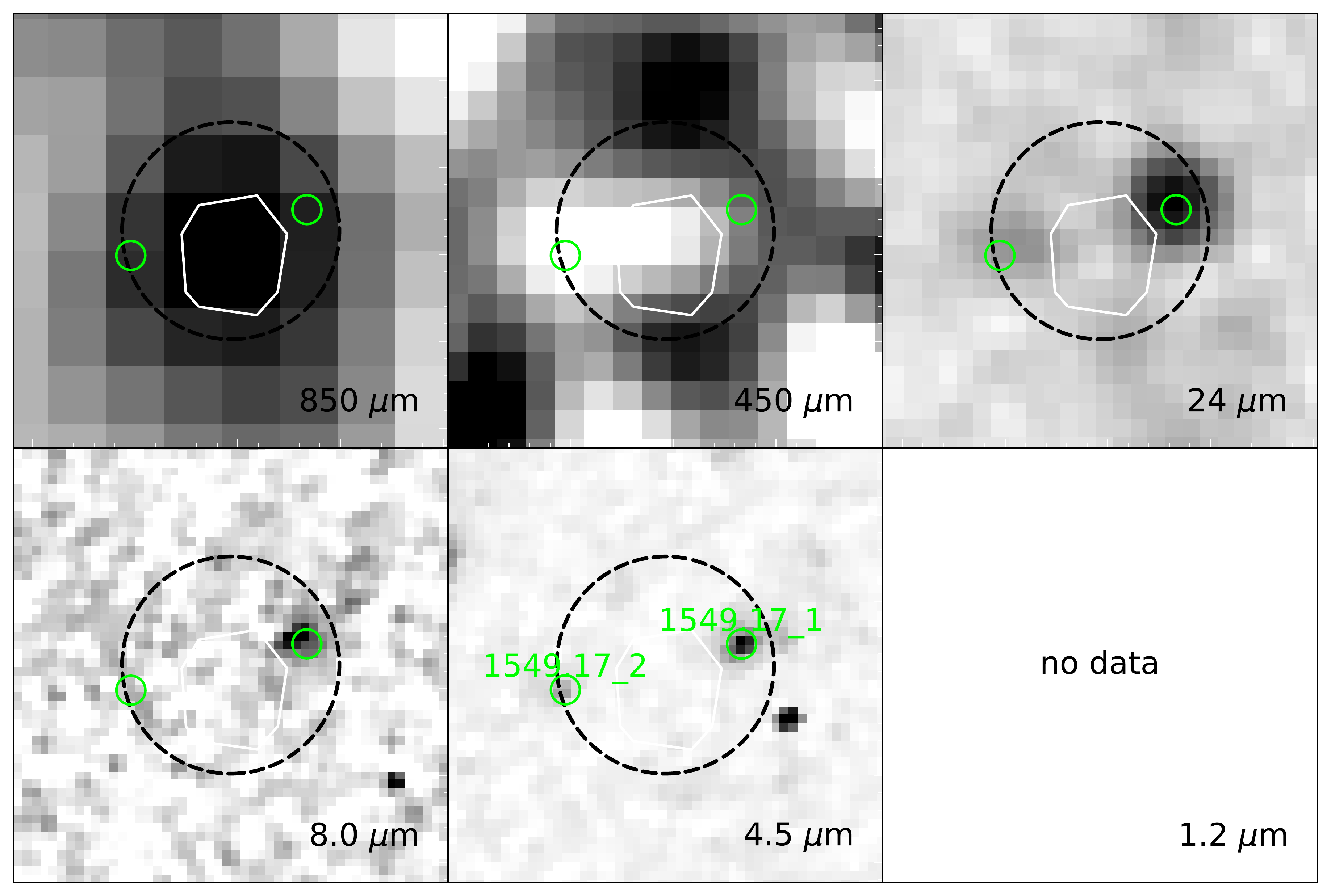}
\includegraphics[width=.49\textwidth]{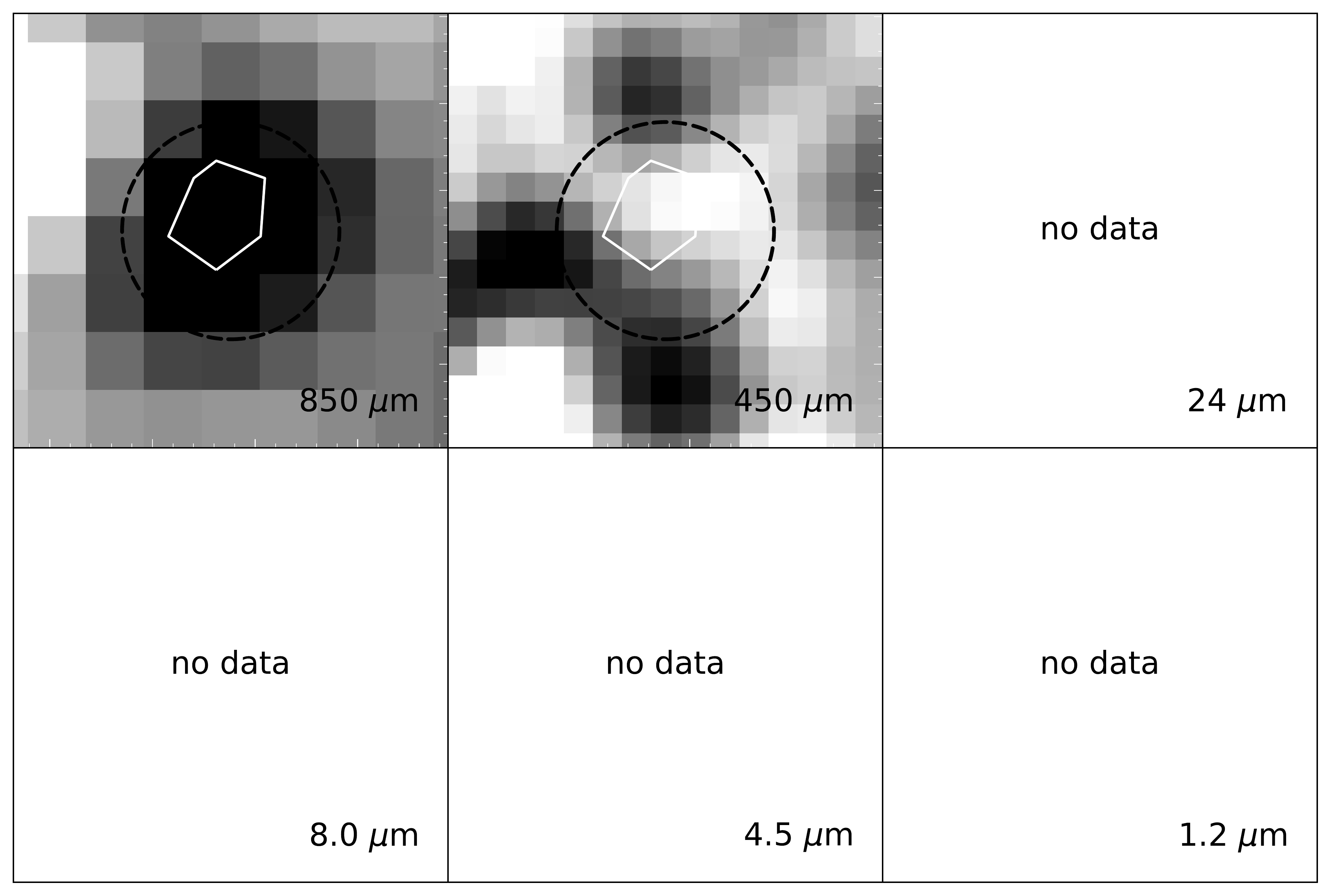}
\emph{1549.19} \hspace{3in} \emph{1549.20}\\
\includegraphics[width=.49\textwidth]{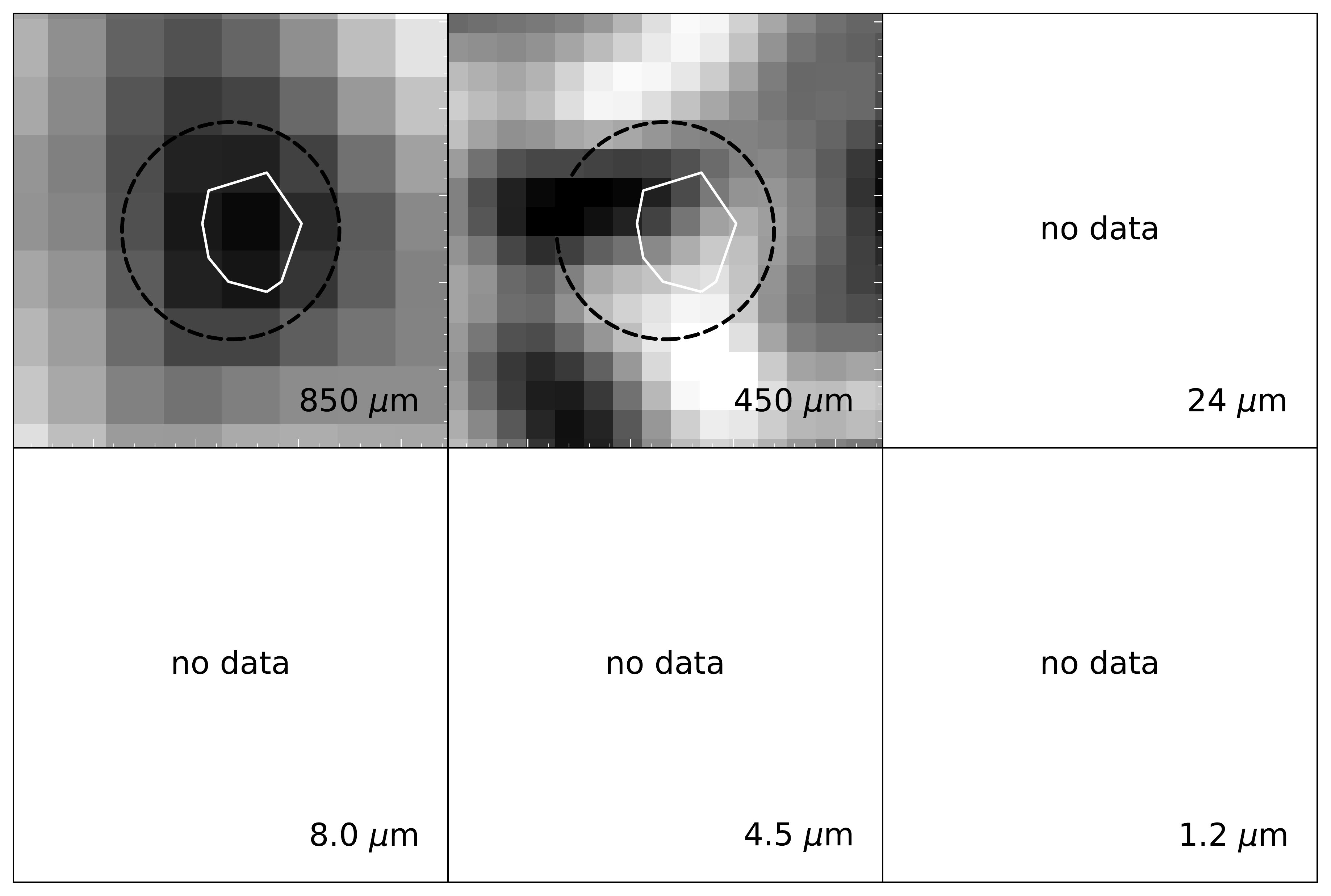}
\includegraphics[width=.49\textwidth]{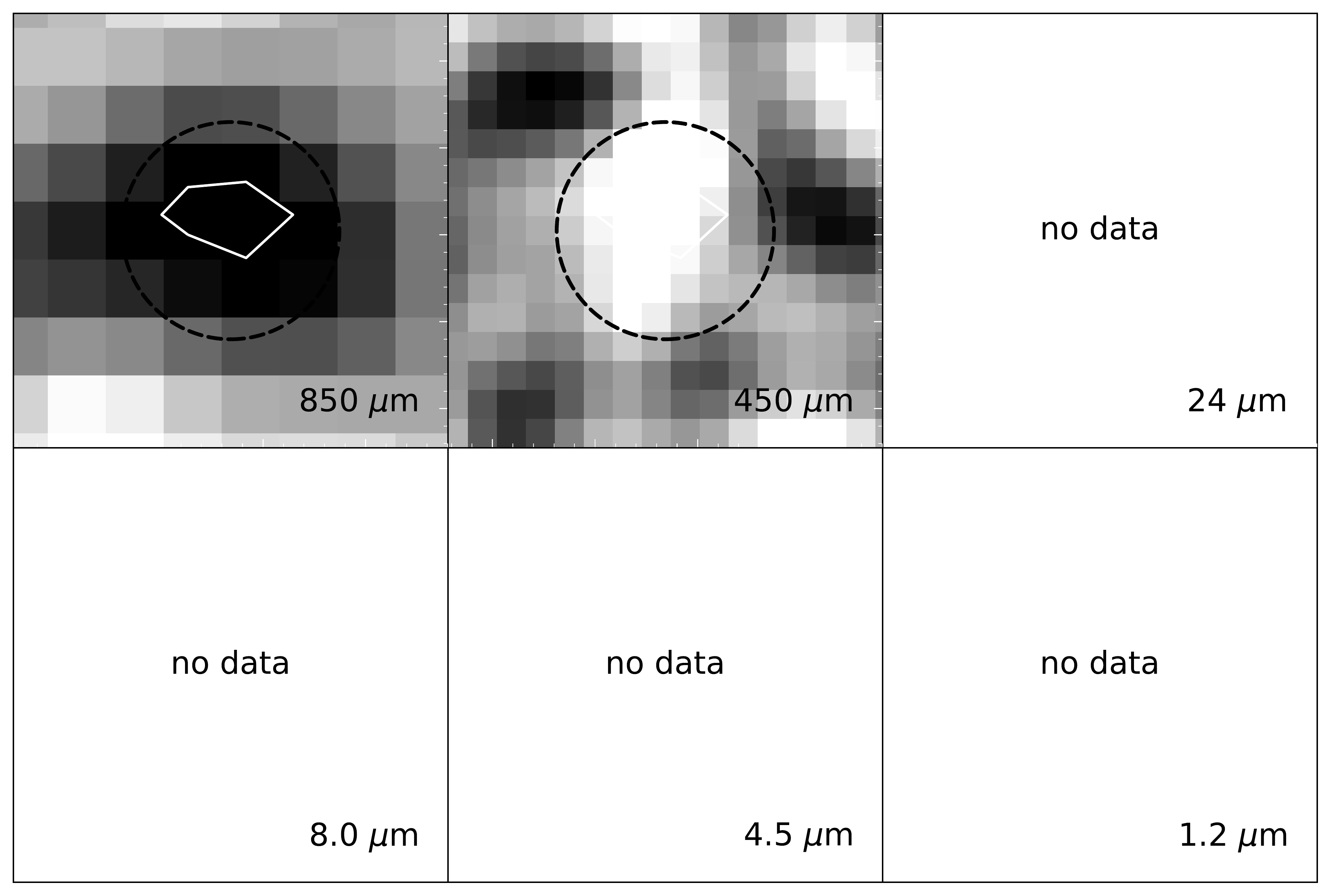}
\emph{1549.21} \hspace{3in} \emph{1549.22}\\
\includegraphics[width=.49\textwidth]{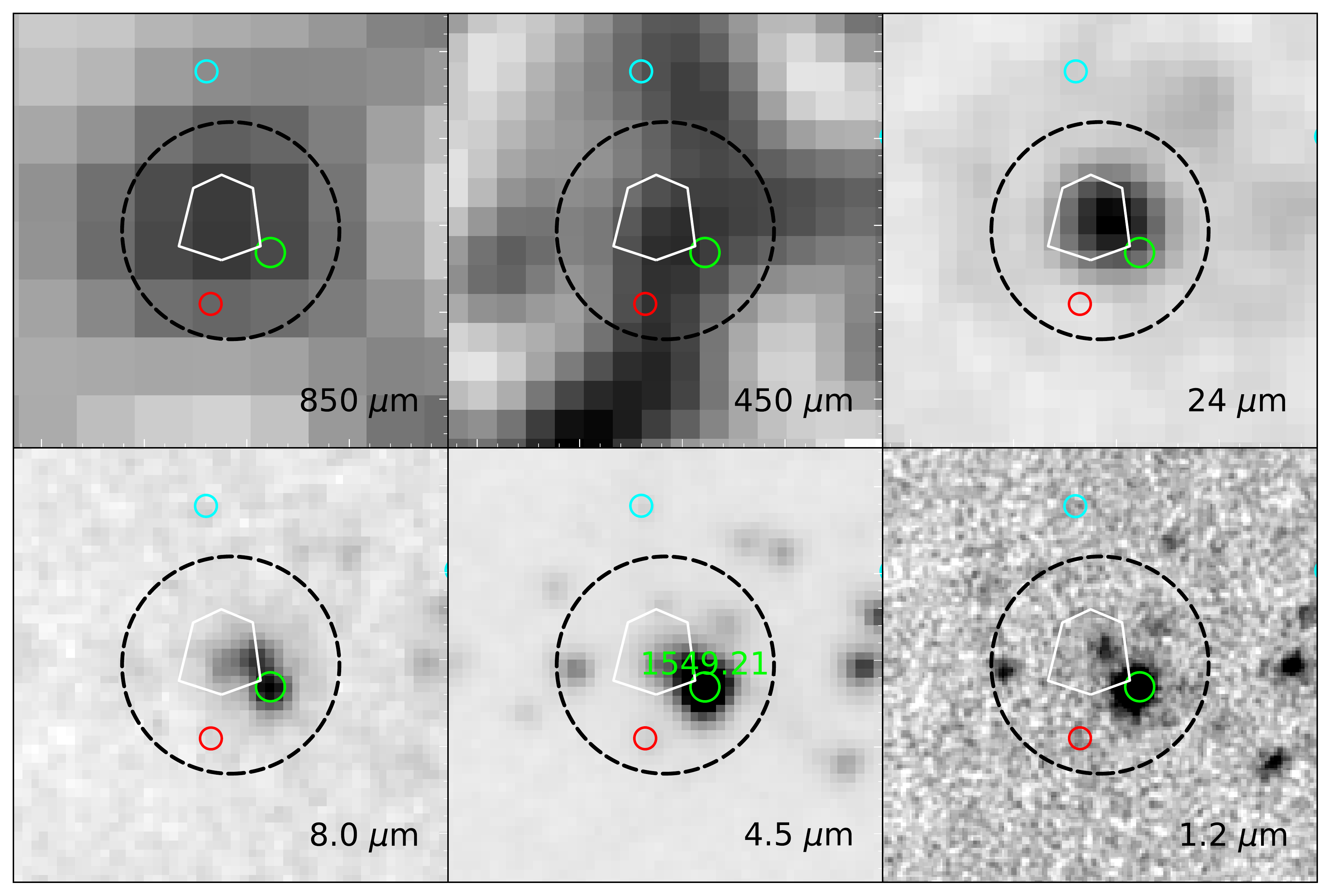}
\includegraphics[width=.49\textwidth]{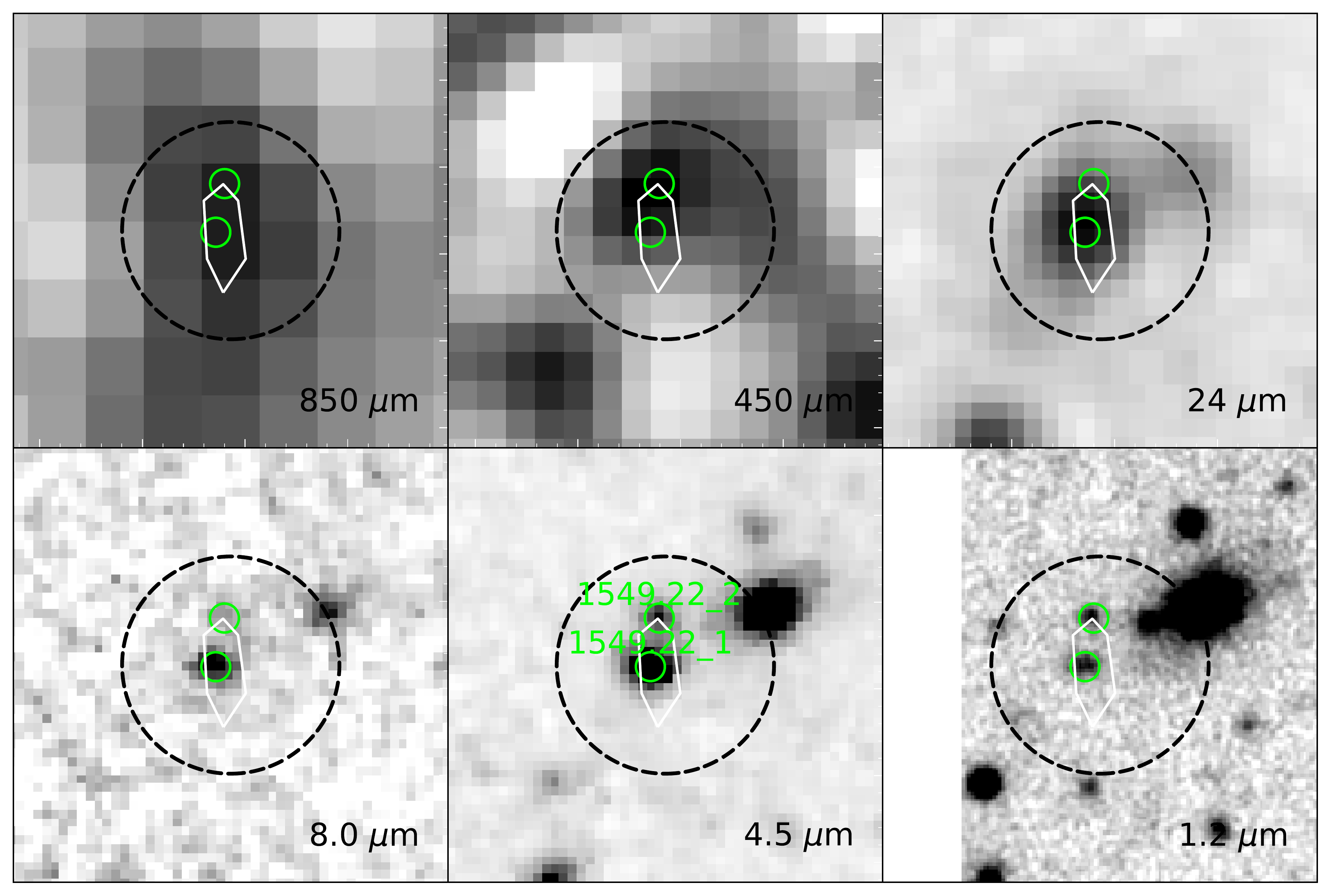}
\emph{1549.23} \hspace{3in} \emph{1549.24}\\
\includegraphics[width=.49\textwidth]{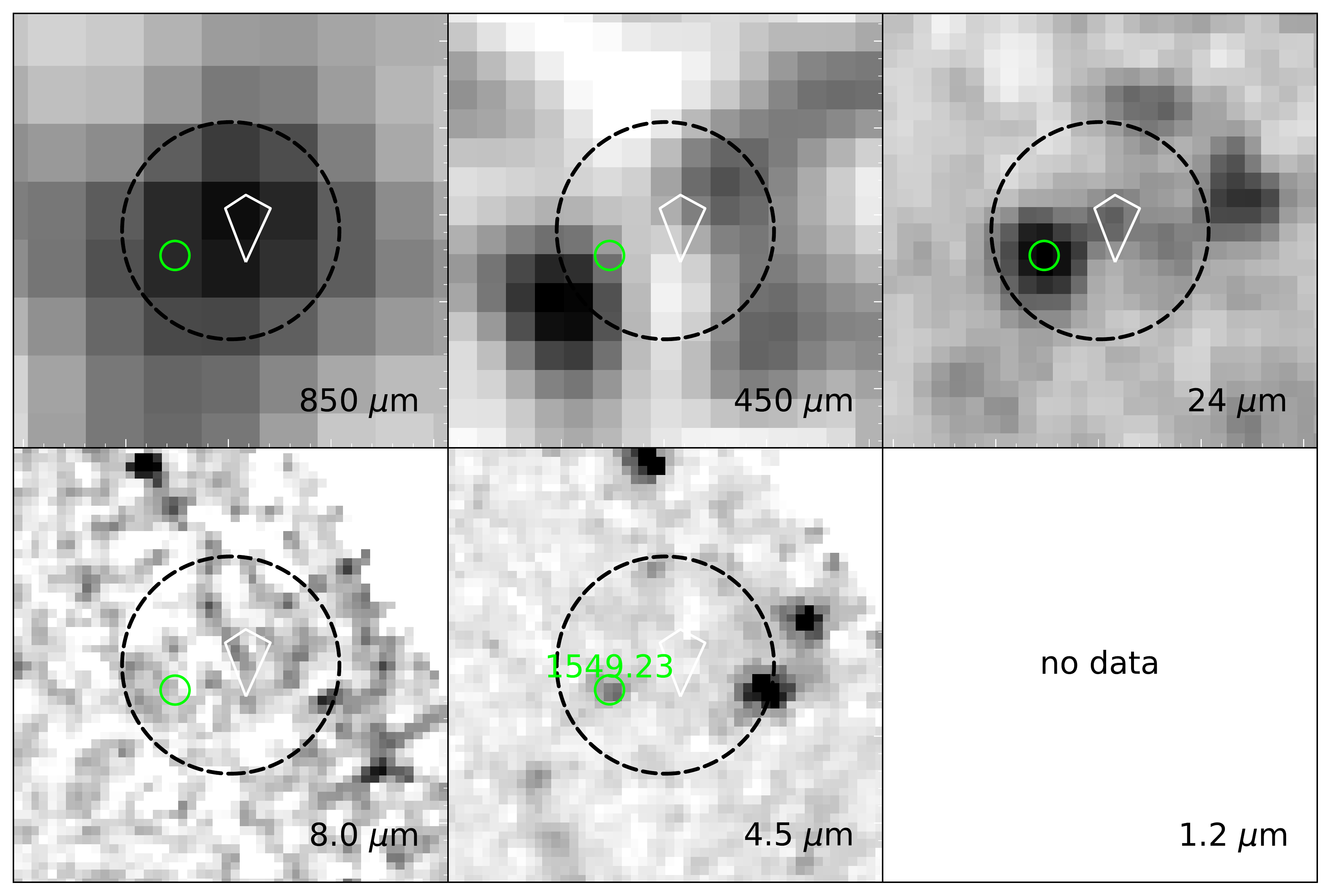}
\includegraphics[width=.49\textwidth]{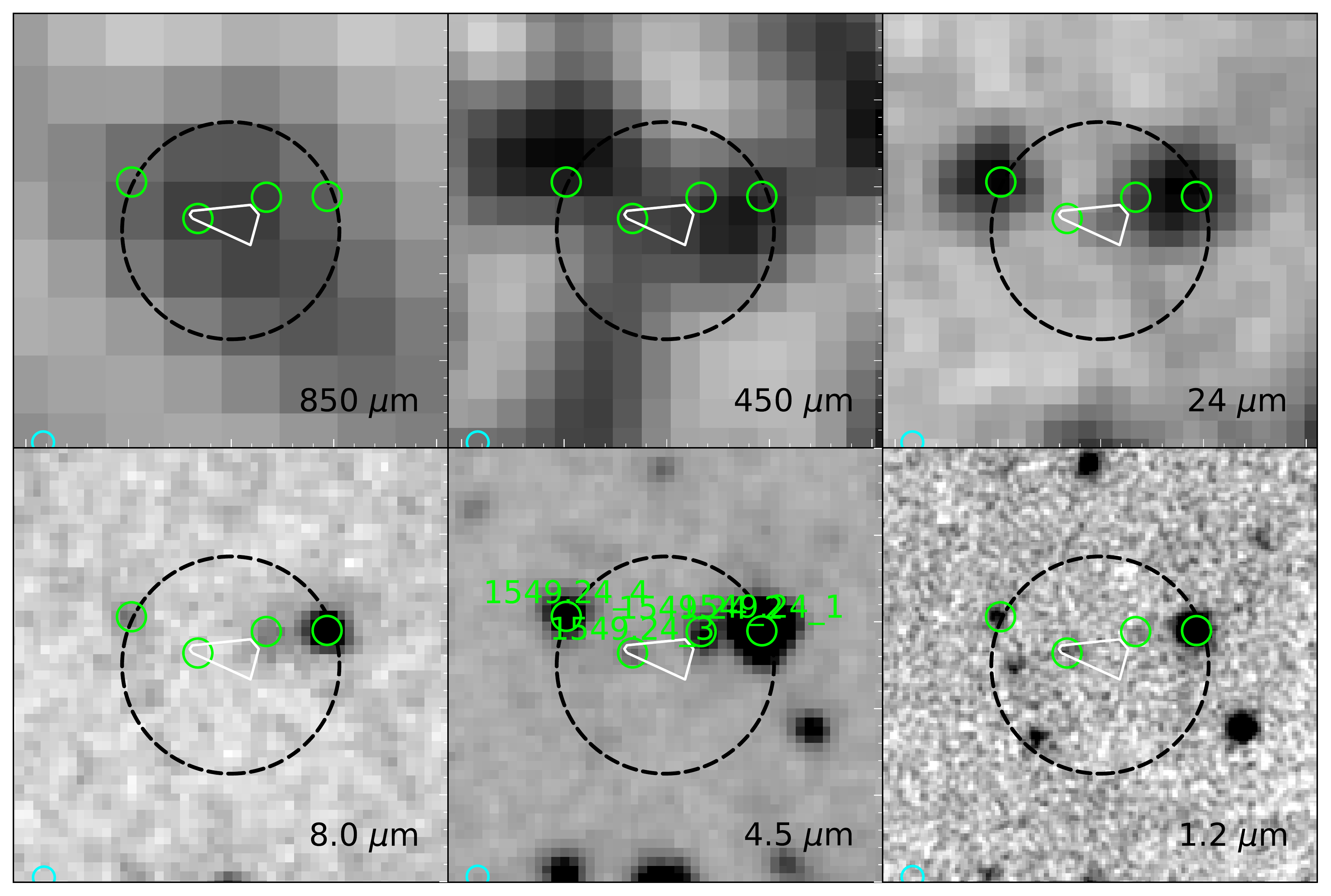}

\end{figure*}
\noindent
\begin{figure*}
\emph{1549.25} \hspace{3in} \emph{1549.26}\\
\includegraphics[width=.49\textwidth]{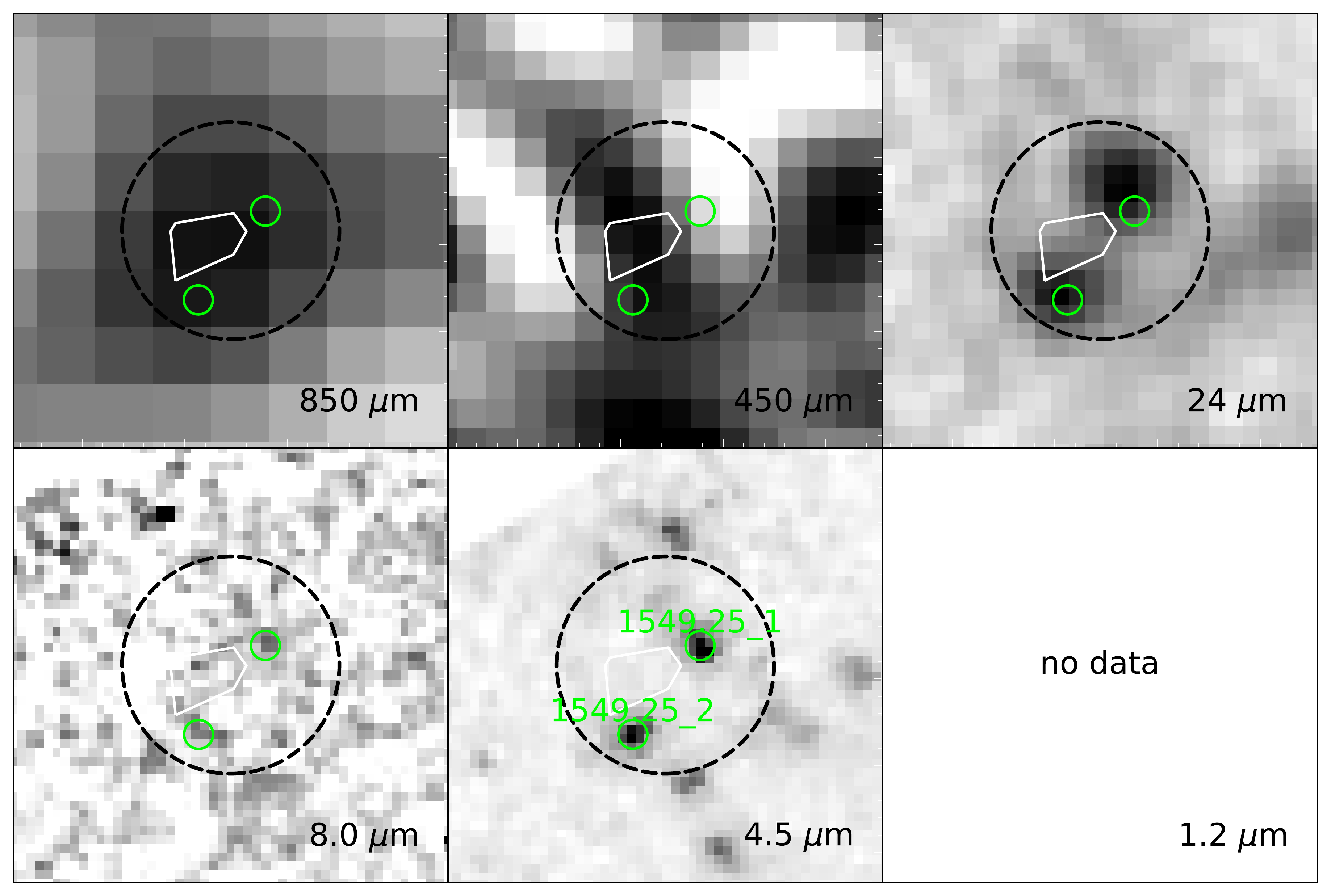}
\includegraphics[width=.49\textwidth]{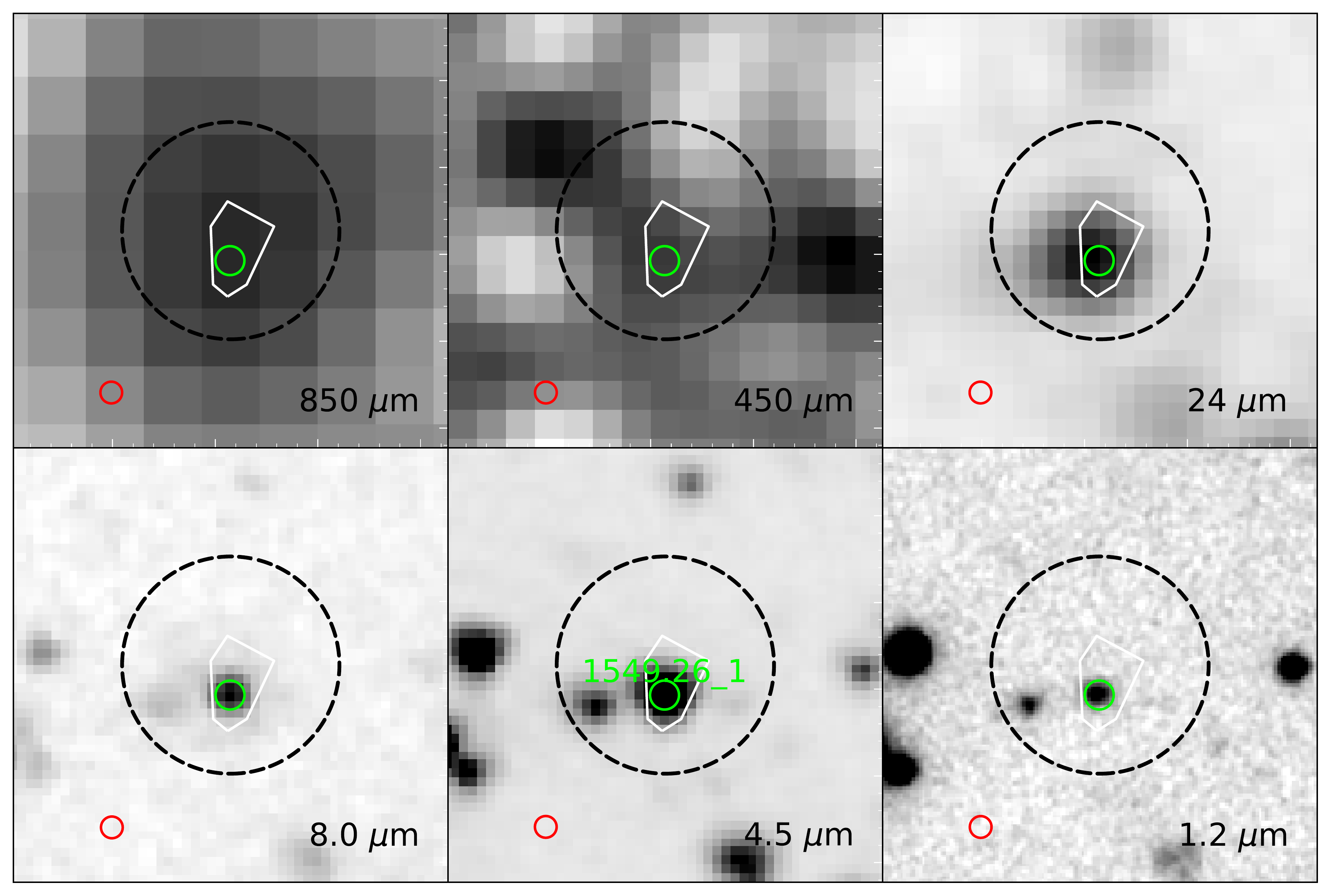}
\emph{1549.27} \hspace{3in} \emph{1549.28}\\
\includegraphics[width=.49\textwidth]{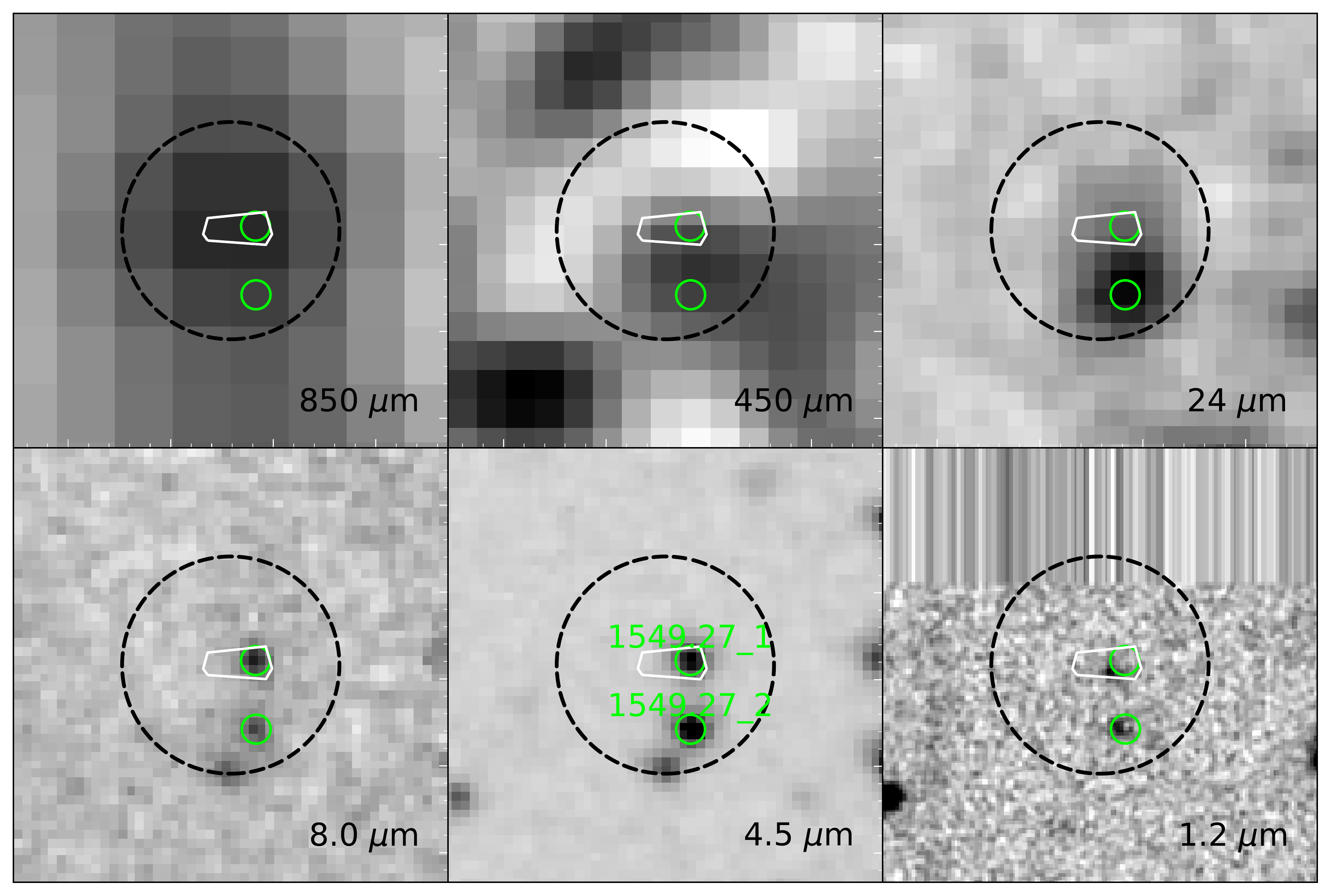}
\includegraphics[width=.49\textwidth]{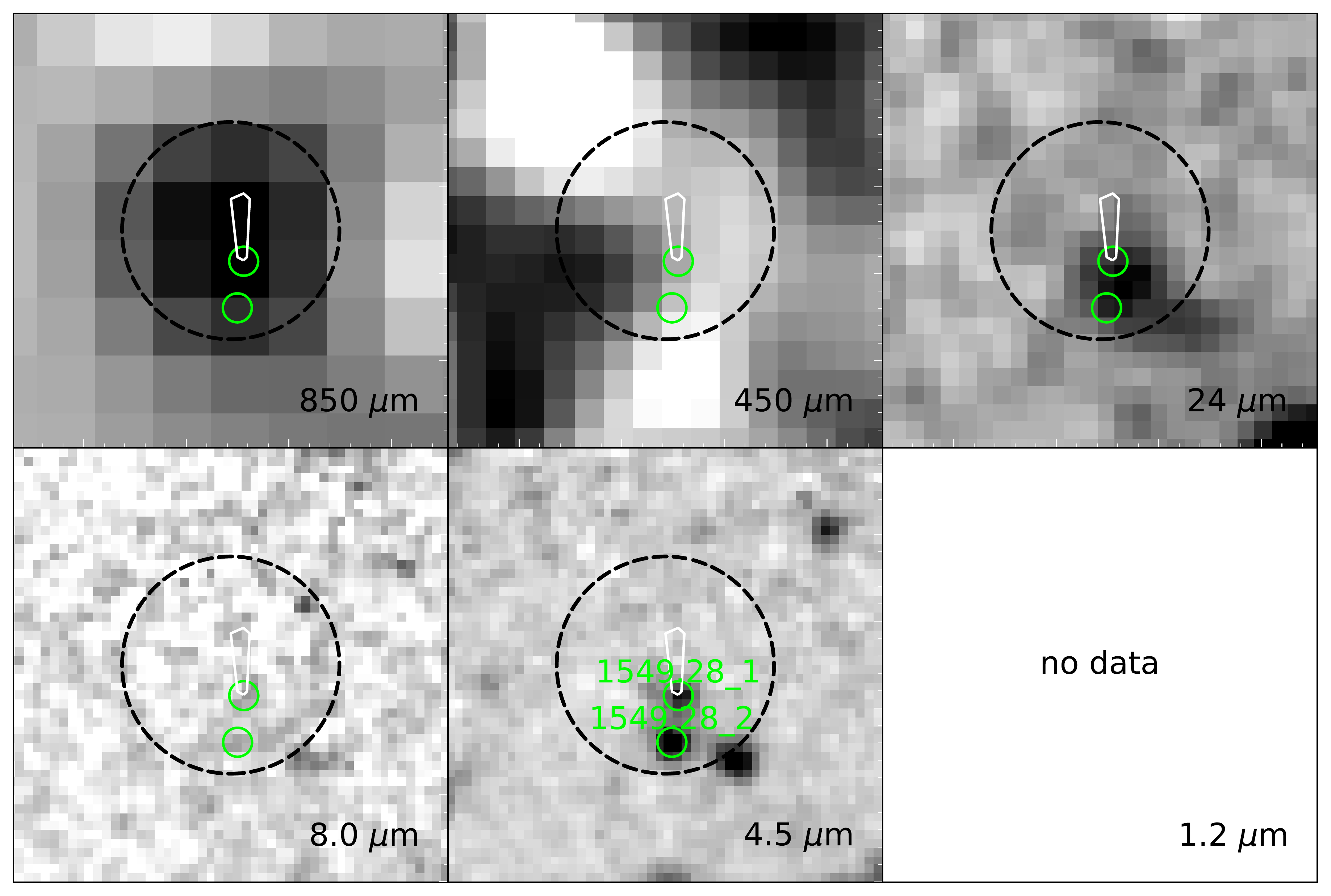}
\end{figure*}

\begin{center}
\begin{figure*}
\hspace{-3.45in}\emph{1549.29}\\
\hspace{-3.45in}\includegraphics[width=.49\textwidth]{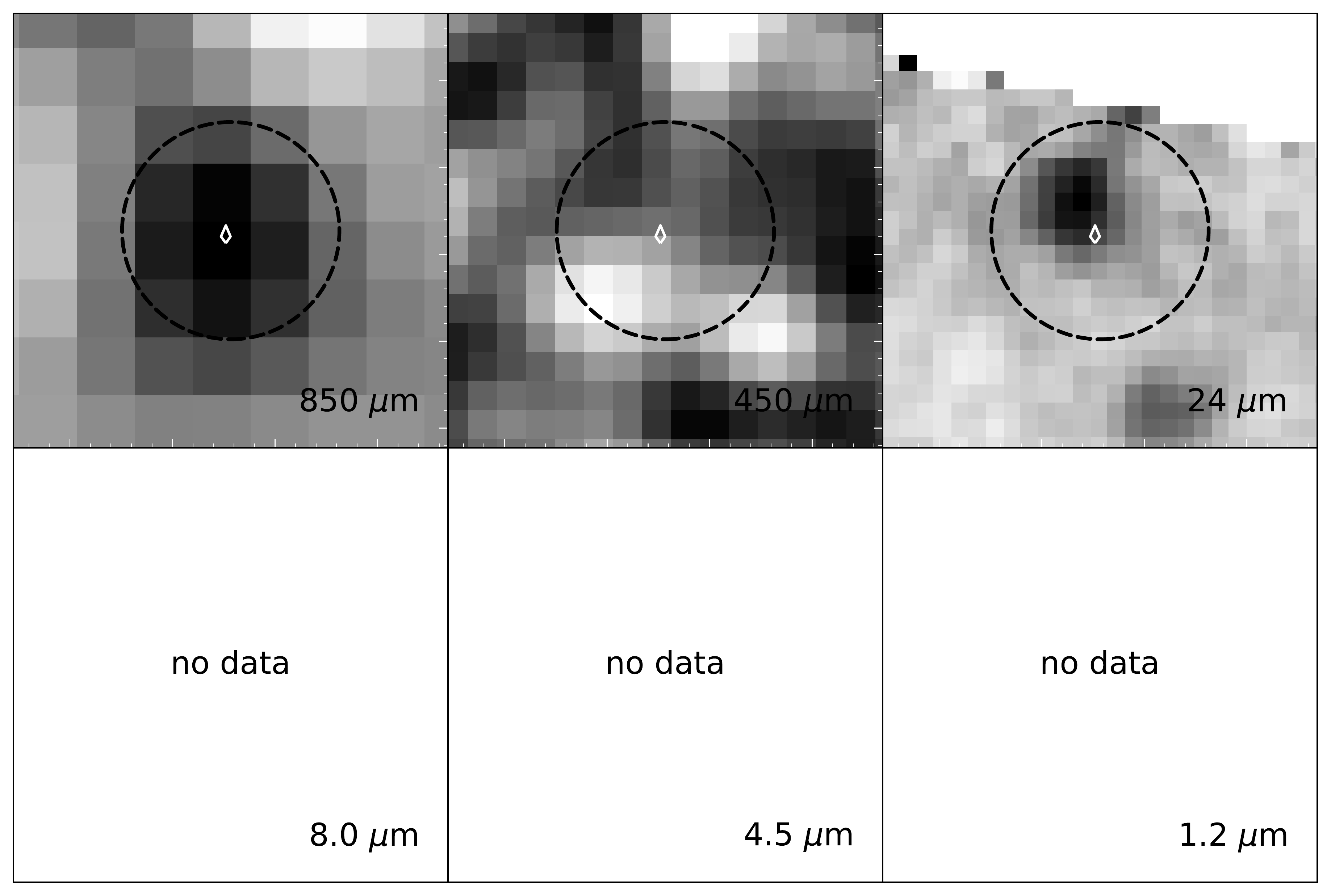}
\end{figure*}
\end{center}

\newpage

\noindent
\begin{figure*}
\emph{1700.1} \hspace{3in} \emph{1700.2}\\
\includegraphics[width=.49\textwidth]{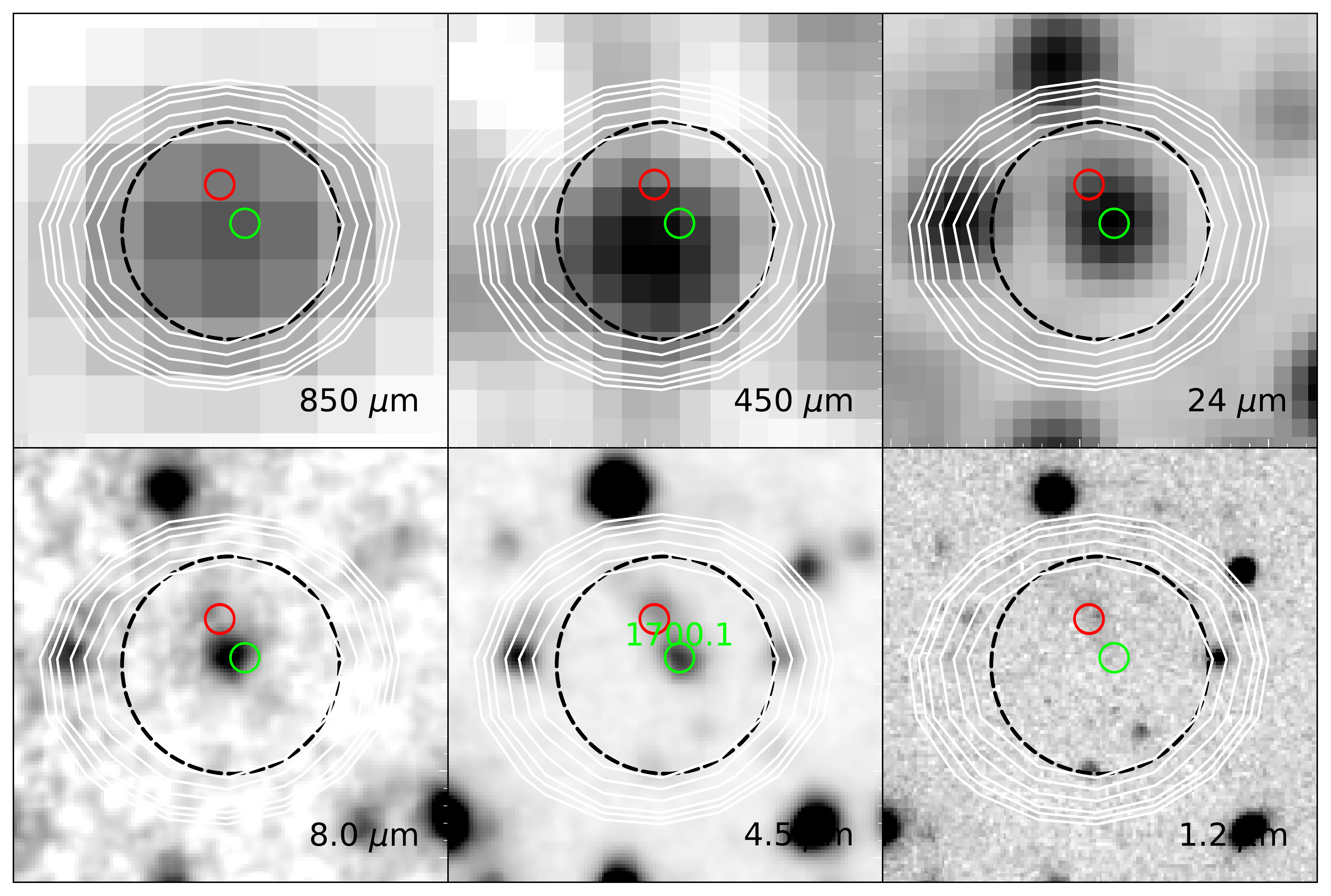}
\includegraphics[width=.49\textwidth]{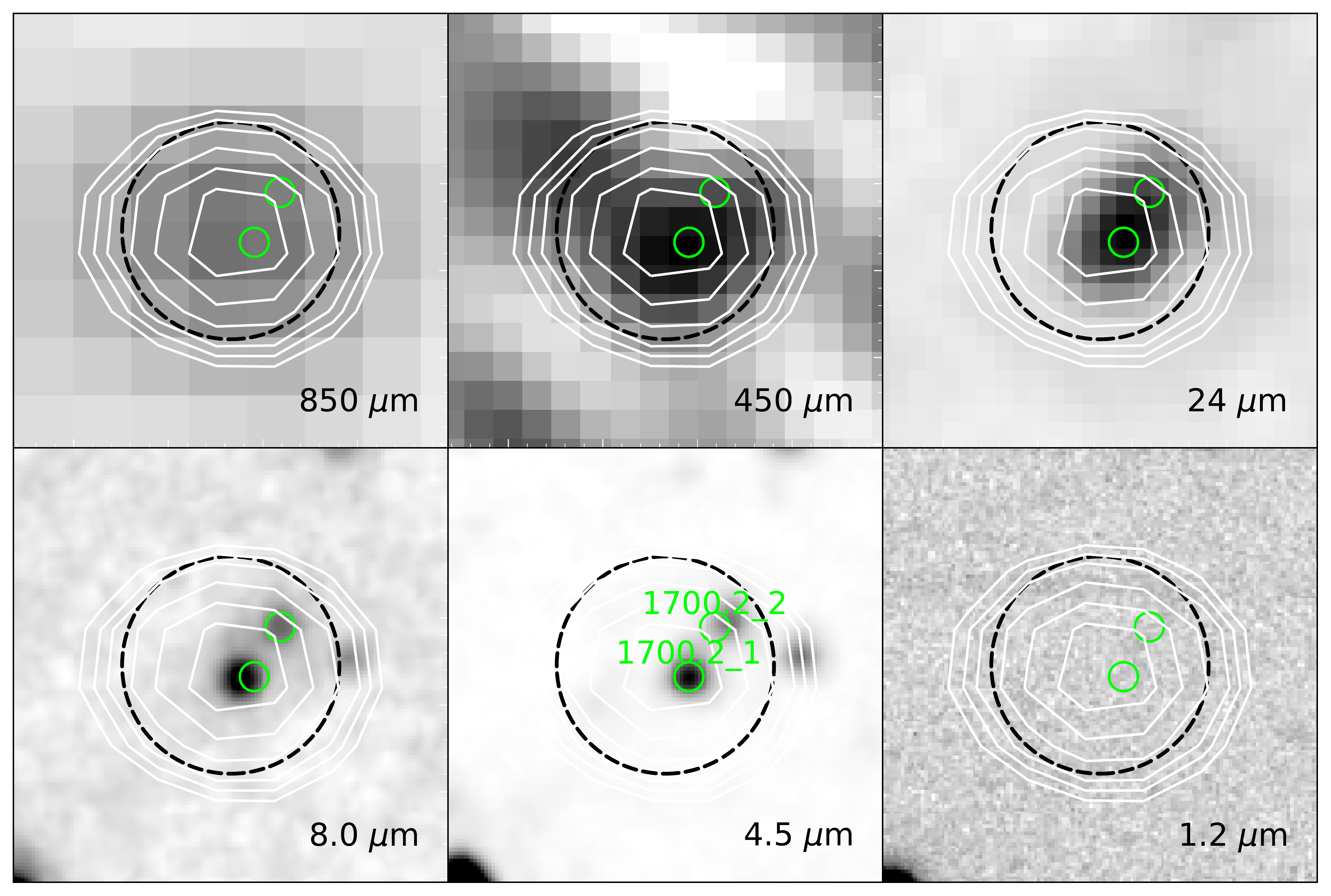}
\emph{1700.3} \hspace{3in} \emph{1700.4}\\
\includegraphics[width=.49\textwidth]{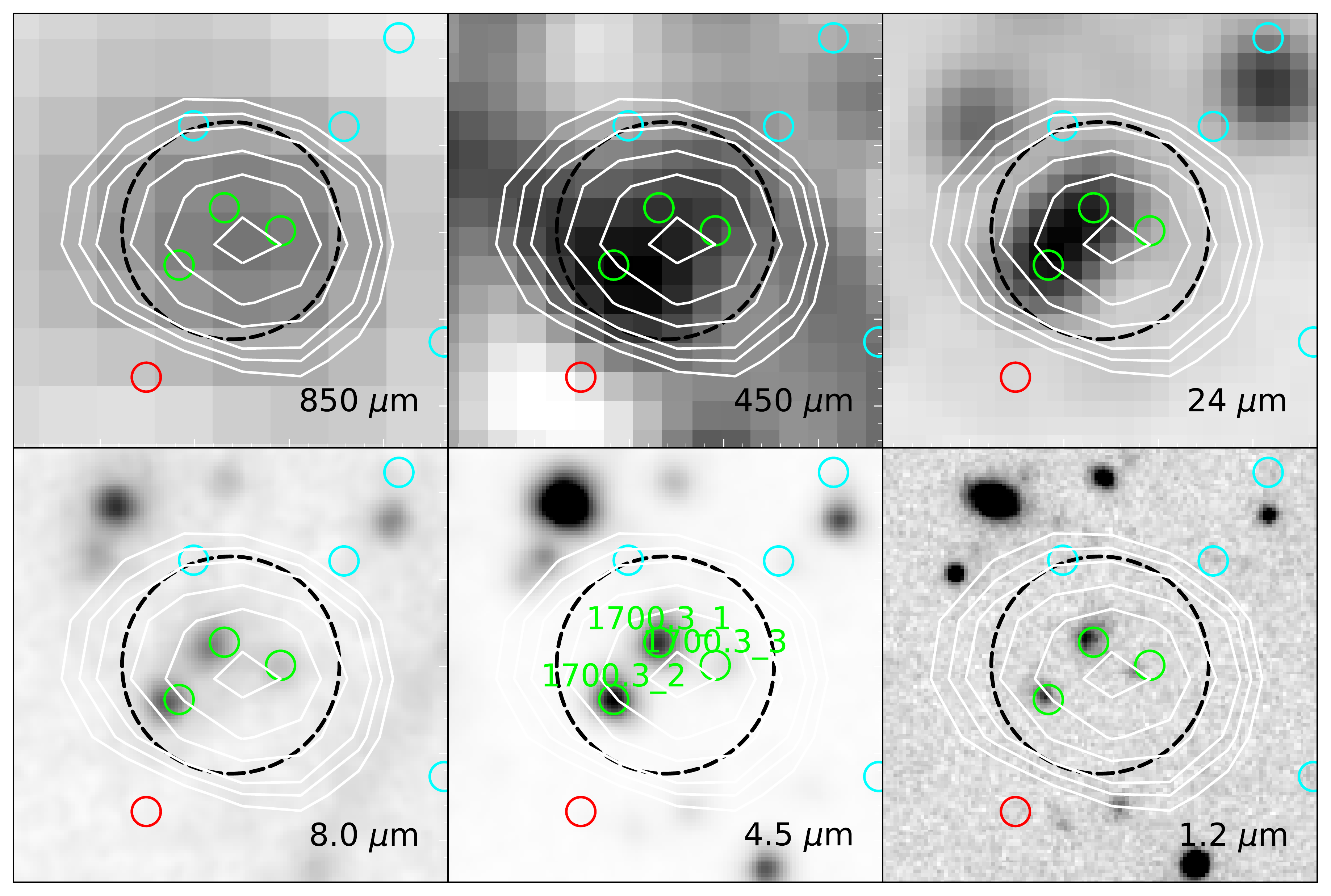}
\includegraphics[width=.49\textwidth]{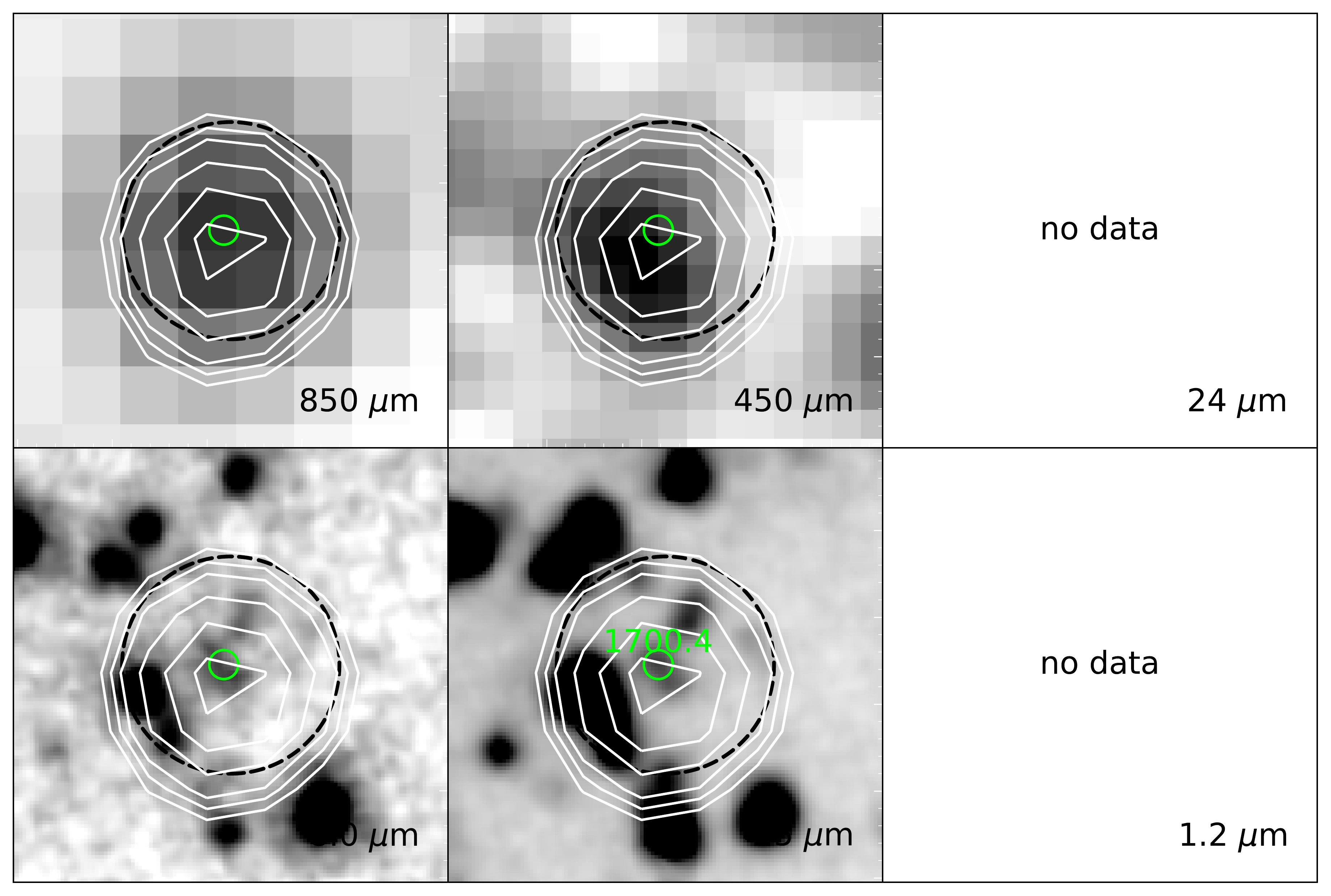}
\emph{1700.5} \hspace{3in} \emph{1700.6}\\
\includegraphics[width=.49\textwidth]{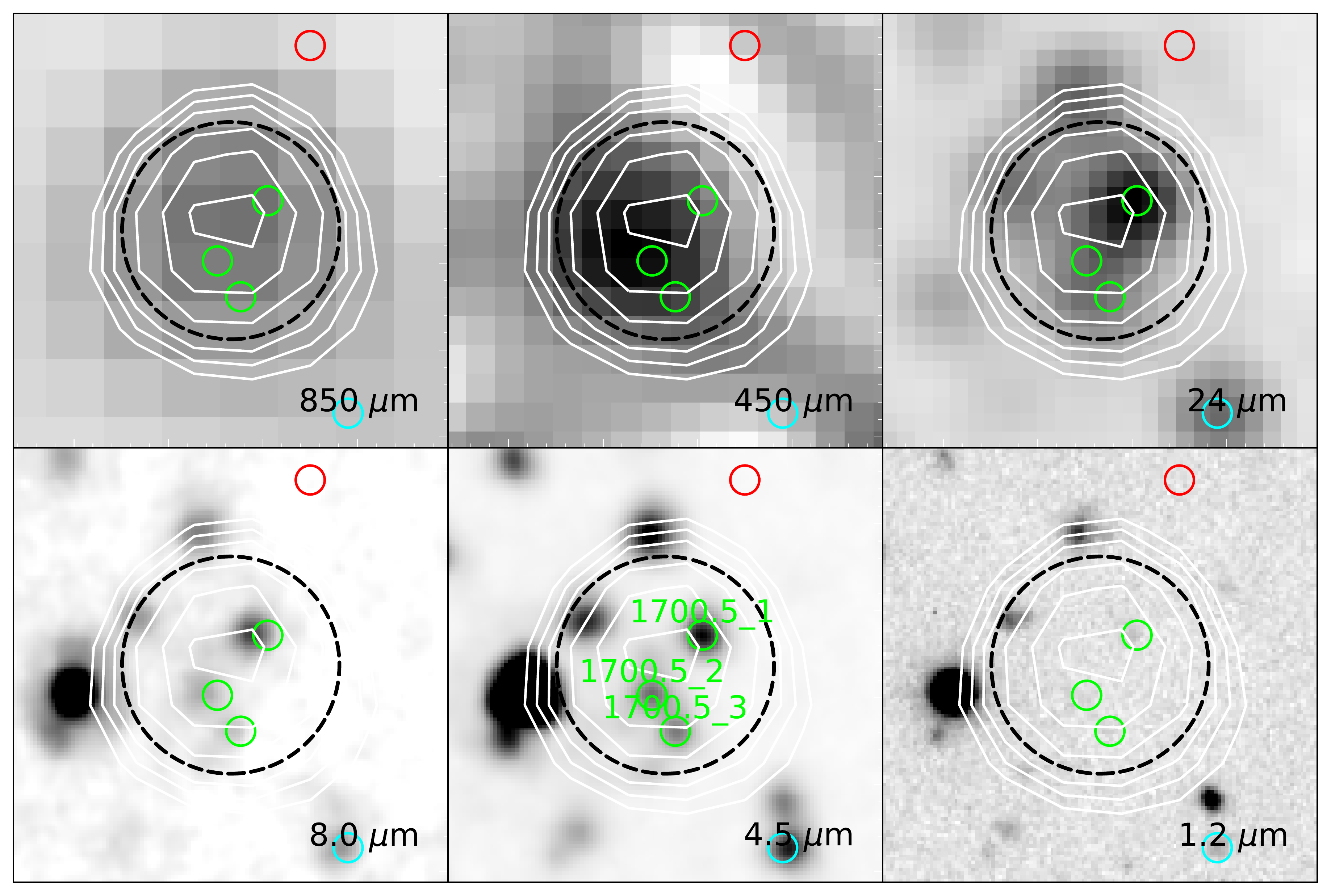}
\includegraphics[width=.49\textwidth]{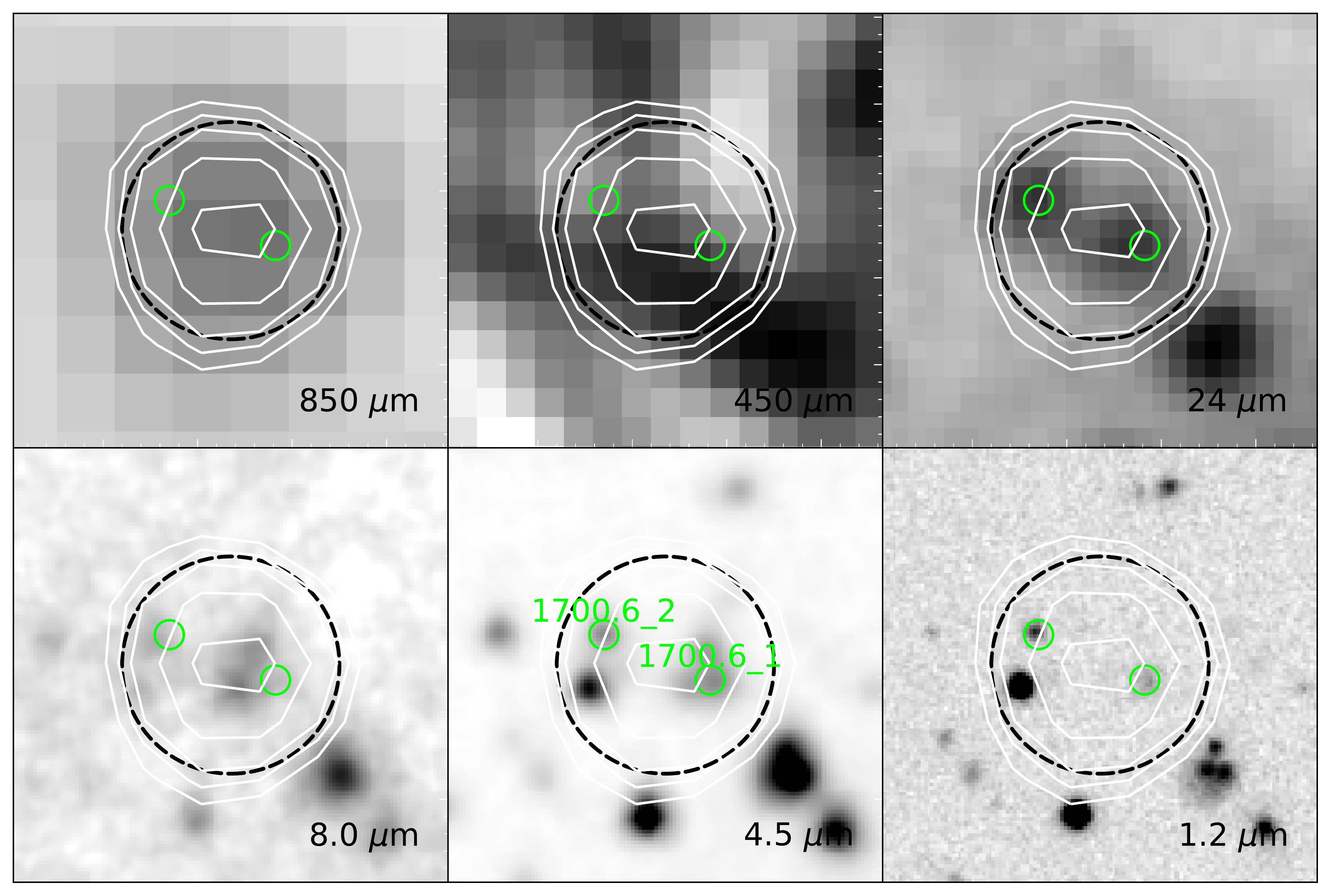}
\emph{1700.7} \hspace{3in} \emph{1700.8}\\
\includegraphics[width=.49\textwidth]{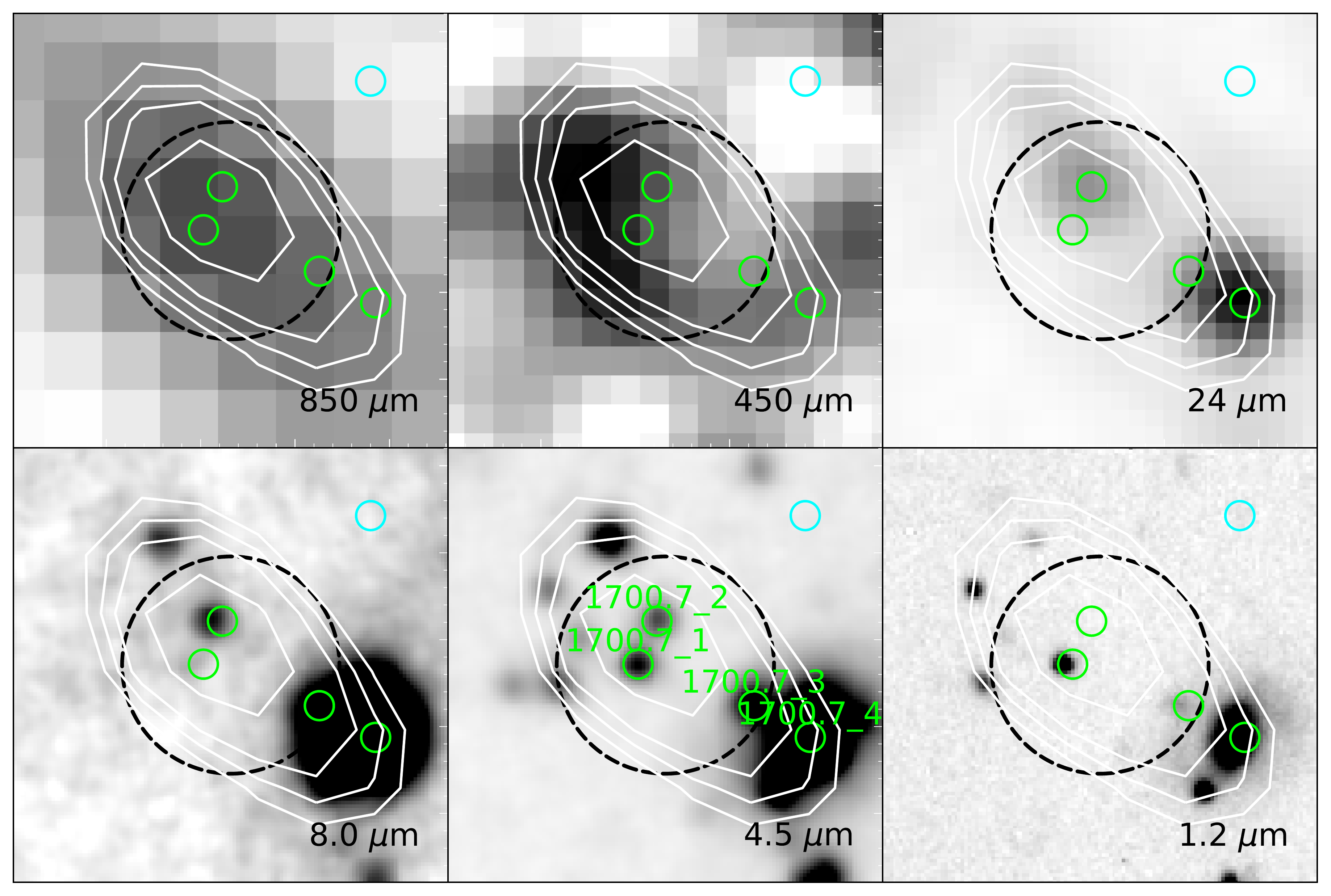}
\includegraphics[width=.49\textwidth]{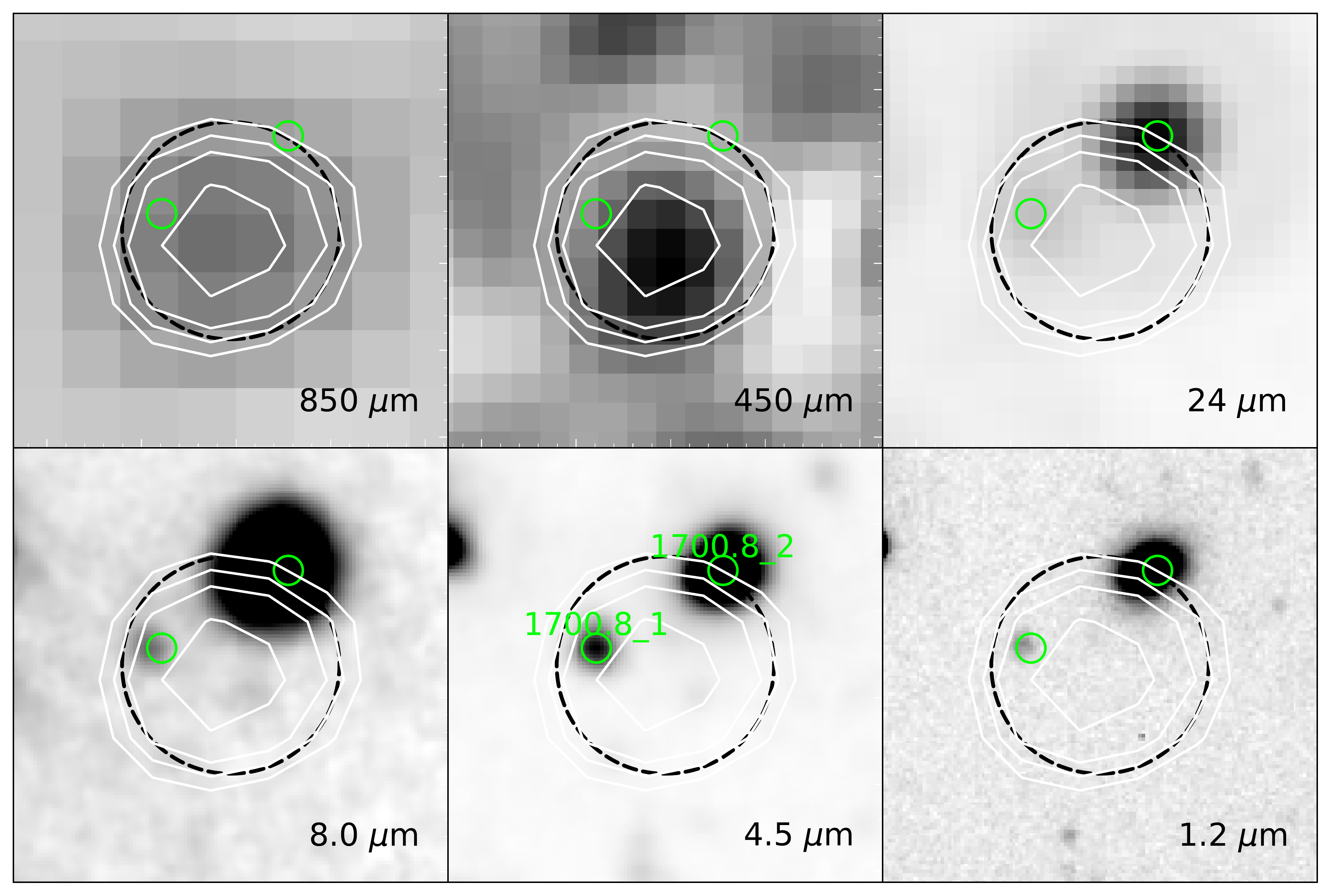}
\end{figure*}
\noindent
\begin{figure*}
\emph{1700.9} \hspace{3in} \emph{1700.10}\\
\includegraphics[width=.49\textwidth]{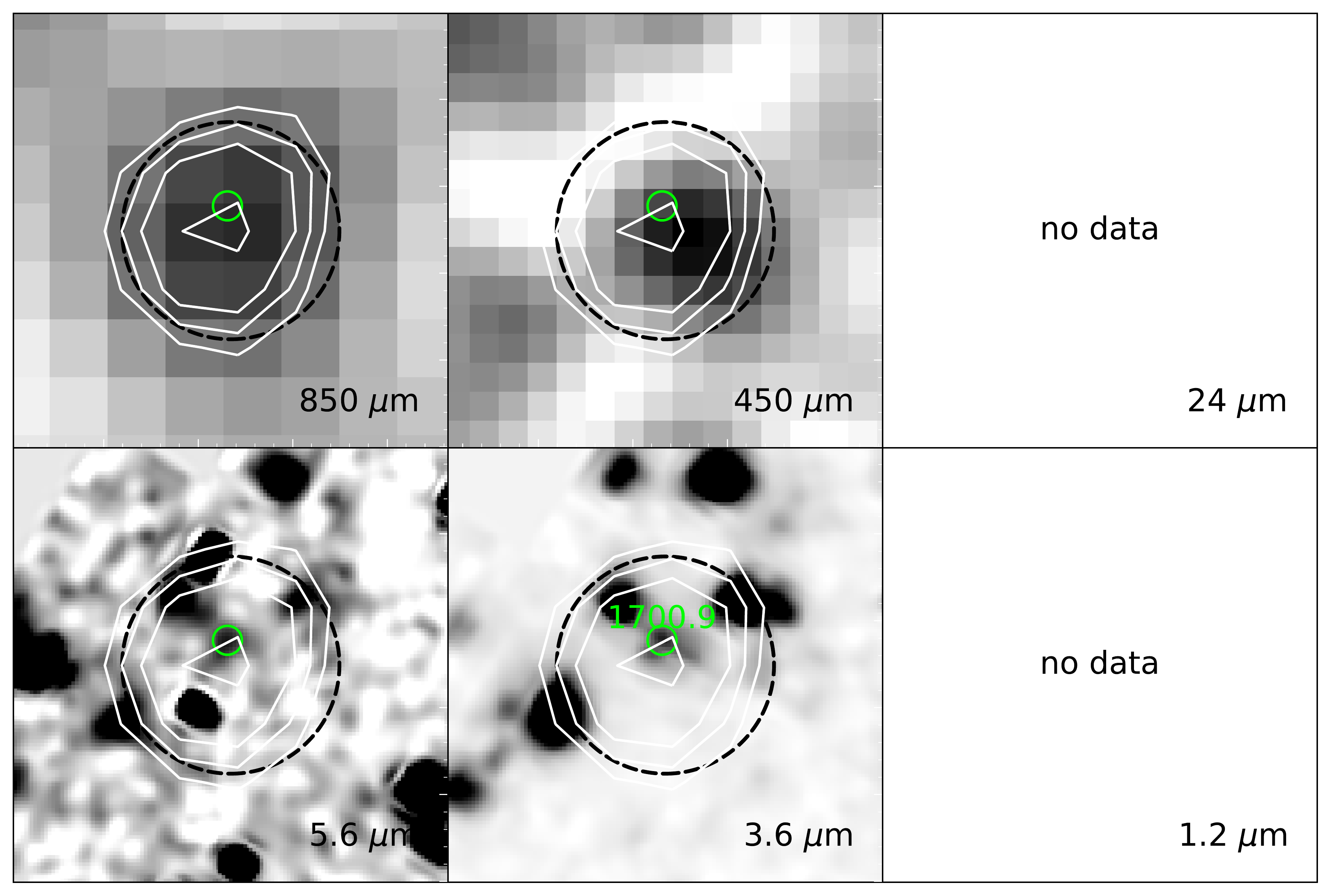}
\includegraphics[width=.49\textwidth]{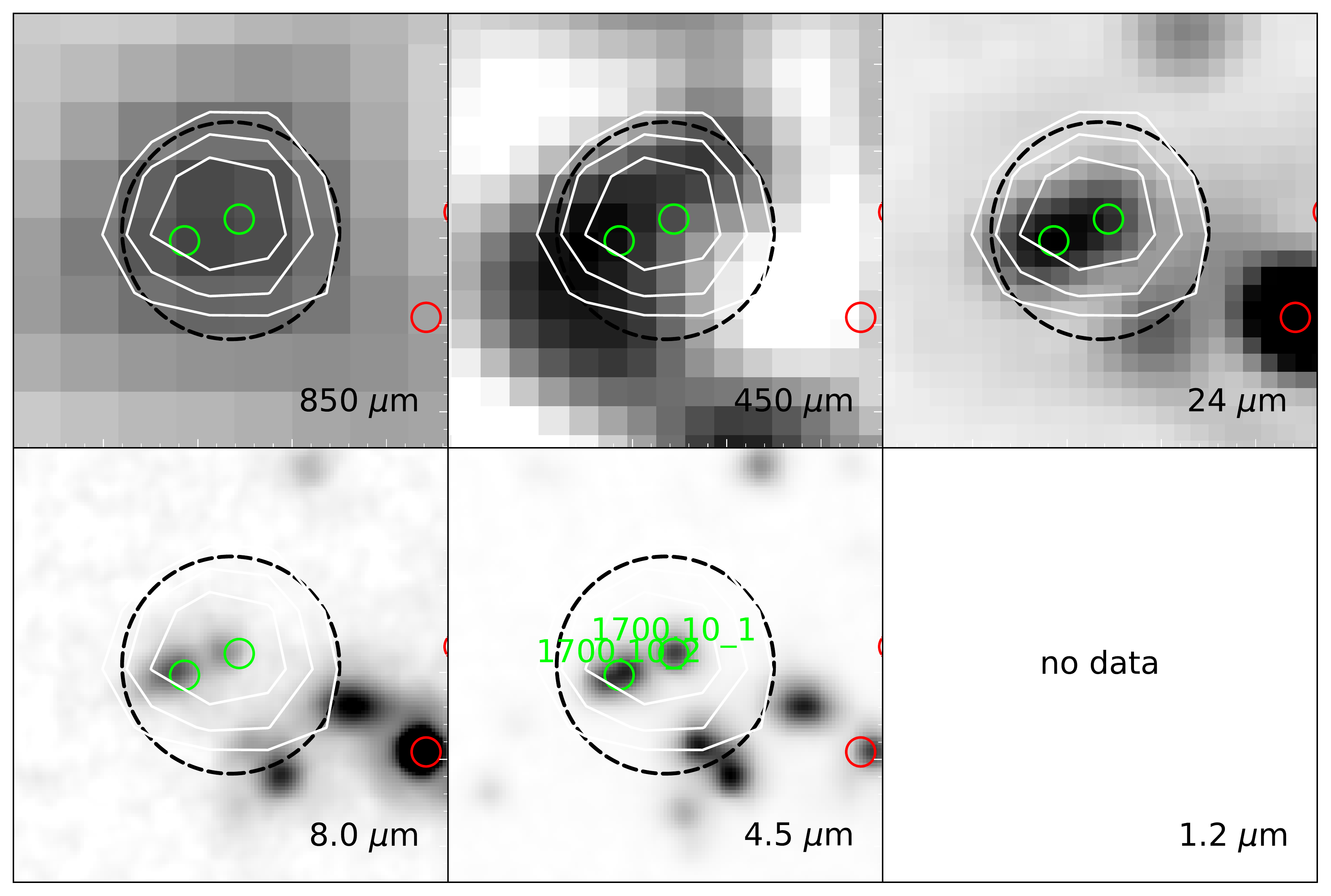}
\emph{1700.11} \hspace{3in} \emph{1700.12}\\
\includegraphics[width=.49\textwidth]{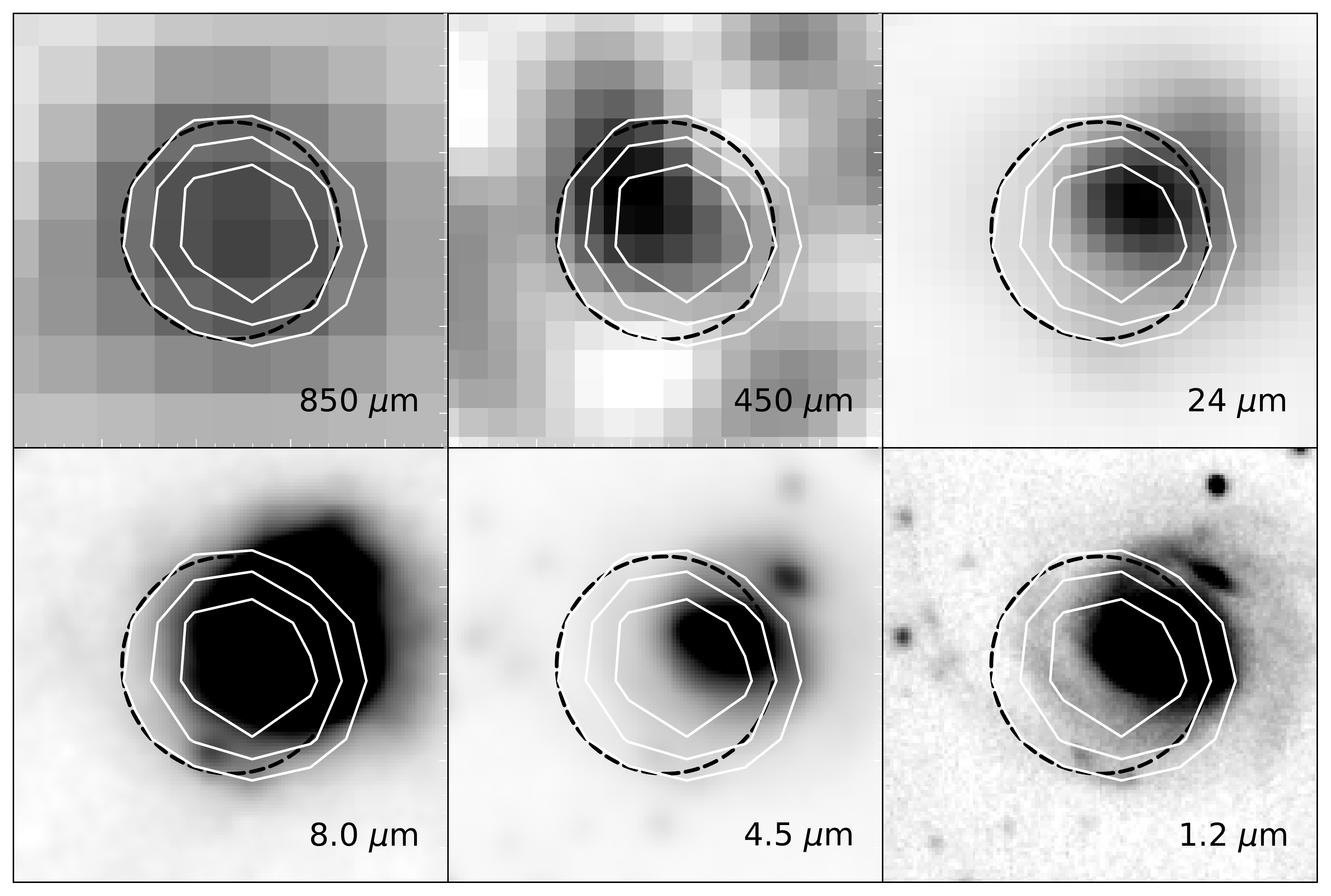}
\includegraphics[width=.49\textwidth]{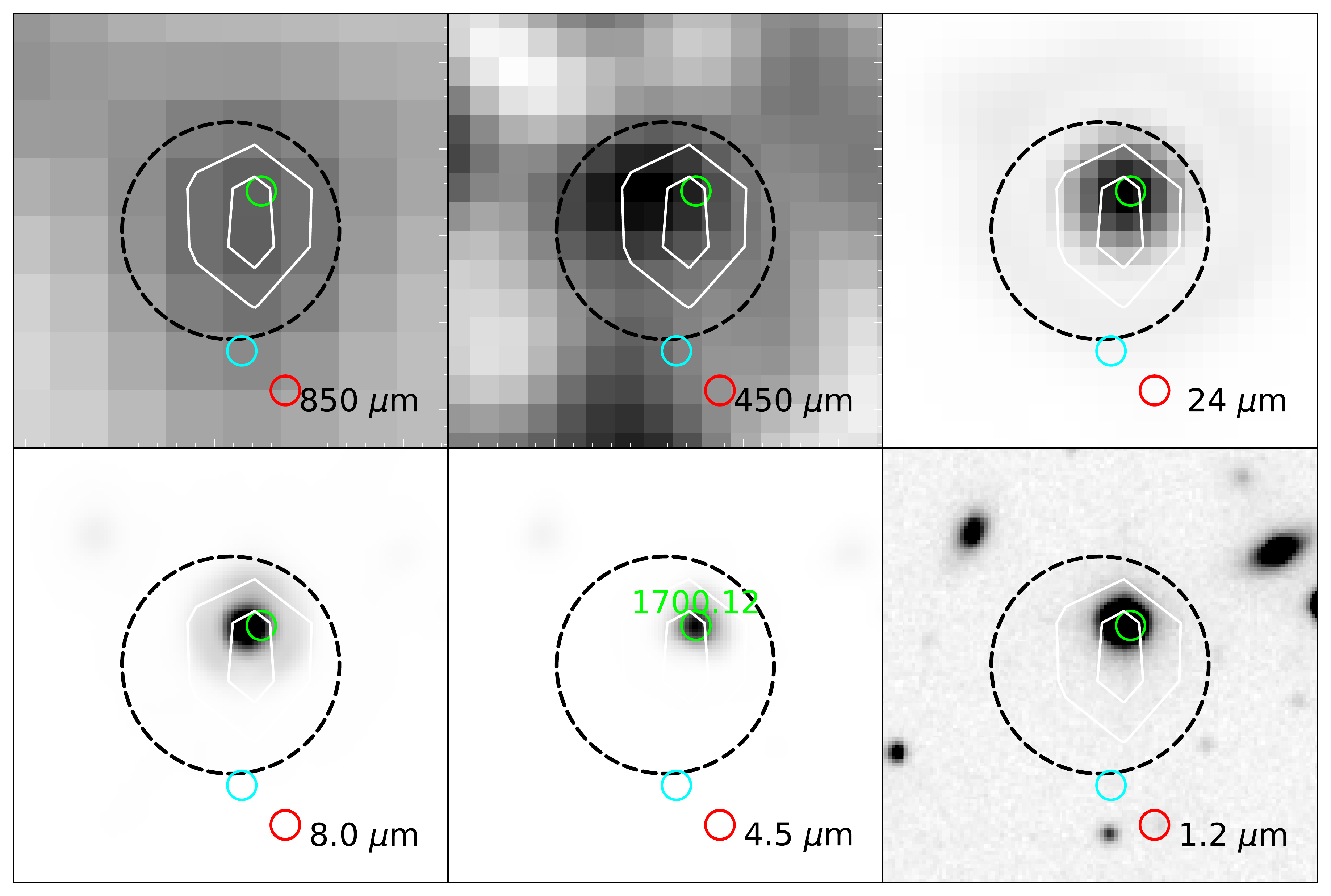}
\emph{1700.13} \hspace{3in} \emph{1700.14}\\
\includegraphics[width=.49\textwidth]{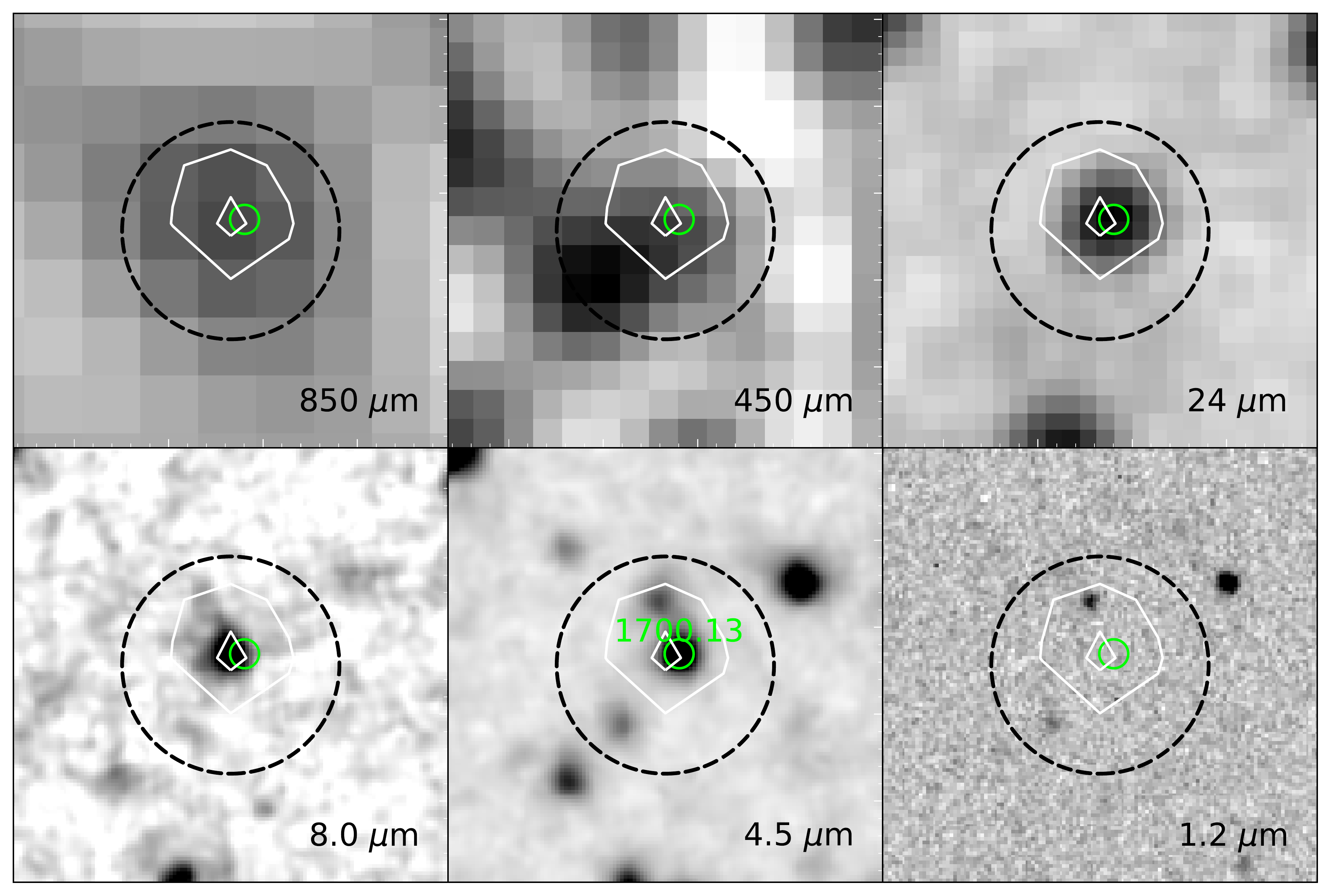}
\includegraphics[width=.49\textwidth]{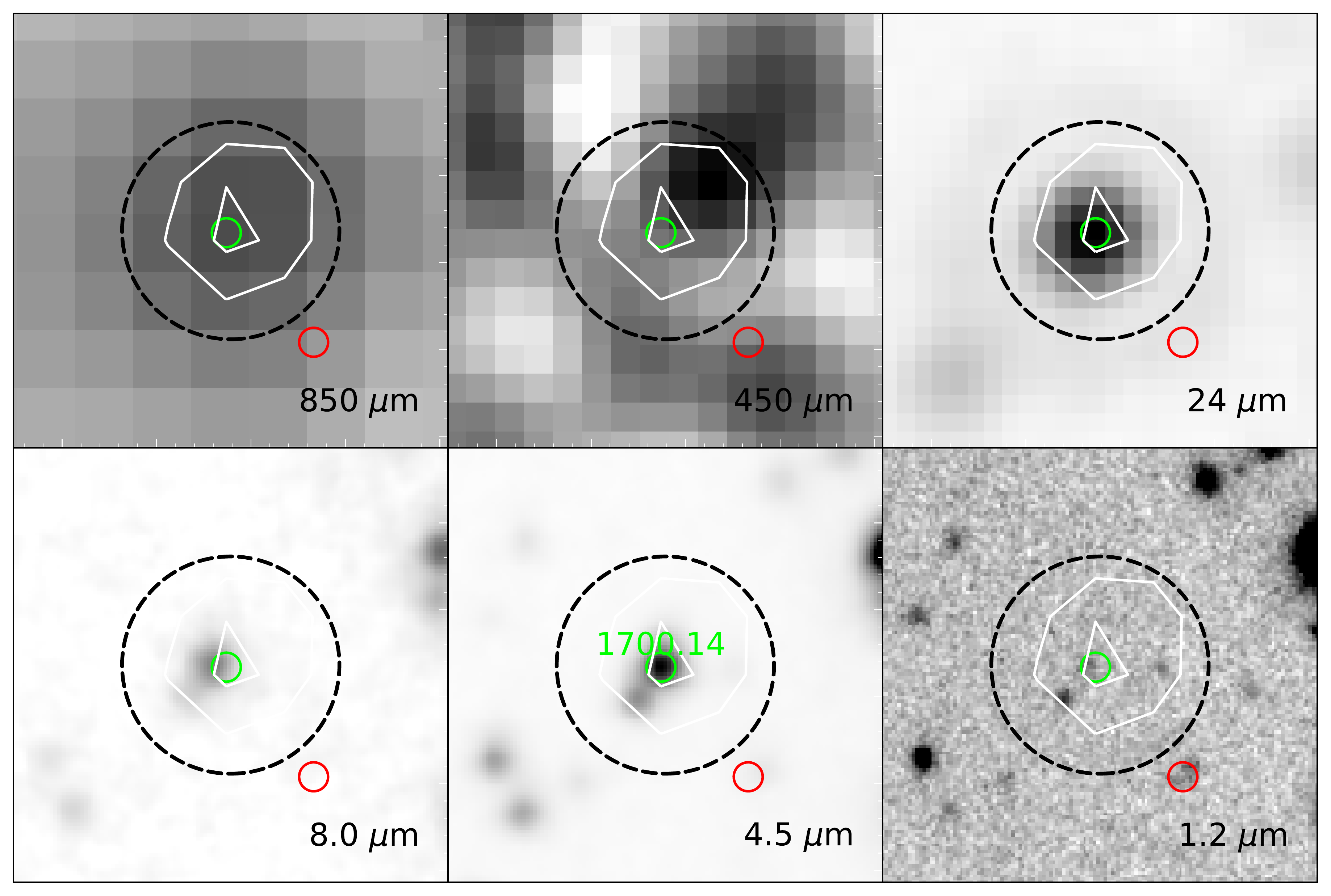}
\emph{1700.15} \hspace{3in} \emph{1700.16}\\
\includegraphics[width=.49\textwidth]{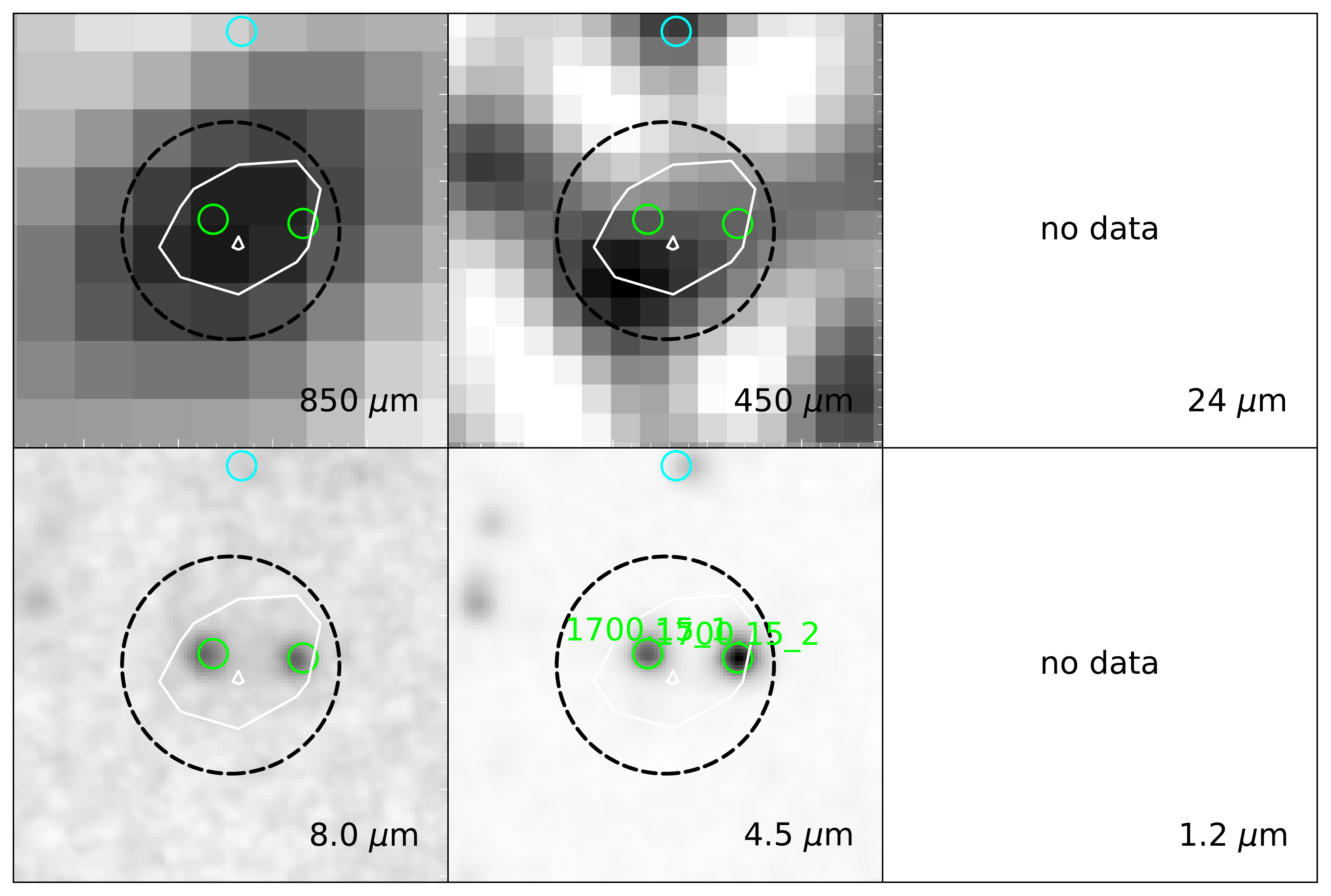}
\includegraphics[width=.49\textwidth]{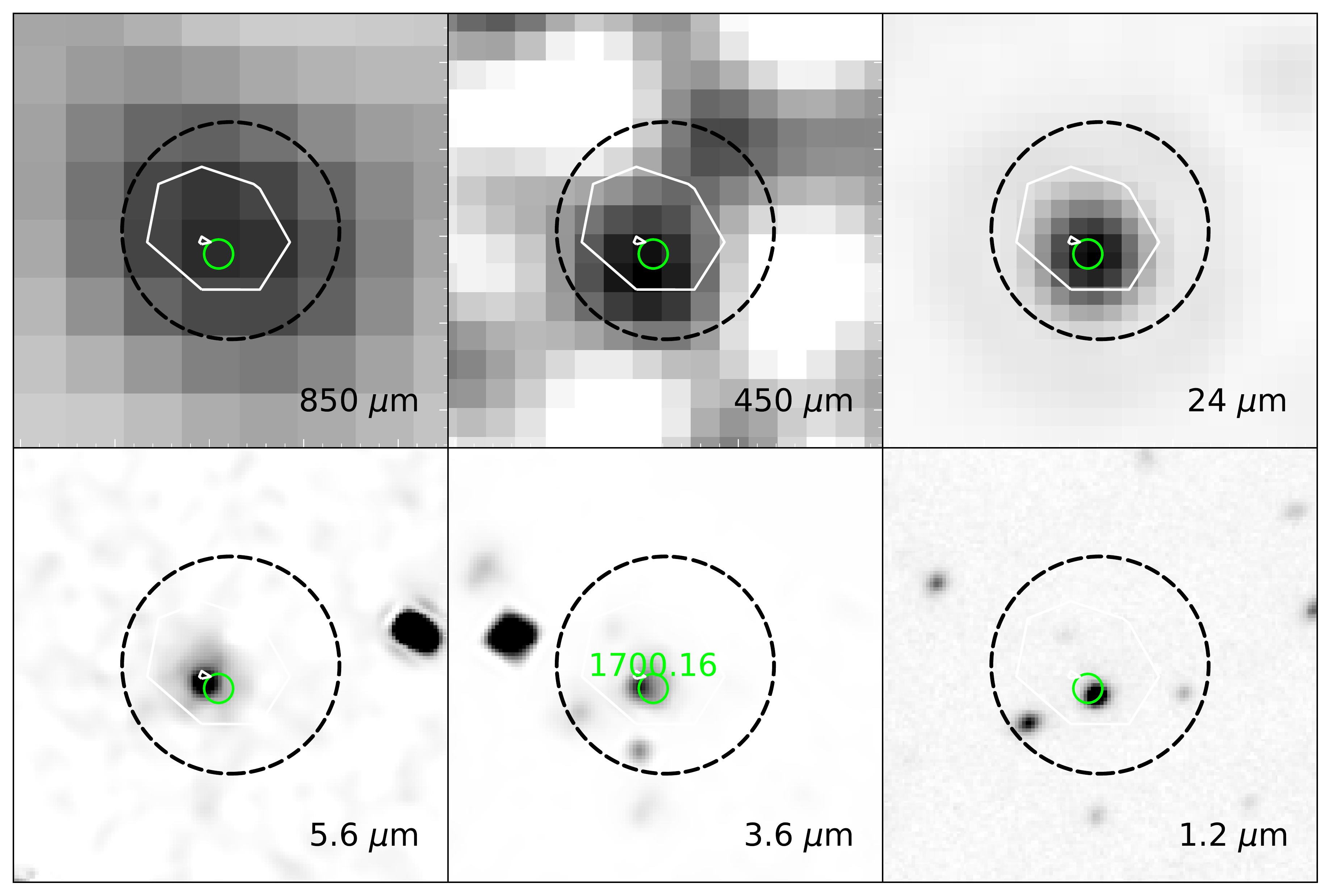}
\end{figure*}
\noindent
\begin{figure*}
\emph{1700.17} \hspace{3in} \emph{1700.18}\\
\includegraphics[width=.49\textwidth]{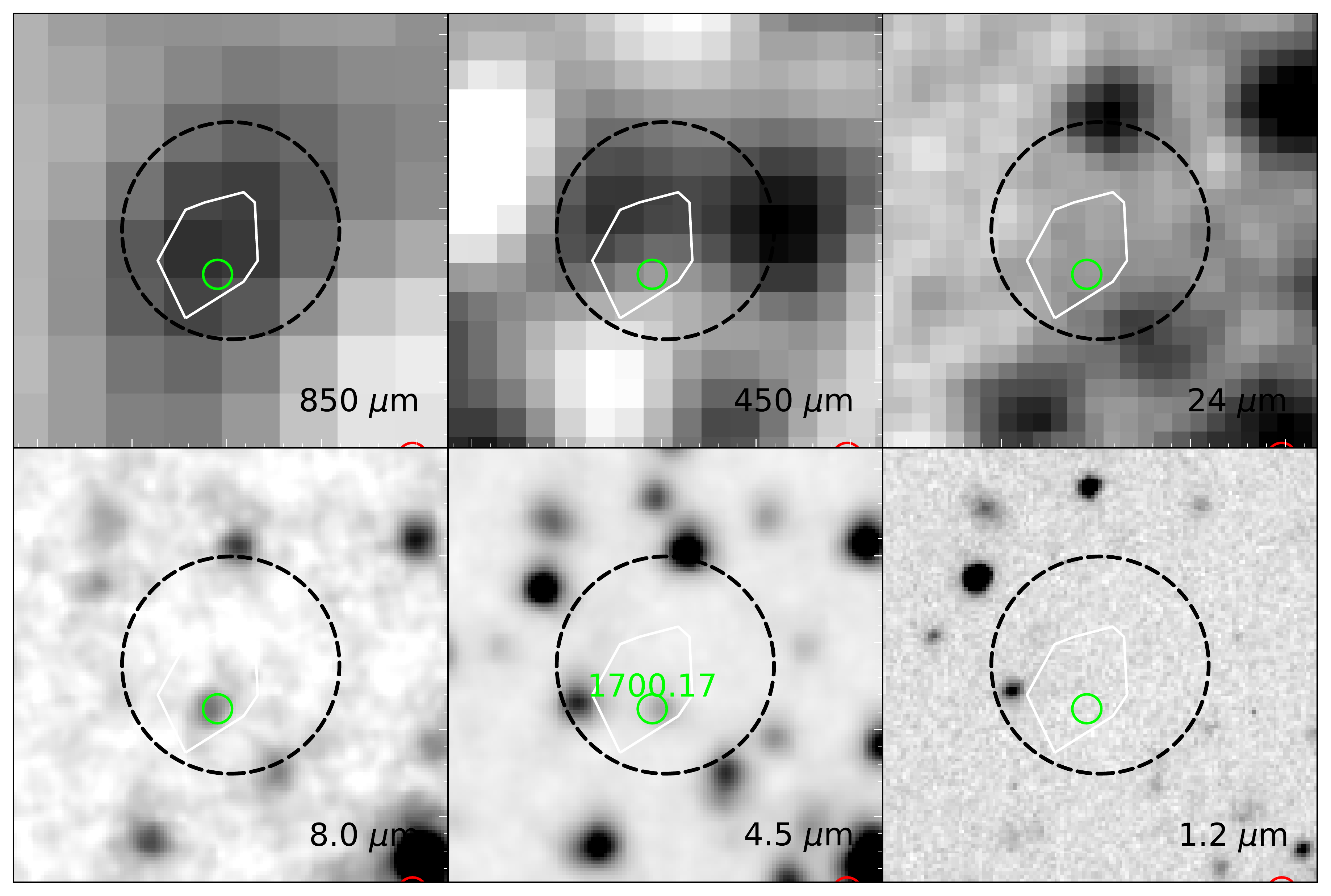}
\includegraphics[width=.49\textwidth]{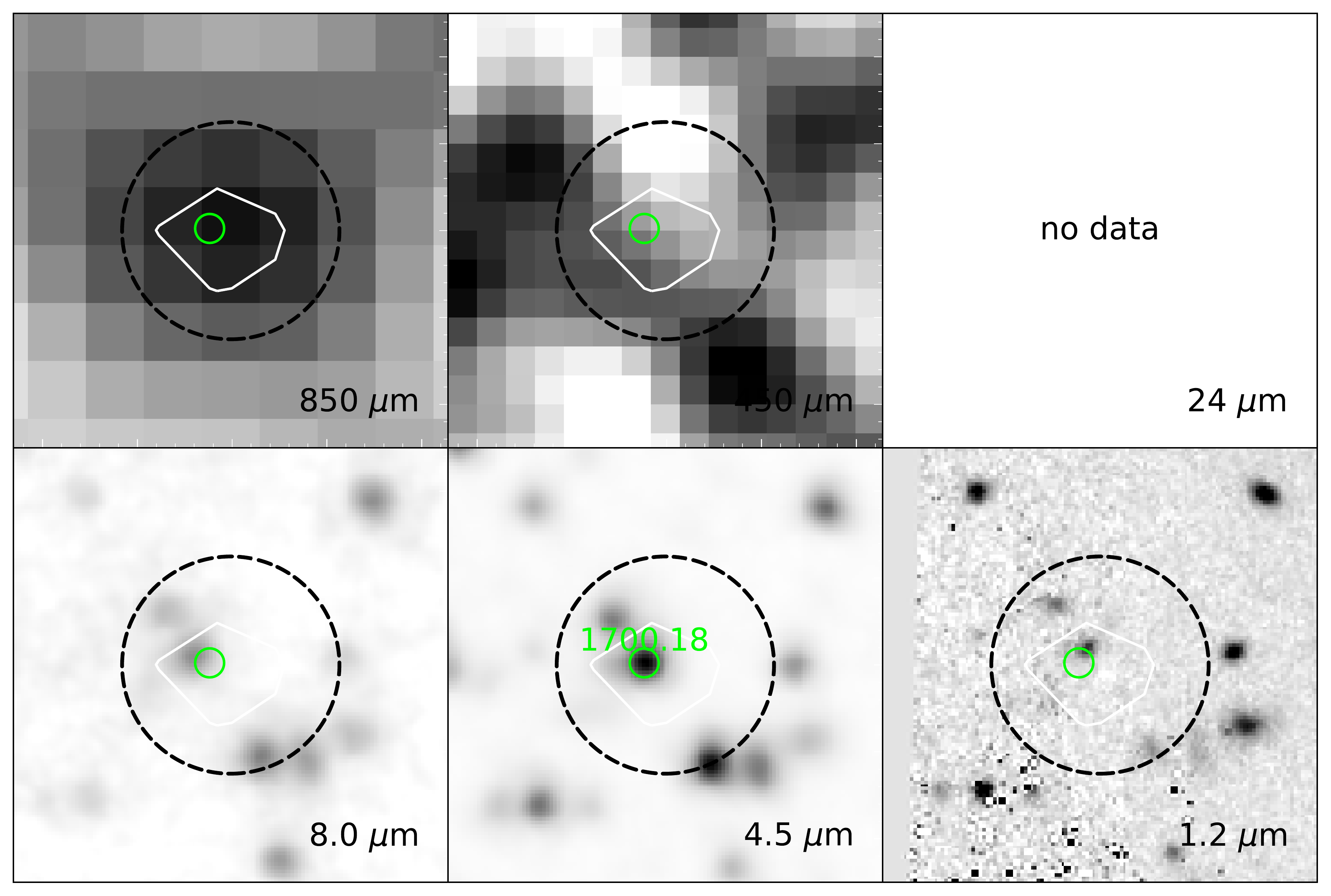}
\emph{1700.19} \hspace{3in} \emph{1700.20}\\
\includegraphics[width=.49\textwidth]{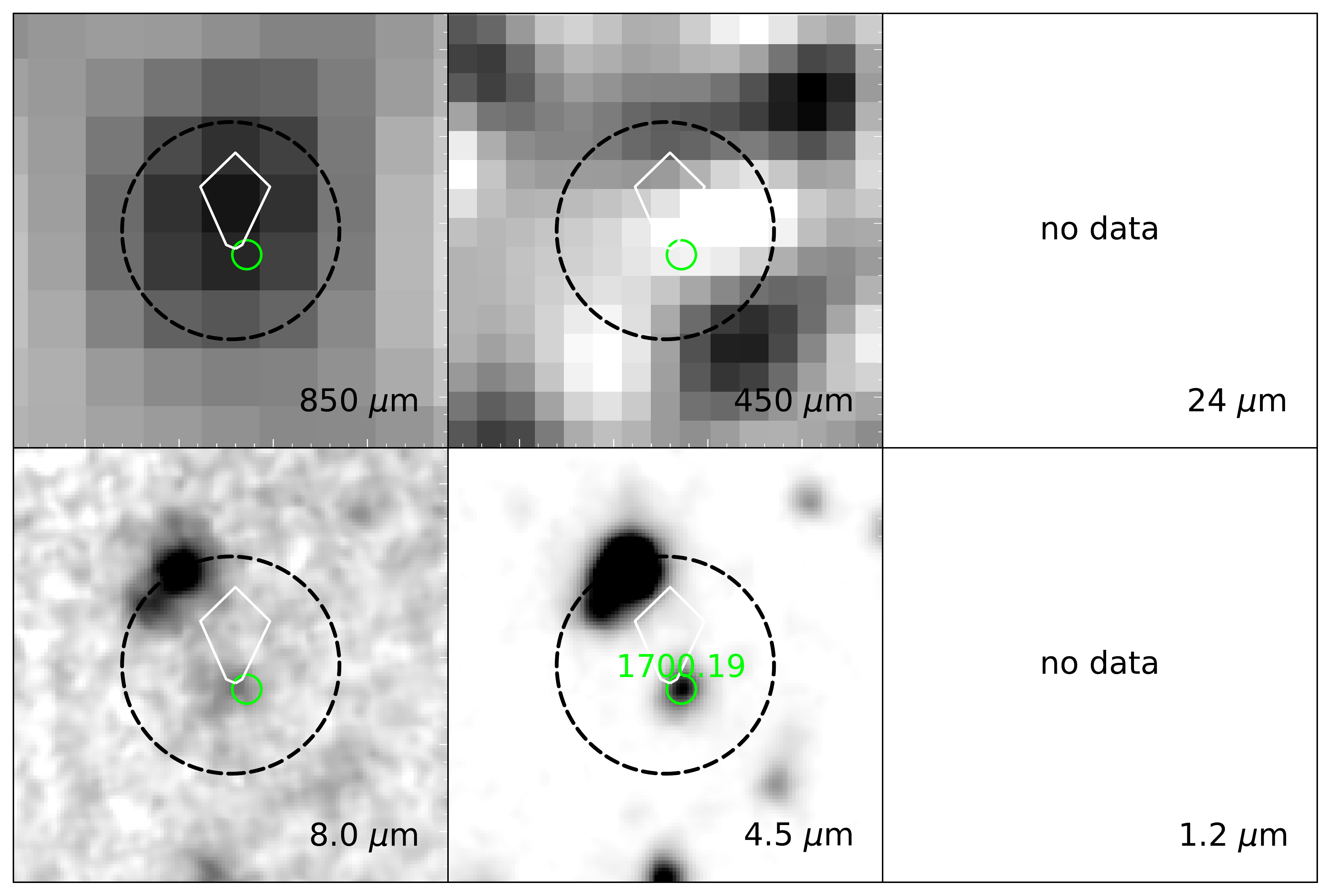}
\includegraphics[width=.49\textwidth]{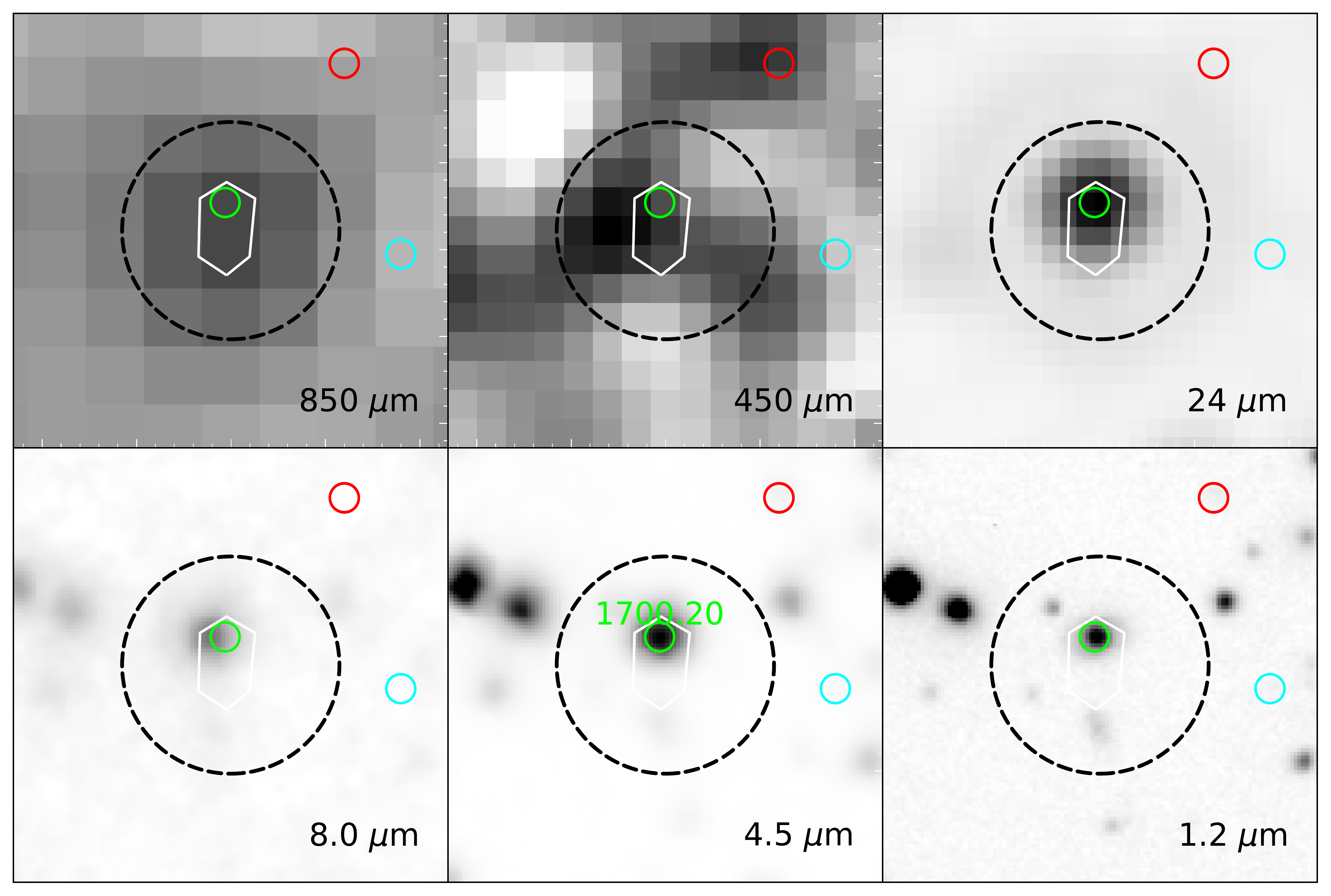}
\emph{1700.21} \hspace{3in} \emph{1700.22}\\
\includegraphics[width=.49\textwidth]{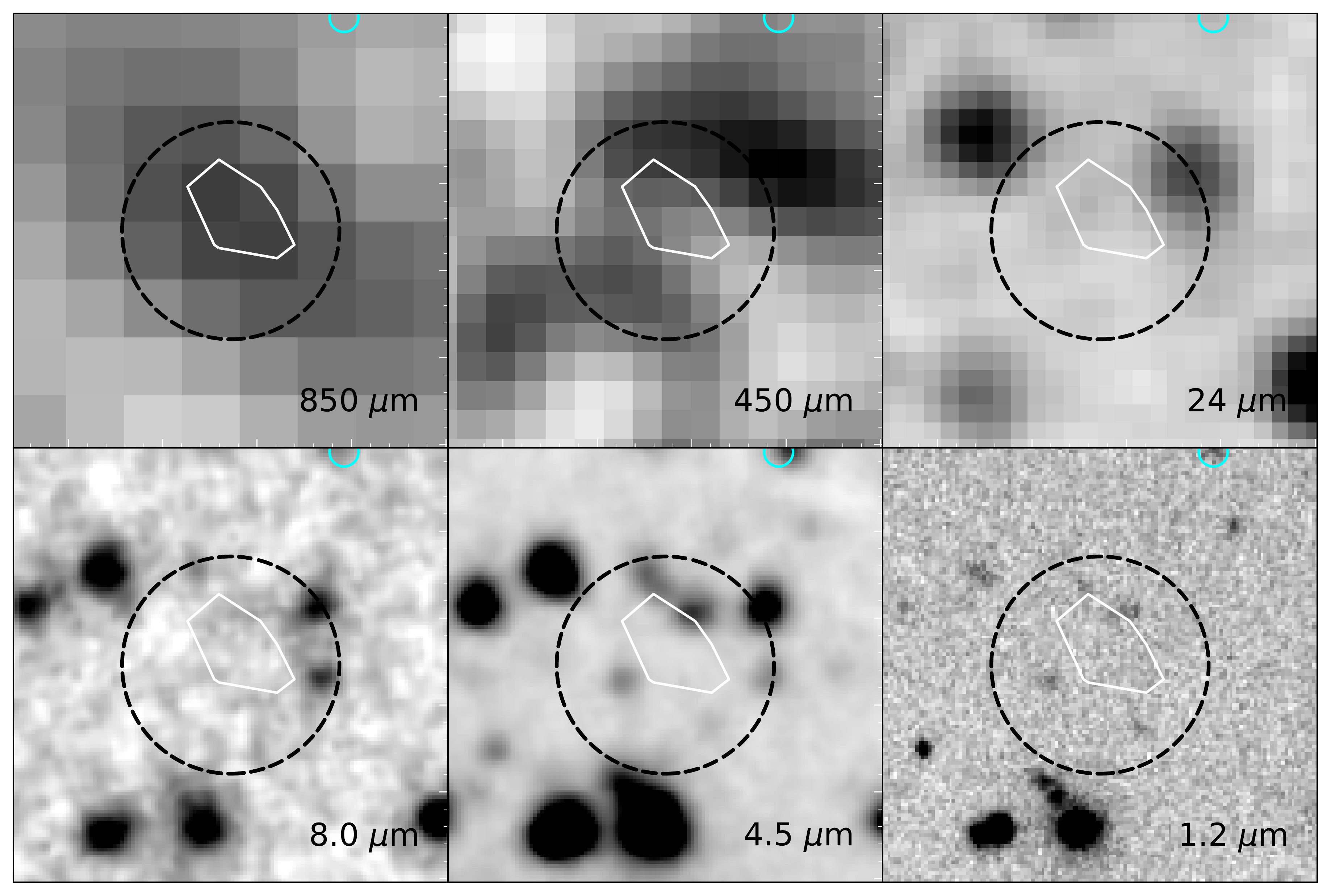}
\includegraphics[width=.49\textwidth]{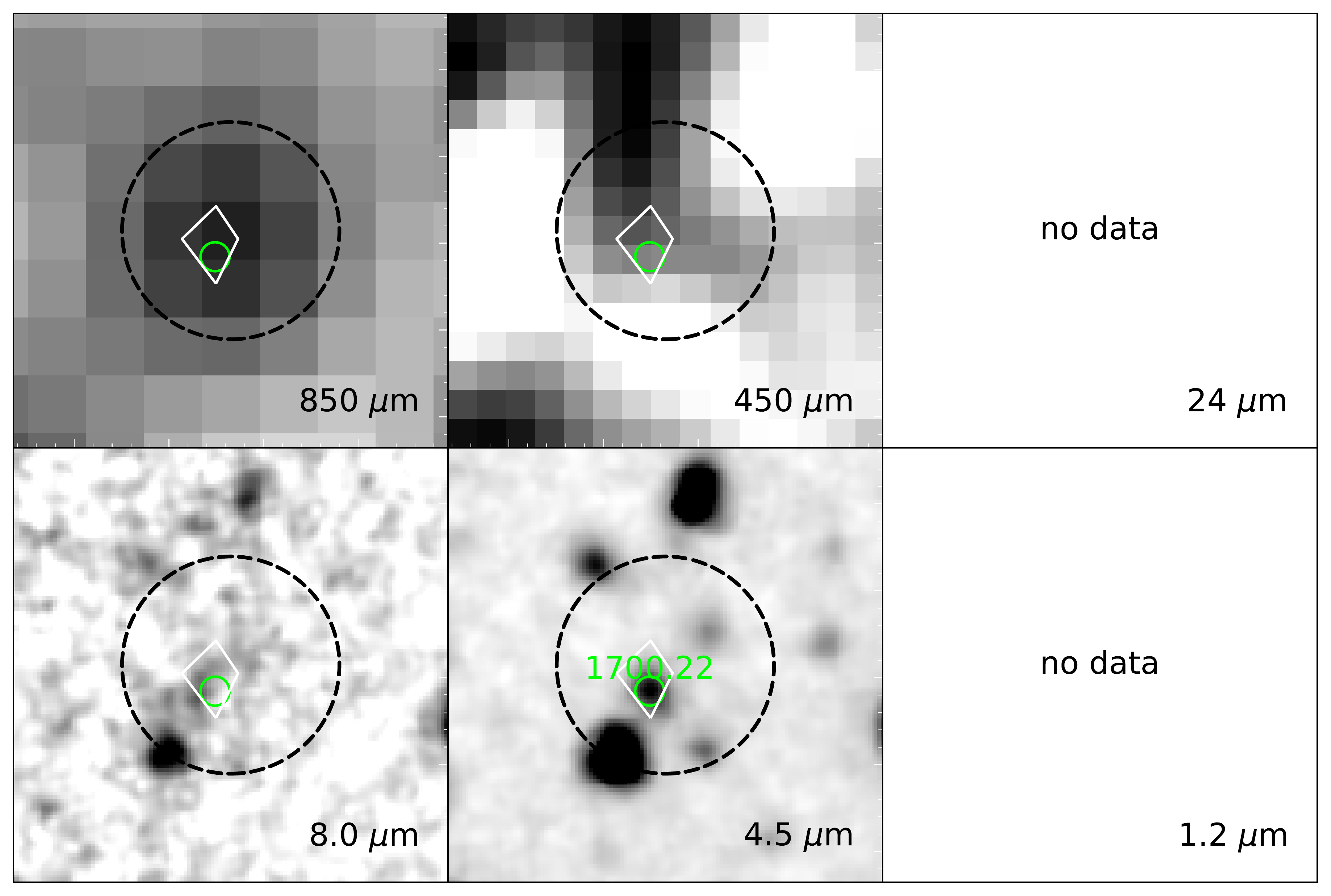}
\emph{1700.23} \hspace{3in} \emph{1700.24}\\
\includegraphics[width=.49\textwidth]{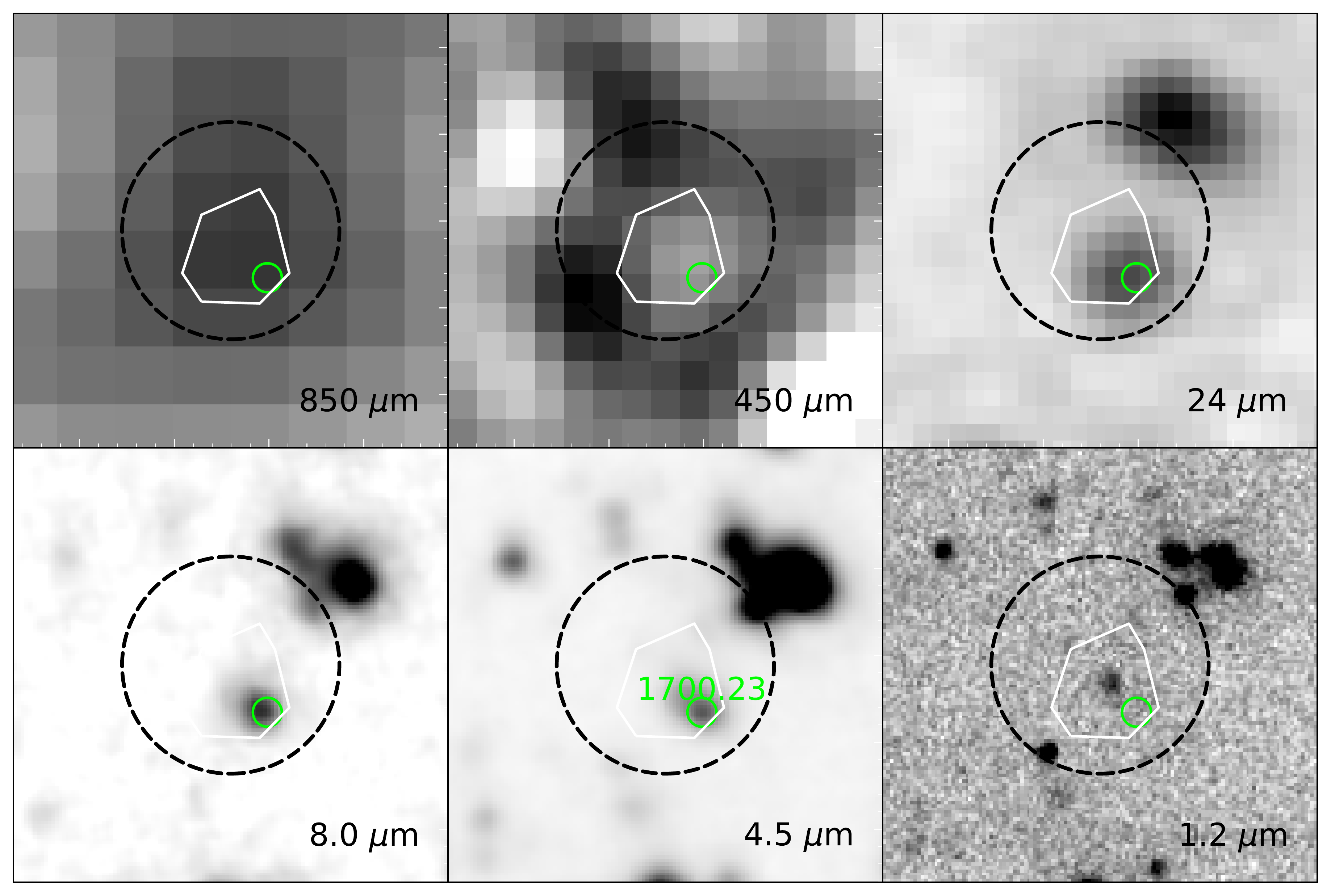}
\includegraphics[width=.49\textwidth]{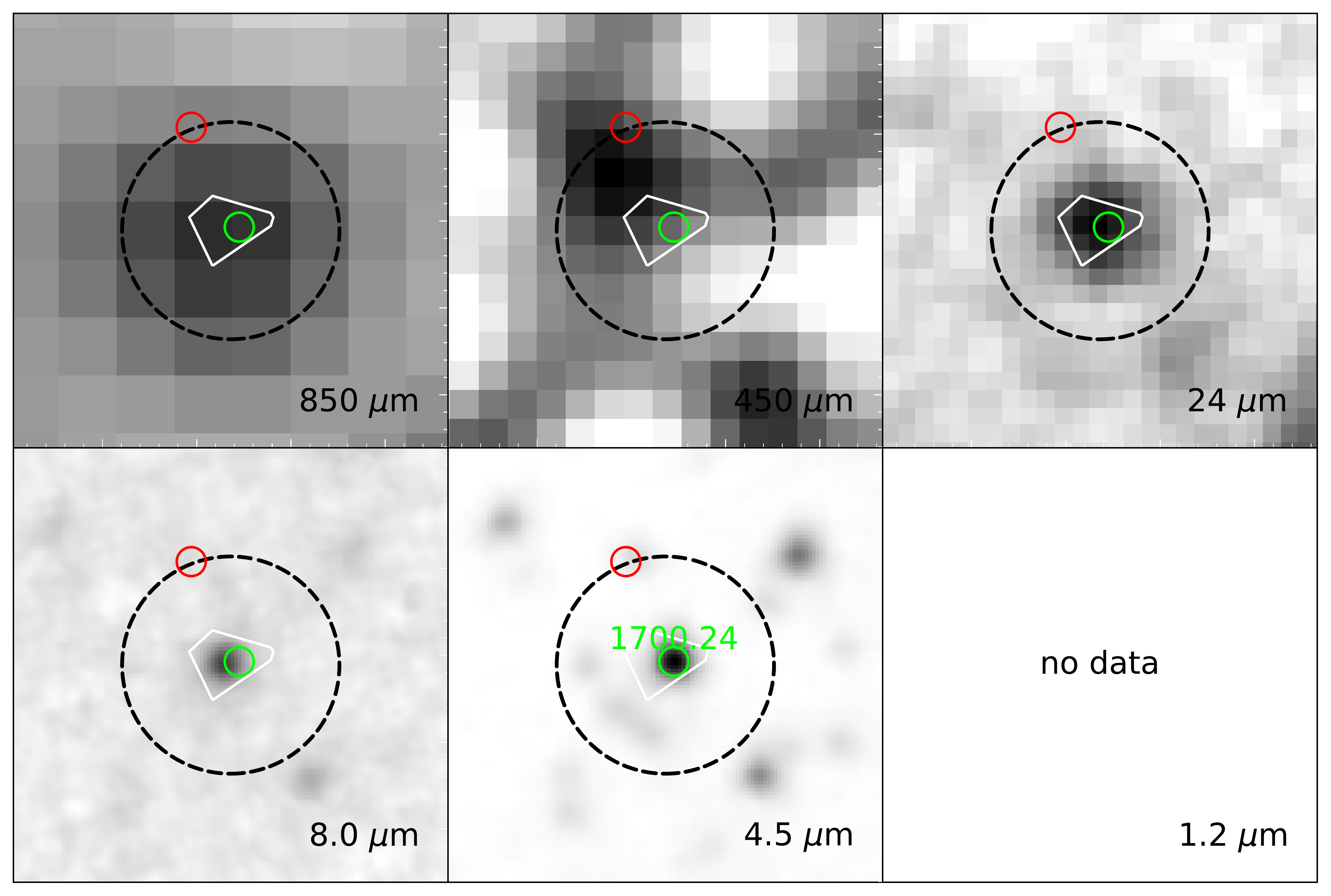}

\end{figure*}
\noindent
\begin{figure*}
\emph{1700.25} \hspace{3in} \emph{1700.26}\\
\includegraphics[width=.49\textwidth]{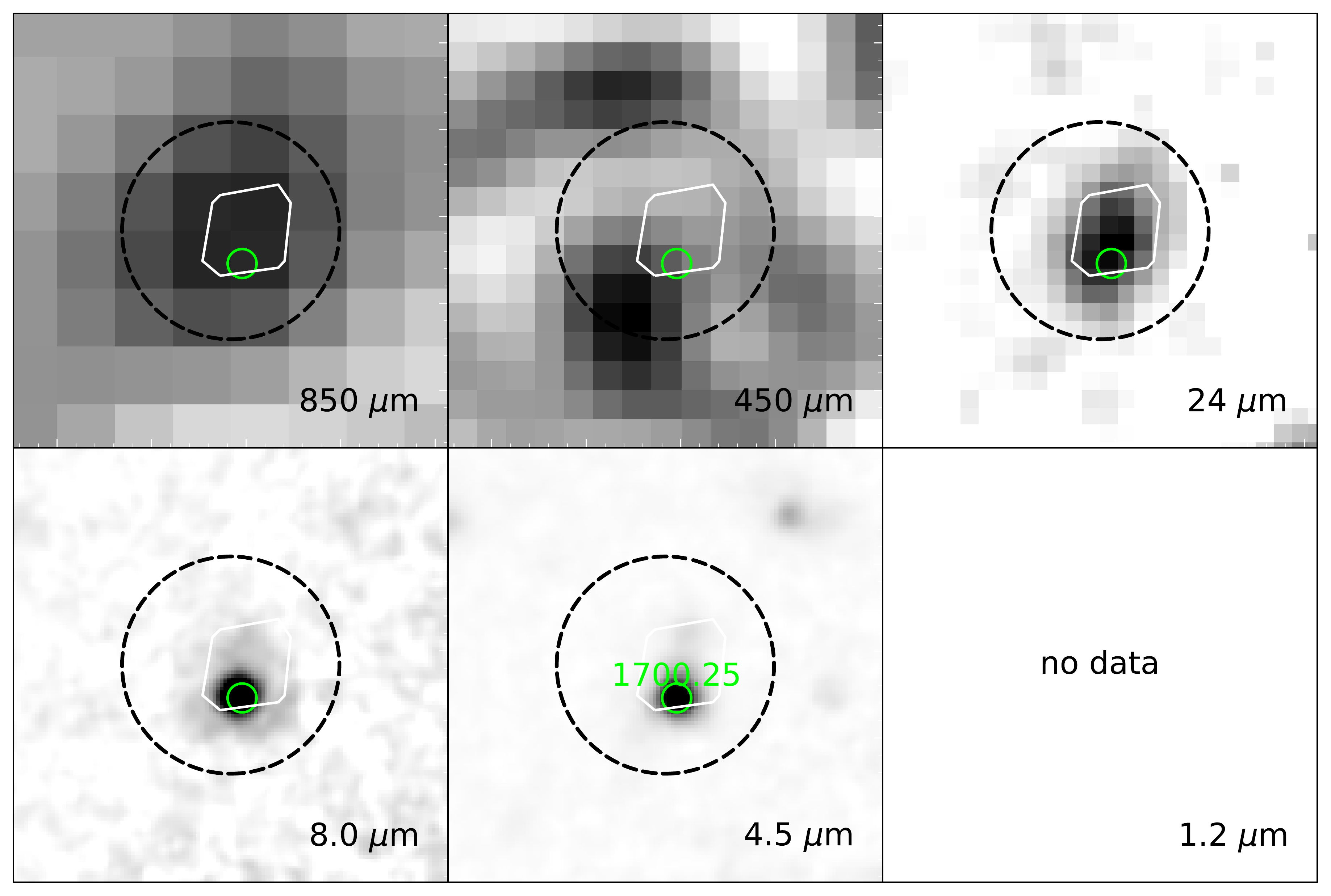}
\includegraphics[width=.49\textwidth]{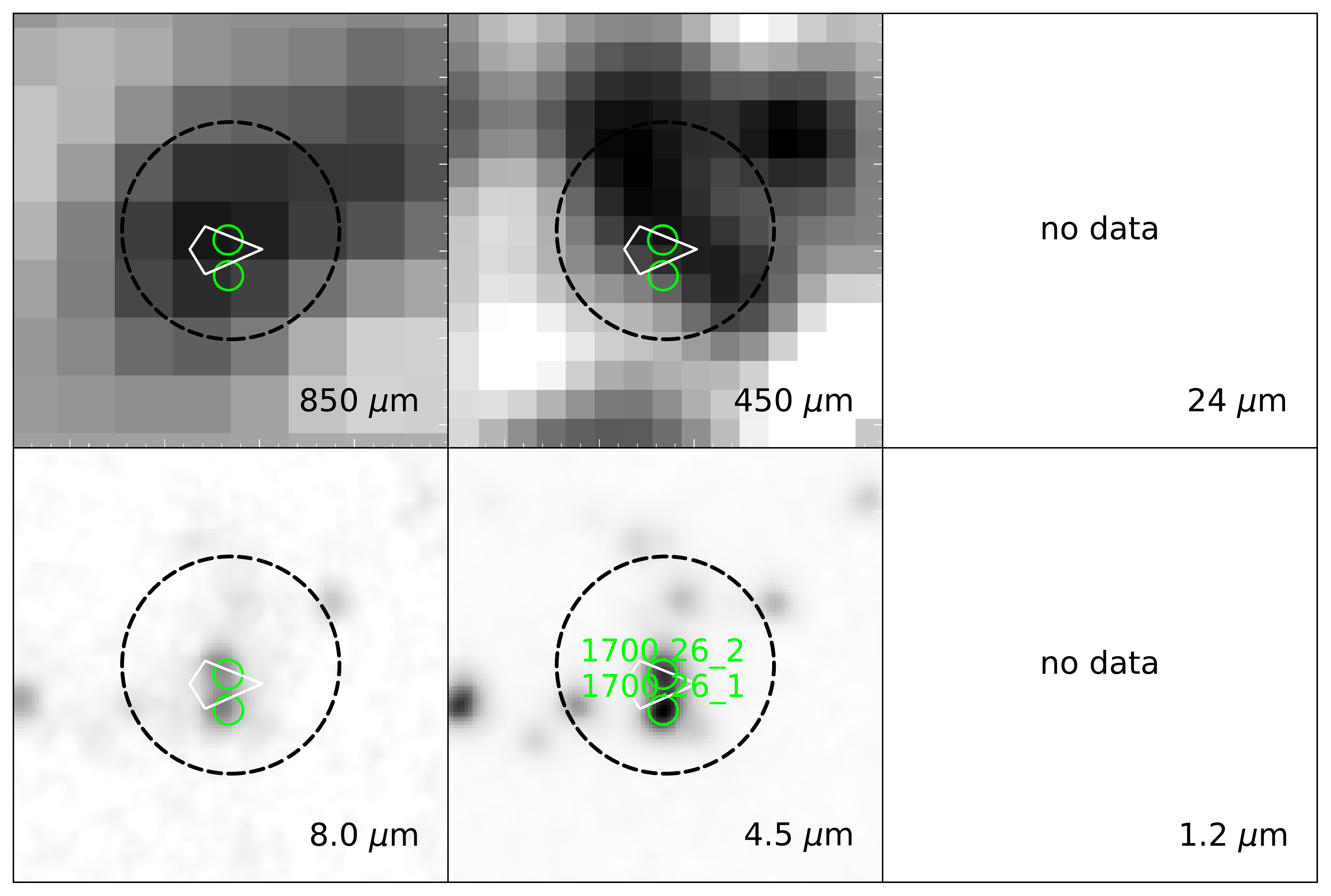}
\end{figure*}
\begin{center}
\begin{figure*}
\hspace{-3.45in}\emph{1700.27}\\
\hspace{-3.45in}\includegraphics[width=.49\textwidth]{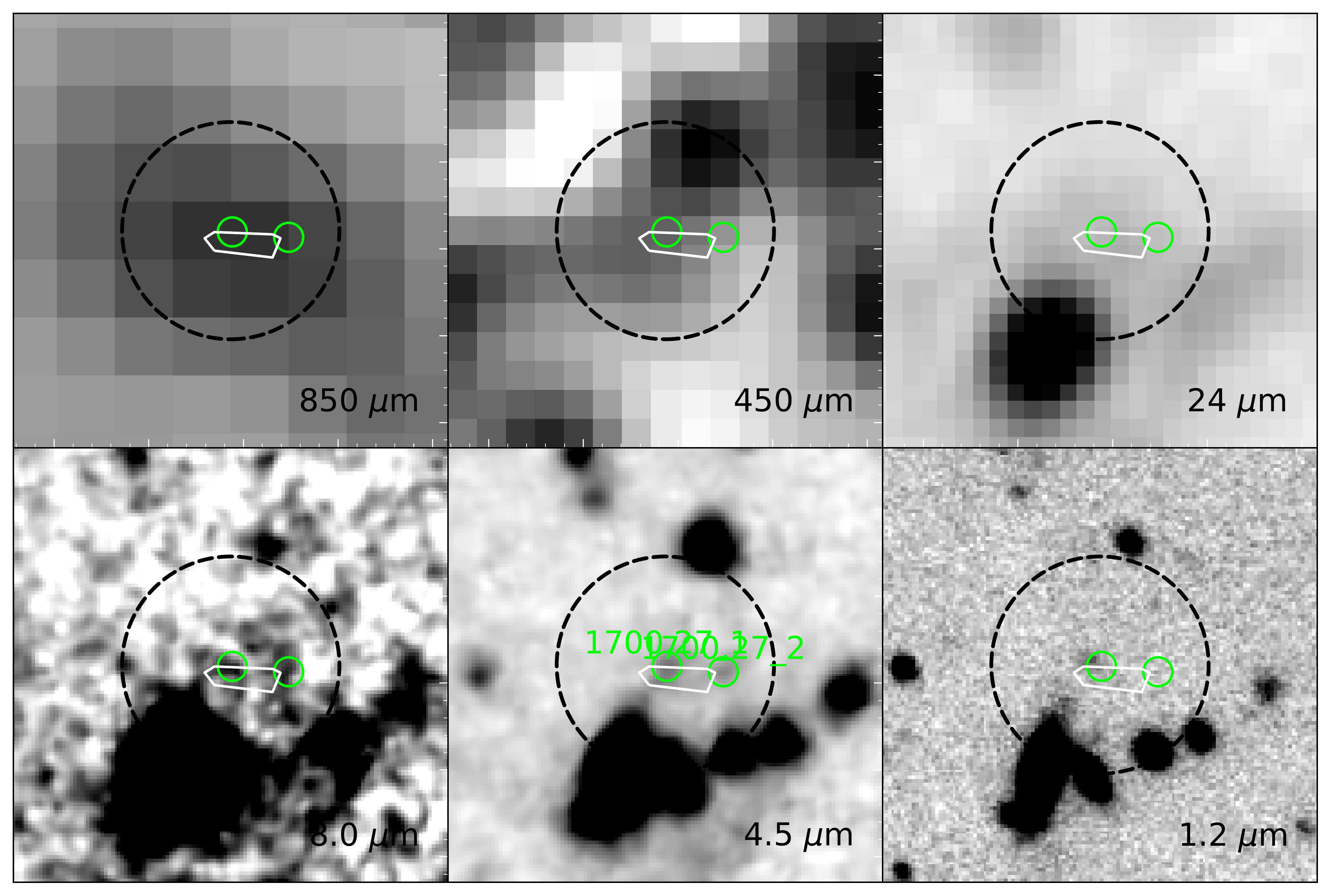}
\end{figure*}
\end{center}


\bsp	
\label{lastpage}
\end{document}